\documentclass[11pt]{article}

\usepackage{jheppub}
\usepackage{amsmath}
\usepackage{bm}
\usepackage{slashed}
\usepackage{graphicx}

\newlength{\dummysp}
\newcommand{\beq}{\begin{eqnarray}}
\newcommand{\eeq}{\end{eqnarray}}

\newcommand{\gappeq}{\mathrel{\rlap {\raise.5ex\hbox{$>$}}
{\lower.5ex\hbox{$\sim$}}}}
\newcommand{\lappeq}{\mathrel{\rlap{\raise.5ex\hbox{$<$}}
{\lower.5ex\hbox{$\sim$}}}}

\newcommand{\ben}{\begin{enumerate}}
\newcommand{\een}{\end{enumerate}}

\newcommand{\bit}{\begin{itemize}}
\newcommand{\eit}{\end{itemize}}

\def\[{\left [}
\def\]{\right ]}
\def\({\left (}
\def\){\right )}
\def\R{{\mathbb R}}
\def\S{{\mathbb S}}
\def\Z{{\mathbb Z}}
\pdfoutput=1

\title{Deconfinement  in $\mathbf{{\cal N}=1}$ super Yang-Mills theory on $\mathbf{\R^3 \times \S^1}$ via
 dual-Coulomb gas and ``affine" XY-model}

\author[a]{Mohamed M. Anber,} \author[a,b]{Scott Collier,} \author[a]{Erich Poppitz,} \author[a]{Seth Strimas-Mackey,}
 \author[a]{Brett Teeple}

\affiliation[a]{Department of Physics,   University of Toronto, 
Toronto, ON M5S 1A7, Canada}
\affiliation[b]{Physics Department, McGill University, 3600 rue University, Montreal, QC H3A 2T8, Canada}
\emailAdd{manber@physics.utoronto.ca}\emailAdd{scollier@physics.mcgill.ca}\emailAdd{poppitz@physics.utoronto.ca}\emailAdd{seth.strimas.mackey@mail.utoronto.ca}\emailAdd{bteeple@physics.utoronto.ca}    
    
\abstract
{We study finite-temperature   ${\cal{N}}=1$ $SU(2)$ super Yang-Mills theory, compactified on a spatial circle of size $L$ with supersymmetric boundary conditions. In the semiclassical small-$L$ regime,  a deconfinement transition occurs at $T_c \ll 1/L$. The transition is due to a competition between  non-perturbative topological ``molecules"---magnetic and neutral bion-instantons---and electrically charged  $W$-bosons and superpartners. Compared to deconfinement in non-supersymmetric QCD(adj) \cite{Anber:2011gn}, the novelty  is the relevance of the light modulus scalar field. It mediates interactions between neutral bions (and $W$-bosons),  serves as an order parameter for the $\Z_2^{(L)}$ center symmetry associated with the non-thermal circle, and explicitly breaks the electric-magnetic (Kramers-Wannier) duality enjoyed by  non-supersymmetric QCD(adj) near $T_c$.  We show that deconfinement can be studied using an effective two-dimensional gas of electric and magnetic charges with (dual) Coulomb and Aharonov-Bohm interactions, or, equivalently, via an XY-spin model with a symmetry-breaking perturbation, where each system  couples to the scalar field. To study the realization of the discrete $R$-symmetry and the   $\Z_2^{(\beta)}$ thermal  and $\Z_2^{(L)}$   non-thermal   center symmetries,  we perform Monte Carlo simulations of both systems. The dual-Coulomb gas simulations are a novel way to analyze deconfinement and provide a new venue to study the phase structure of a class of two-dimensional condensed matter models that can be mapped into dual-Coulomb gases. Our results indicate a continuous deconfinement transition, with $\Z_2^{(L)}$ remaining unbroken at the transition. Thus, the SYM transition appears similar to the one in $SU(2)$ QCD(adj)   \cite{Anber:2011gn}  and is also likely to be characterized by   continuously varying critical exponents.}
  
\begin{document}
\maketitle
\flushbottom

\section{Introduction}
\label{Introduction}

The dynamics of ${\cal{N}}=1$ super Yang-Mills theory (SYM) on $\R^3 \times \S^1$  with supersymmetry-preserving boundary conditions   has been a recurring topic of interest since the late 1990's \cite{Seiberg:1996nz,Aharony:1997bx}. Within the supersymmetric realm, studying SYM (also with matter fields) on this geometry offers a smooth interpolation between three and four dimensional theories and the associated rich web of dualities  \cite{Aharony:2013dha}. Furthermore, at small $\S^1$-size  $L$,   SYM becomes weakly coupled. Semiclassical monopole-instanton calculations are reliable, offering a window of calculability and an explicit check of exact results in supersymmetric theories \cite{Davies:1999uw, Davies:2000nw}.

Surprisingly,  the insight gained from SYM  theory at small-$L$ is relevant not only for studying supersymmetric  gauge theories. It turns out  that, when properly understood, lessons from the small-$L$ SYM dynamics apply to a wide class of nonsupersymmetric gauge theories, as we now briefly review. 
The   exact results of \cite{Seiberg:1996nz,Aharony:1997bx} and the calculation of the mass gap  of \cite{Davies:1999uw, Davies:2000nw} relied heavily on supersymmetry. The underlying mechanism of confinement  and mass gap generation is, however, more general, and its relevance for non-supersymmetric theories (QCD(adj), i.e., YM with adjoint Weyl fermions) was only elucidated in \cite{Unsal:2007jx}. It was shown  that the mass gap is due to the proliferation  in the vacuum of a certain kind of   topological ``molecules" (correlated instanton--anti-instanton events) carrying zero topological charge  and two units of magnetic charge (Ref.~\cite{Unsal:2007jx} called these doubly-charged molecules ``magnetic bions";  we will continue using this name). The mass gap   arises from Debye screening in the magnetic bion gas. Thus, magnetic bion-induced \cite{Unsal:2007jx} screening is a generalization of  the three-dimensional Polyakov mechanism \cite{Polyakov:1976fu} of confinement to a locally four-dimensional theory. In the case of  SYM, it was further argued in \cite{Poppitz:2011wy} that   there is a continuous connection between the monopole-instantons (the constituents of the magnetic bions on $\R^3 \times \S^1$) and the monopole  and dyon particles whose condensation leads to confinement in the four-dimensional Seiberg-Witten theory \cite{Seiberg:1994rs}. In  the case of non-supersymmetric QCD(adj),  various zero-temperature aspects of magnetic bions have been studied in  detail   \cite{Anber:2011de,Argyres:2012ka}.

The lessons learned from SYM on $\R^3 \times \S^1$ did not end with the observation of the role of magnetic bions. More recently, Ref.~\cite{Poppitz:2011wy} identified another kind of topological molecule, called the ``neutral" (or ``center-stabilizing") bion. These are also correlated instanton--anti-instanton events, which now  carry zero topological charge and zero magnetic charge, but two units of scalar charge (which has also been called ``electric" \cite{Diakonov:2007nv} or ``dilaton" \cite{Harvey:1996ur} charge;  both terms have admittedly limited utility). The existence of   neutral bions is significantly more subtle to establish than that of magnetic bions---it requires invoking either supersymmetry, the Bogomolny--Zinn-Justin prescription (see \cite{Poppitz:2011wy} for references),  or  the more phenomenological ``excluded volume" argument \cite{Shuryak:2013tka}. While these arguments are suggestive, ultimately, the necessity of including neutral bion contributions in the path integral can be traced to the divergence of the perturbative series  \cite{Argyres:2012vv,Argyres:2012ka}.\footnote{The  ongoing studies  of ``resurgence"---a generalization of Borel resummation---in field theories in various dimensions are shedding further light on the role  of these molecules and other path integral saddle points, showing  a fascinating interplay  between perturbative and nonperturbative contributions at weak coupling (this is a currently active area of research, see, e.g., the  recent work~\cite{Marino:2012zq,Dunne:2012ae,Cherman:2013yfa,Basar:2013eka,Aniceto:2013fka}  and references therein). }

Magnetic and neutral bions are expected to play an important role in the nonperturbative dynamics of various gauge theories.  A recent example is the  argument that they are  relevant to studies of the thermal deconfinement phase transition in pure YM theory, via the idea of ``continuity".\footnote{\label{two}In \cite{Poppitz:2012sw}, following \cite{Unsal:2010qh}, neutral bions were found to play a role appropriate to their  designation as ``center-stabilizing bions". Upon introducing a small supersymmetry-breaking gaugino mass $m$ in SYM on $\R^3 \times \S^1$, it was found that at small $L$, there is a center-symmetry breaking phase transition, occurring as the (not small) dimensionless parameter $m \over L^2 \Lambda^3$  is varied ($\Lambda$ is the strong scale of SYM). Center-symmetry breaking is  driven by a competition between the nonperturbative contributions of neutral bions and monopole-instantons as well as perturbative contributions.  This small-$L$, small-$m$ quantum phase transition was conjectured to be continuously related to the thermal deconfinement phase transition in pure YM theory, upon increase of $m$. A significant amount of evidence has been accumulated in favour of this conjecture, both in the agreement of the order of the transition for various gauge groups with lattice results \cite{Poppitz:2012nz,Poppitz:2013zqa} and in the $\theta$-angle dependence of the deconfinement temperature \cite{Unsal:2012zj,Bonati:2013tt,Anber:2013sga,D'Elia:2013eua}, see also \cite{Parnachev:2008fy,Thomas:2011ee}.} 
Further, magnetic bions have been shown to be crucial for understanding the deconfinement transition in non-supersymmetric  QCD(adj)  \cite{Anber:2011gn}. We will describe this in some detail as it is important to the subject of this paper.
 
\section{Summary and outline} 
We  consider SYM as a member of a broader framework of non-supersymmetric theories---recall that SYM belongs to the class of QCD(adj) theories, as supersymmetry is automatic for $n_a = 1$ massless Weyl adjoint fermions. Studying the dynamics of these theories on $\R^3 \times \S^1_L$ in the semiclassical small-$L$ regime  is of interest, as it offers a rare opportunity where both the perturbative and nonperturbative effects  that play a role in the physics of confinement and deconfinement are under theoretical control.

In this paper, we focus on the thermal dynamics of SYM. The magnetic and neutral bions described above are expected to be important here as well (let us stress that the current $\R^2 \times \S^1_\beta \times \S^1_L$ setup, with $\S^1_{\beta/L}$ denoting the thermal and non-thermal circles, is different from the one of \cite{Poppitz:2012sw}, where neutral bions also appear; see footnote \ref{two}). 
The thermal deconfinement transition of QCD(adj) at small-$L$, for $n_a>1$, was studied in \cite{Anber:2011gn}.\footnote{Following the earlier work of \cite{Dunne:2000vp,Simic:2010sv}. See \cite{Anber:2013xfa} for a review and  some new results and derivations.} 
It was shown there  that the physics near $T_c$ is  described by a two dimensional Coulomb gas of electric and magnetic charges,
or by an equivalent ``affine" XY-spin model with external-field perturbations. For QCD(adj) with $SU(2)$ and $SU(3)$ gauge groups, the near-critical theory was shown to exhibit an ``emergent" electric-magnetic (Kramers-Wannier) duality, whose presence is intimately tied to the structure of magnetic bions. For $SU(2)$, various properties of the transition could be studied analytically, showing that the transition is continuous and has continuously varying critical exponents, determined by the $W$-boson and magnetic bion fugacities.\footnote{The correlation length critical exponent is $\nu^{-1} = 8 \pi \sqrt{y  \tilde{y} }$, where $y$ and $\tilde{y}$ are the (small) cutoff-scale $W$-boson and magnetic bion  fugacities (we correct for a factor of two in Eq.~(1.3) in \cite{Anber:2011gn}).} For higher-rank groups, $SU(N_c\ge3)$,  the weak-coupling renormalization group methods fail and a  numerical study is required. The Monte Carlo simulation  of \cite{Anber:2012ig} found that  the $SU(3)$ QCD(adj) transition  is 
first order\footnote{An analytical understanding of this is still lacking. The near-$T_c$ Coulomb gas for  $SU(3)$ has a description in terms of a novel self-dual sine-Gordon model associated with the affine roots of the Lie algebra, see Eq.~(3.2) of   \cite{Anber:2012ig}. Any progress on generalizing the studies of  \cite{Lecheminant:2002va,Lecheminant:2006hj} to this case would be of interest.} and is accompanied by a restoration of the discrete anomaly-free chiral symmetry, while the continuous chiral symmetry remains unbroken. This ordering of continuous chiral and deconfinement transitions is the same as in  lattice studies at ``infinite"-$L$, see \cite{Cossu:2008wh} for a recent reference. Recently, a first order phase transition has also been found via simulations   for the $SU(4)/\Z_4$ QCD(adj) theory \cite{Chau}.

Coming back to the $n_a=1$ case of SYM,   we will show that supersymmetry  brings in an extra complication\footnote{While supersymmetric theories are usually more susceptible to theoretical analysis, the present case of SYM vs. QCD(adj) on small $\S^1_L$ is an example where the non-supersymmetric theory is much simpler to study.} related to the role of neutral bions and the associated light modulus scalar, whose mass is effectively protected by supersymmetry at the temperatures of interest. However, despite these subtleties we find that the qualitative properties of the transition remain unchanged with respect to $n_a > 1$ theories.

This paper is organized as follows. We first derive the theory appropriate to the study of the $SU(2)$ SYM thermal transition at small-$L$, in section \ref{Theory and Formulation}. We  show that it is a modification of the  electric-magnetic Coulomb gas, hereafter to be called the ``dual-Coulomb" gas,  and the XY-spin model of \cite{Anber:2011gn}, which requires them to couple to the light modulus field. This coupling explicitly breaks the electric-magnetic duality near criticality. The computation of the effective potential for the  Wilson lines in SYM on $\S^1_\beta \times \S^1_L$, which is crucial for both qualitative and quantitative understanding of the dynamics, is  given in appendix \ref{Computation of the effective potential}. 

We then perform Monte Carlo studies of both the appropriate dual-Coulomb gas (section \ref{Simulations of the Coulomb gas}) and  spin model (section \ref{XY model as a Coulomb gas}). We note that, to the best of our knowledge, simulations of an electric-magnetic Coulomb gas with Aharonov-Bohm interactions have not been performed before. Details relevant to the simulations, notably   the algorithm that was developed, are given in appendices \ref{The discrete dual Sine-Gordon model}, \ref{Monte-Carlo procedure for simulating the double Coulomb gas}.

The results of the Monte Carlo  studies of both systems describing the deconfinement transition are consistent with each other. They reveal a continuous transition, similar to the one in $SU(2)$ QCD(adj) with $n_a>1$. We find that at $T_c$, the discrete chiral $R$-symmetry is restored  and the $Z_2^{(\beta)}$ thermal center symmetry is broken, as appropriate for a deconfinement transition. The $Z_2^{(L)}$ non-thermal center symmetry remains manifest at $T_c$ (it can, however, be argued to break at much higher temperatures, where our effective description is not valid). At the end of the paper, in section \ref{extrapolationsection}, we  also offer some qualitative arguments in favor of the similarity between the transitions in SYM and QCD(adj). 
 We have not calculated critical exponents  but the similarity of our results to $SU(2)$ QCD(adj) with $n_a > 1$ leads us to  conjecture that the transition in $SU(2)$ SYM also has continuously varying critical exponents.
 
We end by noting that there have been attempts to describe the thermal dynamics of QCD on $\R^3 \times \S^1_\beta$ as a plasma of electric and magnetic charges, notably the  study of a  classical   three dimensional plasma of electric and magnetic charges in \cite{Liao:2006ry}. Although this description is based on a largely qualitative picture, its predictions appear to agree with a number of lattice and theoretical results.  In this respect,  our electric-magnetic ``dual-Coulomb" gas description, which is under analytical control at small-$L$ in the  $\R^2 \times \S^1_\beta \times \S^1_L$ geometry, may be the best one can hope for---if one is after a theoretically reliable description of the thermal transition in a small-$N_c$ asymptotically free gauge theory. Our studies can thus be thought of as offering some support to the picture advocated in  \cite{Liao:2006ry}.

\section{Dynamics of SYM on $\mathbf{\R^2 \times \S^1_\beta \times \S^1_L}$}
\label{Theory and Formulation}

This section contains both review of known material (pertaining to the $T=0$ dynamics) as well as some new results. While we have attempted a self-contained description of the $T=0$ case, we assume the reader's familiarity with classical solutions, the semiclassical approximation,  't Hooft vertices and monopole operators. The reader familiar with SYM on $\R^3\times \S^1_L$ is advised to proceed to sections \ref{Perturbative dynamics at non-zero temperature} and \ref{nonperturbative dynamics at finite temperature}, which contain new results pertaining to the perturbative and nonperturbative dynamics at $T>0$. 
\subsection{Perturbative dynamics}
\label{Perturbative dynamics}

We consider ${\cal N}=1$ supersymmetric gauge theory: $SU(2)$ Yang-Mills theory along with a single massless adjoint Weyl fermion, or gaugino. The action of the theory on $\mathbb R^{2}\times \mathbb S^1_{ L}\times \mathbb S^1_{\mathbb \beta}$ is given by
\begin{eqnarray}
S=\int_{\mathbb R^{2}\times \mathbb S^1_{ L}\times \mathbb S^1_{\mathbb \beta}}\mbox{tr}\left[\frac{1}{2g^2}F_{MN}F^{MN}+\frac{2i}{g^2}\bar \lambda \bar \sigma^M D_M\lambda \right]\,,
\label{the total Lagrangian of the theory}
\end{eqnarray}
where $F^{MN}=F^{MN\,a}T^a$ is the field strength tensor, $D_M$ is the covariant derivative, $\lambda=\lambda^a T^a$ is the Weyl fermion, $\sigma_M=\left(i ,\vec \tau\right)$, $\bar\sigma_M=\left(-i,\vec \tau\right)$, $i$ and $\vec \tau$ are respectively the identity (times $i$) and Pauli matrices.  The Lie generators for the $SU(2)$ gauge theory at hand are  $T^a=\tau^a/2$, where $\tau^a$ are the Pauli matrices. Throughout this work we follow the following index convention: the Latin letters $M,N$ run over $0,1,2,3$, the Greek letters $\mu,\nu$ run over $0,1,2$, while the Latin letters $i,j$ run over $1,2$. We also use $\vec x$ to denote two-dimensional vectors in the $1-2$ plane. The components $0$ and $3$ respectively denote the compact temporal and spatial dimensions. Thus, $x^3\equiv x^3+L$, where $L$ is the circumference of the $\mathbb S^1_{L}$  which is taken to be a spatial circle, while  $x^0\equiv x^0+\beta$, where $\beta$ is the circumference of the thermal $\mathbb S^1_{\beta}$ circle and $\beta=1/T$ is the inverse temperature. Both the gauge field and Weyl fermion satisfy periodic boundary conditions around the spatial circle $\mathbb S^1_L$. On the other hand, these fields, respectively, satisfy periodic and anti-periodic boundary conditions around the thermal   $\mathbb S^1_\beta$. 

The quantum theory has a dynamical strong scale $\Lambda_{\mbox{\scriptsize QCD}}$ such that, to one-loop order, we have
\begin{eqnarray}
g^2(\mu)=\frac{8\pi^2}{\beta_0}\frac{1}{\log\left(\mu/\Lambda_{\mbox{\scriptsize QCD}}\right)}\,,
\label{QCD beta function}
\end{eqnarray}
where $\mu$ is the renormalization scale  and $\beta_0=6$. In this paper, we will consider small spatial circle circumference compared to the dynamical strong scale, i.e. $L\Lambda_{\mbox{\scriptsize QCD}}\ll1$. The fermion sector in (\ref{the total Lagrangian of the theory}) enjoys a classical $U(1)$ chiral symmetry. At zero temperature,  the BPST instantons  break this symmetry to its discrete subgroup $\mathbb Z_4$. The $\mathbb Z_2$ subgroup of $\mathbb Z_4$ is the fermion number modulo $2$ which is preserved as long as Lorentz symmetry is respected. Thus,  we are left with the genuine discrete chiral (or $R$-) symmetry  $\mathbb Z_2^{(R)}$, which can be potentially broken at low temperatures. 

Now, we turn to  important gauge invariant variables which are the Polyakov/Wilson loops, or holonomies, defined as the path ordered exponent in the $\mathbb S_L^1$ and $\mathbb S^1_\beta$ directions:
\begin{eqnarray}
\nonumber
\Omega_L(\vec x, x^0)&=&Pe^{i\oint_{\mathbb S_L^1}A^3(\vec x, x^0,x^3 )}\,,\\
\Omega_\beta(\vec x, x^3)&=&Pe^{i \oint_{\mathbb S_\beta^1}A^0(\vec x, x^0,x^3 )}\,,
\end{eqnarray}
where $\vec x \in \mathbb R^2$. The holonomies transform under $x$-dependent gauge transformations as $\Omega_L (\vec x, x^0) \rightarrow U_L^{-1}(\vec x, x^0)\Omega_L (\vec x, x^0) U_L(\vec x, x^0)$, and $\Omega_\beta (\vec x, x^3) \rightarrow U_\beta^{-1}(\vec x, x^3)\Omega_\beta (\vec x, x^3) U_\beta(\vec x, x^3)$. Hence, the eigenvalues of $\Omega_L$ and $\Omega_\beta$ are gauge invariant and in turn the gauge invariant trace of the holonomies $\mbox{tr}\Omega_L$ and $\mbox{tr}\Omega_\beta$ work as order parameters for the two global center symmetries $\mathbb Z_2^{(L)}$, and $\mathbb Z_2^{(\beta)}$. Under the action of these symmetries we have $\mbox{tr}\Omega_L \stackrel{\mathbb Z_2^{(L)}}{\rightarrow} e^{i\pi k}\mbox{tr}\Omega_L$ and $\mbox{tr}\Omega_\beta \stackrel{\mathbb Z_2^{(\beta)}}{\rightarrow} e^{i\pi k} \mbox{tr}\Omega_\beta$, with $k=1,2$. In the following, we explore the fate of $\mathbb Z_2^{(L,\;\beta)}$ after carefully examining the ingredients of the theory at zero and finite temperatures.

\subsubsection{Perturbative dynamics at zero temperature}
\label{Perturbative dynamics at zero temperature}

First, we consider the theory at zero temperature. Taking $L\Lambda_{\mbox{\scriptsize QCD}}\ll1$, we can perform reliable loop calculations to integrate out the heavy Kaluza-Klein modes along $\mathbb S^1_L$. This  amounts to finding the Coleman-Weinberg effective potential $V_{\mbox{\scriptsize eff, pert}}(\Omega_L)$. However, supersymmetry guarantees the vanishing of this perturbative  potential to all orders in perturbation theory. In other words, the supersymmetric theory possesses a moduli space of vacua. Despite this, there are non-perturbative contributions to the effective potential, $V_{\mbox{\scriptsize eff, non-pert}}(\Omega_L)$, which we calculate in the next section. At zero temperature, the leading two-derivative terms in the action read:
\begin{eqnarray}
S_{\beta\rightarrow \infty}&=&\int_{\mathbb R^{2}\times \mathbb S^1_\beta}\frac{L}{g^2}\mbox{tr}\left[-\frac{1}{2}F_{\mu\nu}F^{\mu\nu}+\left(D_\mu A_3\right)^2+2i \bar \lambda \left(\bar \sigma^\mu D_\mu \lambda-i\bar\sigma_3\left[A_3,\lambda\right]\right) \right]\,.
\end{eqnarray}
Thus, the gauge field component along the  $\mathbb S^1_L$ direction is an adjoint compact Higgs field $A_3$. As we shall see in the following section, due 
 to non-perturbative contributions, the minimum of the effective potential is located at
\begin{eqnarray}
\langle \Omega_L \rangle=\mbox{diag}\left(e^{\frac{i\pi}{2}}, e^{-\frac{i\pi}{2}}\right)\,,\quad \mbox{or}\quad \langle A_3 \rangle\equiv \langle A_3^3 \rangle\; T^3=\frac{\pi}{L}\; T^3\,. 
\label{vacuum of the theory}
\end{eqnarray}
Since $\mbox{tr}\langle \Omega_L \rangle=0$, the $\mathbb Z_2^{(L)}$ center symmetry is preserved at zero temperature. In the vacuum (\ref{vacuum of the theory}), the $SU(2)$ gauge theory is broken by the Higgs field $\langle A_3 \rangle$ spontaneously to $U(1)$. Because the Higgs field is in the adjoint representation, two of its three components (in the color space) are eaten by the gauge fields which become massive with mass $M_W=\frac{\pi}{L}$. The heavy gauge fields are the $W$-bosons.  What remains is the $A_3$ component along the third color direction (the Cartan subalgebra direction), $A_3^3$ (which acquires an exponentially small mass $\sim e^{-\frac{8\pi^2}{g^2}}$, as we will see in the next section, due to non-perturbative effects). In addition, the color space components of the fermions $\lambda$ that do not commute with $\langle A_3\rangle$ acquire a mass $M_W=\frac{\pi}{L}$. These are the $W$-boson superpartners or winos. Therefore, at zero temperature and for distances $\gg L$, the $3$D low energy Euclidean action is
\begin{eqnarray}
S_{\mbox{\scriptsize low energy}\,,\,\,\beta \rightarrow \infty}&=&L\int_{\mathbb R^2\times \mathbb S^1_\beta}\frac{1}{4g^2}\left(F^3_{\mu\nu}\right)^2+\frac{1}{2g^2}\left(\partial_\mu A_3^3\right)^2+\frac{i}{g^2}\bar \lambda^3 \bar \sigma^\mu \partial_\mu \lambda^3\,.
\label{final perturbative zero T action}
\end{eqnarray}
Hence, the effective perturbative theory describes the free  fields $A^3_\mu$, $A_3^3$, and $\lambda^3$.  Because of the absence of any coupling between the electromagnetic field and fermions, the coupling constant $g$ ceases to run at energy scales $<1/L$. Since we consider $L\Lambda_{\mbox{\scriptsize QCD}}\ll1$, the frozen value of the coupling, $g^2(\mu \sim 1/L)$ is small because of asymptotic freedom (\ref{QCD beta function}).

 Now, using the abelian duality 
 \begin{equation}
 \label{sigmadef}
 \epsilon_{\mu\nu\lambda}\partial_\lambda \sigma=\frac{4\pi L}{g^2}\;F^3_{\mu\nu} ,
 \end{equation}
 we can map the gauge field to a spin-zero dual photon $\sigma$ (recall that, 
in the $SU(2)$ theory, charge and flux quantization imply that the dual photon is a compact scalar field with period $2 \pi$, see, e.g., \cite{Anber:2011gn}). 
We also define the field $\phi$ as
\begin{eqnarray}
\label{phidef}
\phi \equiv \frac{4\pi L}{g^2}\;A_3^3-\frac{4\pi^2}{g^2}\,,
\end{eqnarray}
such that the point $\phi=0$ corresponds to having an exact $\mathbb Z_2^{(L)}$ center symmetry. Then,  the free bosonic part of (\ref{final perturbative zero T action}) reads
\begin{eqnarray}
{\cal L}_{\mbox{\scriptsize free bosonic}\,,\,\,\beta \rightarrow \infty}=\frac{1}{2}\frac{g^2}{(4\pi)^2 L}\left[\left(\partial_\mu \sigma\right)^2+\left(\partial_\mu \phi\right)^2 \right]\,.
\label{phikin}
\end{eqnarray}
One can assemble the bosonic kinetic terms (\ref{phikin}) and the gaugino kinetic term from (\ref{final perturbative zero T action}) using the K\" ahler potential
\begin{eqnarray}
\label{kahler}
K=\frac{g^2}{2(4\pi)^2 L} \; \mathbf{B}^\dagger \mathbf{B}\,,
\end{eqnarray} 
where $\mathbf{B}$ is a dimensionless chiral superfield whose lowest component is $\phi-i\sigma$.\footnote{For a full component expression, albeit in a different $\sigma$-matrix basis, see the appendix of \cite{Intriligator:2013lca}. For completeness, we also note that there are corrections to the dependence of $\mathbf{B}$ on $\phi$ due to the non-canceling fermion and boson determinants in the monopole-instanton backgrounds, which we henceforth ignore, see appendix A of \cite{Poppitz:2012sw} for details.}

\subsubsection{Perturbative dynamics at finite temperature}
\label{Perturbative dynamics at non-zero temperature}

Unlike the zero temperature case, the perturbative Coleman-Weinberg potential does not vanish at finite temperatures since the boundary conditions along $\S^1_\beta$ break supersymmetry. We can calculate the one-loop induced effective potential $V_{\mbox{\scriptsize eff, pert}}\left(A_0^3,A_3^3\right)$ by starting from the full Lagrangian (\ref{the total Lagrangian of the theory}) and integrating out the heavy Kaluza-Klein modes along the two cycles of the flat torus $\mathbb T^2=\mathbb S^1_L\times \mathbb S^1_\beta$. This boils down to evaluating the determinant of the operator $D_M^2$ on $\mathbb R^2\times \mathbb S^1_L\times \mathbb S^1_\beta$ that is common to the gauge and gaugino fluctuations in the background of constant holonomies along the cycles of the  torus $\mathbb T^2$. The calculation is explained in detail  in appendix \ref{Computation of the effective potential}. The holonomies along the two circles should  commute in order to  minimize the classical action; in other words,  we  take only the Cartan subalgebra components, $A_0^3$ and $A_3^3$, to be nonzero.  Therefore,  the effective action at finite temperature reads
\begin{eqnarray}
S_{\mbox{\scriptsize low energy},\,\beta }&=&L\int_{\mathbb R^2\times \mathbb S^1_\beta}\frac{1}{4g^2}\left(F^3_{\mu\nu}\right)^2+V_{\mbox{\scriptsize eff, pert}}\left(A_0^3,A_3^3\right)+\frac{1}{2}\left(\partial_\mu A_3^3\right)^2+i\bar \lambda^3 \bar \sigma^\mu \partial_\mu \lambda^3,
\end{eqnarray}
where
\begin{eqnarray}
\label{full potential on T2} 
&&V_{\mbox{\scriptsize eff, pert}}\left(A_0^3,A_3^3\right)
= \\
&&-2\sum_{n=-\infty}^\infty \sum_{p=1}^\infty \left[1-(-1)^p\right]\frac{e^{-2\pi p\left|\frac{n\beta}{L}+\frac{ \beta A_3^3}{2\pi}\right|}}{\pi \beta^3 L p^3}\left(1+2\pi p \left|\frac{n\beta }{L}+\frac{\beta A_0^3}{2\pi}  \right| \right)\cos\left(p\beta A_0^3\right)\,.
\nonumber
\end{eqnarray}
This potential is periodic in $A_0^3$ and $A_3^3$, with respective periods $2\pi/L$ and $2\pi/\beta$, and it encodes the information about the two center symmetries $\mathbb Z_2^{(L)}$ and $\mathbb Z_2^{(\beta)}$. At low temperatures, below the deconfinement phase transition, our simulations indicate that both $\mathbb Z_2^{(L)}$ and $\mathbb Z_2^{(\beta)}$ are respected. The center symmetry $\mathbb Z_2^{(L)}$ remains unbroken even for temperatures  above the deconfinement temperature. However, for temperatures $\gtrsim M_W$, where $M_W={\pi \over L}$ is the $W$-boson mass, an analysis of the potential (\ref{full potential on T2}) shows that the $\mathbb Z_2^{(L)}$ symmetry breaks.  

In the deconfinement transition analysis, we will be interested only in  temperatures much smaller than $M_W$, i.e.,  $LT\ll 1$. Hence,  we can retain only the $p=1$ term in (\ref{full potential on T2}), as higher-$p$ terms have extra  exponential suppression $\sim e^{- p {2 \pi  \over LT}}$. Thus, for the purpose of the deconfinement transition studies, we have to a very good accuracy
\begin{eqnarray}
V_{\mbox{\scriptsize eff, pert}}\left(A_0^3,A_3^3\right)
\cong-\frac{4}{\pi \beta^3 L}\sum_{n=-\infty}^{\infty}e^{-2\pi \left|\frac{n\beta}{L}+\frac{ \beta A_3^3}{2\pi}\right|}\left(1+2\pi  \left|\frac{n\beta }{L}+\frac{\beta A_0^3}{2\pi}  \right| \right)\cos\left(\beta A_0^3\right)\,.
\label{one loop approximate perturbative potential}
\end{eqnarray}

Let us end  the discussion of our finite-$T$ perturbative analysis with a few comments.\footnote{\label{zeromodeatt} For completeness, note that the effective potential (\ref{one loop approximate perturbative potential}) only receives contributions from loops of the heavy Kaluza-Klein modes on $\mathbb T^2$. In the presence of relevant interactions, loops of the zero-modes can also contribute; these will be discussed in section \ref{nonperturbative dynamics at finite temperature}.} Most importantly, note that at  temperatures $TL\ll 1$, so that (\ref{one loop approximate perturbative potential}) can be used, the $W$-boson (and superpartner)  loop contribution to the mass of the Coulomb-branch modulus $\phi$ (or $A_3^3$, recall (\ref{phidef})) is exponentially suppressed, $\sim e^{- {M_W \over T}} = e^{- {\pi \over LT}}$, as is clear from the first term in $V_{\mbox{\scriptsize eff, pert}}$.
If $\phi$ was an exact modulus in the $T=0$ theory (as  in theories with extended supersymmetry), the finite-$T$ loop contribution to its effective potential would largely determine the behavior of the theory; see the finite-$T$ study of ${\cal{N}}=2$ SYM theory  \cite{Paik:2009iz}. 
Here, however, there is a more interesting story to tell, thanks to the existence of nonperturbative saddle points to the Yang-Mills equations, which lift the Coulomb branch  of ${\cal{N}}=1$ SYM theory. The $\phi$ ``modulus" is not massless, but acquires an exponentially small mass ($\sim e^{- {4 \pi^2 \over g^2}}$), which dominates over the  thermal contribution  (\ref{one loop approximate perturbative potential}) ($\sim e^{- {\pi \over LT}}$) for sufficiently low temperatures. 

Finally, let us point out that the analysis of the thermal effect  of the heavy $W$-bosons and gauginos, summarized in 
Eq.~(\ref{one loop approximate perturbative potential}), will be quite a bit more subtle than simply minimizing the perturbative potential  (or free energy) $V_{\mbox{\scriptsize eff, pert}}$. This is because of the existence of  a nonperturbative sector of the theory carrying magnetic charges. The description of the thermal dynamics will necessarily involve the coupling of the thermally excited  electric charges (the $W$-bosons and superpartners) to the nonperturbative magnetic sector. For now, we only note that Eq.~(\ref{one loop approximate perturbative potential}), properly interpreted, will still play an important role.

\subsection{Nonperturbative dynamics}
\label{Nonperturbative dynamics}

The Lagrangian (\ref{the total Lagrangian of the theory}) admits monopole-instantons as well as magnetic  and neutral bion-instantons.  These are, respectively, self-dual and non-self-dual nonperturbative solutions to the equations of motion of finite action. According to the path integral formalism, one has to include both perturbative and non-perturbative contributions to the path integral. In this section, we briefly review these solutions and elucidate the method we follow to include their effects in the partition function.

\subsubsection{Nonperturbative dynamics at zero temperature}
\label{{Nonperturbative dynamics at zero temperature}}

In addition to the perturbative excitations described above, the equations of motion of  (\ref{the total Lagrangian of the theory})  admit various nonperturbative solutions. On $\mathbb R^3 \times \mathbb S^1_L$, the simplest of these objects are BPS monopole-instantons allowed by the non-trivial homotopy $\pi_2(SU(2)/U(1))= \pi_1(U(1)) = \mathbb Z$. Due to the compact nature of the $x^3$ coordinate, the equations of motion also admit another class of solutions known as {\it twisted}  or Kaluza-Klein (KK) monopole-instantons.\footnote{These were most clearly identified in Ref.~\cite{Lee:1997vp} using D-branes. Within field theory, Ref.~\cite{Kraan:1998pm,Lee:1998bb} identified them as constituents of periodic instantons with non-vanishing holonomy (``calorons").} Monopole-instantons are (anti)self-dual ``particle-like" objects localized in space and time (spacetime events), have internal structure and are sources of a long range field, thanks to the unbroken $U(1)$. 

In the following, we use the semi-classical approximation to include the effect of these  objects in the sum over histories. In this approximation, the instantons are dilute enough such that their internal structure does not play any role, and therefore we can replace the non-abelian field solution with an effective abelian one.
The abelian field of a single BPS ($\overline{\mbox{BPS}}$) monopole-instanton localized at the origin $x_0=x_1=x_2=0$, in the stringy gauge, is given by
\begin{eqnarray}
\nonumber
A_0^{3,\mbox{\scriptsize BPS}, \overline{\mbox{\scriptsize BPS}}}&=&\mp\frac{x_1}{r\left(r+x_2\right)}\,,\\
\nonumber
A_1^{3,\mbox{\scriptsize BPS}, \overline{\mbox{\scriptsize BPS}}}&=& \pm\frac{x_0}{r\left(r+x_2\right)}\,,\\
\nonumber
A_2^{3,\mbox{\scriptsize BPS}, \overline{\mbox{\scriptsize BPS}}}&=&0\,,\\
A_3^{3,\mbox{\scriptsize BPS}, \overline{\mbox{\scriptsize BPS}}}&=&\frac{\pi}{L}-\frac{1}{r}\,,
\label{BPS monopole field}
\end{eqnarray}
where the superscript $3$ indicates that it is only the third color component that has an abelian field;  $x_{1,2}$ and $x_0$ are respectively the spatial and Euclidean time coordinates, and $r=\sqrt{x_0^2+x_1^2+x_2^2}$ is the Euclidean spherical-polar radius. The abelian field of a KK ($\overline{\mbox{KK}}$) monopole reads

\begin{eqnarray}
\nonumber
A_0^{3,\mbox{\scriptsize KK}, \overline{\mbox{\scriptsize KK}}}&=&\pm\frac{x_1}{r\left(r+x_2\right)}\,,\\
\nonumber
A_1^{3,\mbox{\scriptsize KK}, \overline{\mbox{\scriptsize KK}}}&=& \mp\frac{x_0}{r\left(r+x_2\right)}\,,\\
\nonumber
A_2^{3,\mbox{\scriptsize KK}, \overline{\mbox{\scriptsize KK}}}&=&0\,,\\
A_3^{3,\mbox{\scriptsize KK}, \overline{\mbox{\scriptsize KK}}}&=&\frac{\pi}{L}+\frac{1}{r}\,.
\label{KK monopole field}
\end{eqnarray}
The monopole-instantons carry magnetic charge, $Q_m$, which is defined as the surface integral of the monopole-instanton magnetic field over a 2-sphere:
\begin{eqnarray}
\int_{S_\infty^2} dS_\mu B^{3}_\mu=4\pi Q_m\,,
\label{definition of magnetic charge}
\end{eqnarray}
where $B^3_\mu=\epsilon_{\mu\nu\alpha}\partial_\nu A^3_\alpha=Q_m\frac{x_\mu}{r^3}$. In addition to the magnetic force these instantons can experience, they also attract or repel each other due to the exchange of a long range scalar field, the $A^3_3$ component. The fact that the monopole-instantons can exchange a long range scalar is attributed to the nature of these instantons which saturate the BPS bound, thanks to supersymmetry. This effect is absent in QCD(adj) with $n_a>1$ adjoint fermions.  Further, these objects carry fractional topological charge $Q_T$ defined as
\begin{eqnarray}
Q_T=\frac{1}{32\pi^2}\int_{\mathbb R^3 \times \mathbb S^1_L} F^{a}_{MN}F^a_{MN}\,.
\label{definition of topological charge}
\end{eqnarray}
Hence, we can use (\ref{definition of magnetic charge}) and (\ref{definition of topological charge}) to read off the  magnetic and topological charges, $(Q_m,Q_T)$, of the various self-dual solutions as follows:
\begin{eqnarray}
\mbox{BPS}\,\, (+1,1/2)\,,\quad \overline{\mbox{BPS}}\,\, (-1,-1/2)\,,\quad \mbox{KK}\,\, (-1,1/2)\,,\quad \overline{\mbox{KK}}\,\, (+1,-1/2)\,.
\end{eqnarray}

Due to the presence of the gaugino, the Nye-Singer index theorem \cite{Nye:2000eg} (a physicist's derivation appears in \cite{Poppitz:2008hr}) implies that each of the monopole-instantons has two fermionic zero modes. Using the fields $\phi$  instead of $A_3^3$, see (\ref{phidef}), and $\sigma$ instead of $A_\mu^3$, see (\ref{sigmadef}),  the BPS ($\overline{\mbox{BPS}}$)  and KK ($\overline{\mbox{KK}}$) monopole-instantons along with the attached zero modes can be represented using the following 't Hooft vertices:\footnote{To explain the appearance of $\bar\lambda\bar \lambda$ in the self-dual---and thus chiral---BPS and KK 't Hooft vertices, we note that the fermion component of the chiral superfield $\mathbf{B}$ is 
$\sim  {L \over g^2} \theta \sigma^3 \bar\lambda  $, in the  notation of \cite{Wess:1992cp}.}
\begin{eqnarray}
\label{fermionvertices}{\cal M}_{\mbox{\scriptsize BPS}}=e^{-\frac{4\pi^2}{g^2}}e^{-\phi+i\sigma}\bar\lambda \bar\lambda\,, \quad {\cal M}_{\mbox{\scriptsize KK}}=e^{-\frac{4\pi^2}{g^2}}e^{\phi-i\sigma}\bar\lambda \bar\lambda\,,\\
\overline{{\cal M}}_{\overline{\mbox{\scriptsize BPS}}}=e^{-\frac{4\pi^2}{g^2}}e^{-\phi-i\sigma} \lambda \lambda\,, \quad \overline{{\cal M}}_{\overline{\mbox{\scriptsize KK}}}=e^{-\frac{4\pi^2}{g^2}}e^{\phi+i\sigma} \lambda  \lambda\,.\nonumber
\end{eqnarray}
The exponential factors $e^{\pm \phi \pm i \sigma}$ appearing in (\ref{fermionvertices}) encode the long-range fields  (\ref{BPS monopole field}) and (\ref{KK monopole field}) of the relevant solutions. Insertions of the local operators (\ref{fermionvertices}) in the partition function corresponds to the contribution of a monopole-instanton along with its fermion zero modes and long-range fields.
Since the monopole-instantons are attached to fermionic zero modes, including only the objects (\ref{fermionvertices}) in the partition function   will not alter the vacuum structure of the theory, as a potential for $\sigma$ or $\phi$ will not be generated and  the dual photon $\sigma$ will remain massless. We also note that invariance of the 't Hooft vertices (\ref{fermionvertices}) under the anomaly-free chiral $\Z_2^{(R)}$ symmetry implies that the dual photon shifts, $\sigma  \stackrel{\mathbb Z_2^{(R)}}{\rightarrow} \sigma + \pi$.
This intertwining of topological shift symmetries and anomaly-free chiral symmetries is common  for theories on $\R^3 \times \S^1$ \cite{Aharony:1997bx}. 

In addition to the self-dual solutions, the Yang-Mills vacuum allows for non self-dual objects to form. These are composite molecules (correlated events) which consist of various combinations of  monopole-instantons with magnetic and topological charges, $(Q_m,Q_T)$, and 't Hooft vertices, as follows:
\begin{eqnarray}
\begin{array}{cccc}
\mbox{molecule}\quad & \mbox{composite} & (Q_m,Q_T) & \mbox{amplitude}\\\\
\mbox{neutral bion}\quad & {\cal M}_{\mbox{\scriptsize BPS}}\overline{{\cal M}}_{\overline{\mbox{\scriptsize BPS}}} & (0,0) & e^{-\frac{8\pi^2}{g^2}}e^{-2\phi}\\\\
\overline{\mbox{neutral bion}} \quad & {\cal M}_{\mbox{\scriptsize KK}}\overline{{\cal M}}_{\overline{\mbox{\scriptsize KK}}} & (0,0) & e^{-\frac{8\pi^2}{g^2}}e^{2\phi}\\\\
\mbox{magnetic bion}\quad & {\cal M}_{\mbox{\scriptsize BPS}}\overline{{\cal M}}_{\overline{\mbox{\scriptsize KK}}} & (+2,0) & e^{-\frac{8\pi^2}{g^2}}e^{2i\sigma}\\\\
\overline{\mbox{magnetic bion}}\quad & {\cal M}_{\mbox{\scriptsize KK}}\overline{{\cal M}}_{\overline{\mbox{\scriptsize BPS}}} & (-2,0) & e^{-\frac{8\pi^2}{g^2}}e^{-2i\sigma}
\end{array}\,.
\label{the molecules}
\end{eqnarray}
Magnetic bions, with 't Hooft vertices $~e^{ \pm 2 i \sigma}$, are stable objects in the sense that the repulsion force (due to magnetic and scalar interactions) between the constituent monopole-instantons is balanced by the attraction due to fermionic zero-modes hopping between the constituent monopoles. The far field of a magnetic bion ($\overline{\mbox{bion}}$) at distances $\gg r_*={4 \pi L \over g^2}$ (where $r_*$ is the magnetic bion radius, see \cite{Poppitz:2012sw} and footnote \ref{tzerobionsize})  can be obtained by directly summing the contributions from (\ref{BPS monopole field}) and  (\ref{KK monopole field}), which we present here for a later reference (using the original, rather than dual  (\ref{phidef}, \ref{sigmadef}) variables): 
\begin{eqnarray}
\nonumber
A_0^{3\,\mbox{\scriptsize bion}}&=&-2\frac{x_1}{r\left(r+x_2\right)}\,,\\
\nonumber
A_1^{3\,\mbox{\scriptsize bion}}&=& 2\frac{x_0}{r\left(r+x_2\right)}\,,\\
\nonumber
A_2^{3\,\mbox{\scriptsize bion}}&=&0\,,\\
A_3^{3\,\mbox{\scriptsize bion}}&=&\frac{2\pi}{L}\,.
\label{magnetic bion field}
\end{eqnarray}
Hence, the magnetic bions do not source a macroscopic scalar field like their monopole constituents. 
Furthermore, because the magnetic bions do not have fermionic zero modes, the inclusion of these objects in the path integral can dramatically change the nature of the vacuum: now the dual photon $\sigma$ acquires a mass and the theory confines the electric charges. 

On the other hand, the neutral bions,\footnote{We do not go into the details of the ``force" balance leading to the formation of the neutral bions---as opposed to the discussion of magnetic bions after Eq.~(\ref{the molecules}), see also footnote \ref{tzerobionsize}. This is because all forces between their constituents are attractive and special attention is needed to argue that there is a stable ``molecule", see \cite{Poppitz:2011wy, Argyres:2012ka}.}  which source a long-distance scalar field (their 't Hooft vertex, see Eq.~(\ref{the molecules}), is $\sim e^{ \pm 2 \phi}$), generate a potential that stabilizes the center symmetry. Since the perturbative potential vanishes, it is only this neutral bion-induced potential that provides the required stabilization mechanism. The total potential $V_{\mbox{\scriptsize non-pert}}(\phi, \sigma)$ can be obtained by summing up the amplitudes in  (\ref{the molecules}). 

Alternatively, a neat way of obtaining this potential is via the use of supersymmetry \cite{Seiberg:1996nz,Aharony:1997bx,Davies:1999uw, Davies:2000nw}. The monopole-instantons (\ref{fermionvertices}) carry two fermion zero-modes and hence generate a superpotential, which is given by
\begin{eqnarray}
W_{\mathbb R^3 \times \mathbb S^1}=\frac{2 }{g^2 L^2}\; e^{-\frac{4\pi^2}{g^2}}\; \cosh  \mathbf{B}\,,
\end{eqnarray}
where the 4D gauge coupling is normalized at the scale $1/L$.
Then, the scalar potential can be easily found:
\begin{eqnarray}
V_{\mbox{\scriptsize non-pert}}(\phi, \sigma)=K^{-1}_{\mathbf{B}^\dagger \mathbf{B}}\left|\frac{\partial W}{\partial \mathbf{B}}\right|^2=\frac{64\pi^2  e^{-\frac{8\pi^2}{g^2}} }{g^6L^3}\left(\cosh 2\phi-\cos 2\sigma  \right)\,,
\end{eqnarray}
where $K_{\mathbf{B}^\dagger \mathbf{B}}$ is the mixed second derivative of K\" ahler potential (\ref{kahler}). Finally, the full zero-temperature Lagrangian reads
\begin{eqnarray}
\nonumber
{\cal L}_{\beta \rightarrow \infty}&=&\frac{1}{2}\frac{g^2(L)}{(4\pi)^2 L}\left[\left(\partial_\mu \sigma\right)^2+\left(\partial_\mu \phi\right)^2 \right]+i\frac{L}{g^2}\bar \lambda \sigma_\mu\partial_\mu \lambda +{\alpha \over g^4} e^{-\frac{4\pi^2}{g^2(L)}}\left[\left(e^{-\phi+i\sigma}+e^{\phi-i\sigma}\right)\bar\lambda \bar\lambda+\mbox{c.c.} \right]\\
&&+\frac{64\pi^2  e^{-\frac{8\pi^2}{g^2}} }{g^6L^3}\left(\cosh 2\phi-\cos 2\sigma  \right)\; ,
\label{full zero temperature Lagrangian}
\end{eqnarray}
where $\alpha$ is an inessential numerical factor (it can be determined by supersymmetry from the other terms). 

The conclusion for the zero-$T$ realization of $\Z_2^{(R)}$ and $\Z_2^{(L)}$---the two discrete global symmetries of (\ref{full zero temperature Lagrangian}), acting as $\phi  \stackrel{\mathbb Z_2^{(L)}}{\rightarrow} - \phi, \sigma  \stackrel{\mathbb Z_2^{(L)}}{\rightarrow} - \sigma$ and $\sigma  \stackrel{\mathbb Z_2^{(R)}}{\rightarrow} \sigma + \pi, \lambda  \stackrel{\mathbb Z_2^{(R)}}{\rightarrow} i \lambda$---is that the $\Z_2^{(L)}$ center symmetry is unbroken, as $\langle \phi \rangle = 0$ minimizes the potential in (\ref{full zero temperature Lagrangian}), while the discrete chiral $\Z_2^{(R)}$ symmetry is broken by the expectation value of the dual photon, whose potential is minimized at $ \langle \sigma \rangle = 0$ or $ \langle \sigma \rangle= \pi$. There is a mass gap (the dual photon, the $\phi$ and $\lambda$ fields have equal masses), and the theory confines electric charges. A calculation of the string tension can be found, e.g., in \cite{Anber:2013xfa}.

\subsubsection{Nonperturbative dynamics at finite temperature}
\label{nonperturbative dynamics at finite temperature}

In this section, we study the finite temperature version of the Lagrangian (\ref{full zero temperature Lagrangian}). First, we note that the dual photon mass squared, $m_{ph}^2$, and the $\phi$ mass squared, $m_\phi^2$  , are given by $m_{ph}^2=m_\phi^2 \sim L^{-2} e^{-\frac{8 \pi^2}{g^2}}$, with exponential accuracy. 

At any finite temperature $T$, the fermions  acquire a thermal mass $\sim T$. The light gauginos do not directly participate in the deconfinement transition since they do not carry electric or magnetic charges (while the effect of the heavy ones is similar to the $W$-bosons and will be accounted for).  However, the fermions  indirectly participate, as they facilitate  the formation of magnetic bions, as we now describe. 
The deconfinement temperature in SU(2) QCD(adj) \cite{Anber:2011gn} is   $T_c \simeq {g^2 \over 8 \pi L}$ (this estimate will be also seen to hold  for SYM---it  is the temperature where the electric $W$-boson and magnetic bion fugacities are of the same order, see section \ref{Simulations of the Coulomb gas}). Thus, $T_c$ is  smaller than the inverse bion radius, which for SYM (see \cite{Poppitz:2012sw} and comments after Eq.~(\ref{finiteTbion})) is ${1\over r_*} = {g^2\over 4 \pi L}$. However, because  $T_c r_* \sim 1/2$ one should consider   the finite-$T$ modification of the fermion propagator at  scales of order  the bion size and the generation of an attractive potential between the magnetic bion constituents.\footnote{In QCD(adj), there is  a ``parametric" suppression, $T_c r_* \sim 1/n_a$ by the number of adjoint flavors \cite{Anber:2011gn} (in reality, though, $n_a < 6$ for asymptotic freedom).} 
A  calculation of the temperature-dependent potential between the BPS and anti-KK monopole instantons (the magnetic bion constituents) induced by boson and fermion exchange shows that the fermions induce an attractive potential, whose properties near the minimum are not significantly altered for temperatures $T$ about several  times larger than $T_c$. 

\begin{figure}[h]
\centering 
\includegraphics[width=.9
\textwidth]{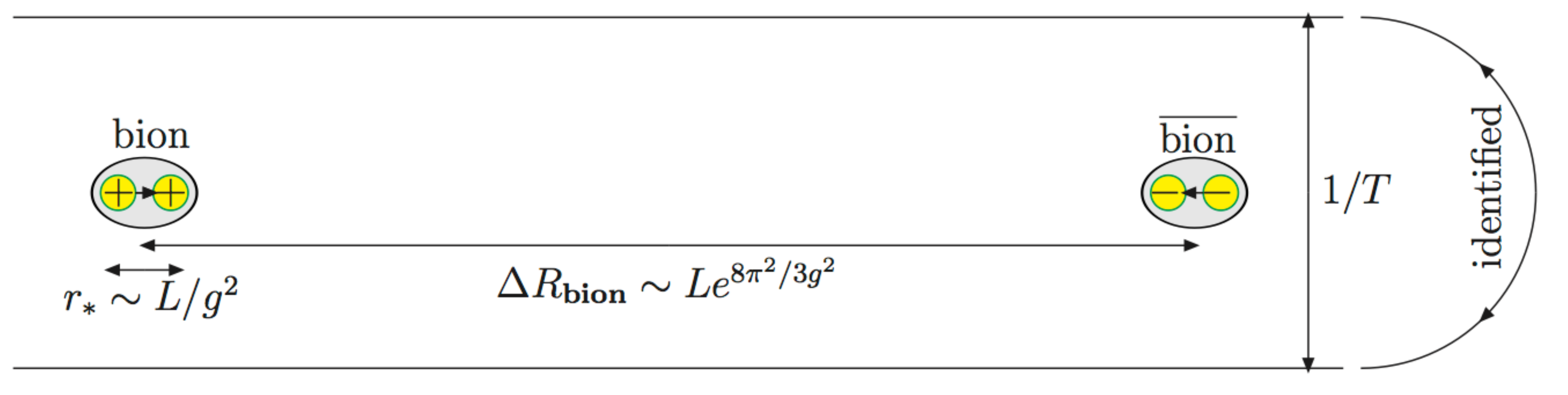}
\caption{\label{fig:scales} The scales in the finite-temperature problem. The bion size is much smaller than the inverse temperature, which, in turn, is much smaller than inter-bion separation, i.e., $ r_{*} <  \beta \ll {\Delta R}_{\rm bion} $. }
\end{figure}

We now present the results of such a calculation.
Recall that the size of the cores of the monopole-instantons is $L$ and that we are working at $TL\ll 1$, thus treating these objects as pointlike is justified (images have to be counted when computing their interactions, see Fig.~\ref{fig:scales}). 
Explicitly, the finite-$T$ potential between a BPS and an anti-KK monopole-instanton, as  a function of the $\R^2$ distance $x$ between their centers,  is given by:\footnote{\label{tzerobionsize}At $T=0$, the potential is  $V(r)_{BPS-\overline{KK}} = {8 \pi L \over g^2 r} + 2 \log {r\over L}$, where $r$ is the $\R^3$ distance between their centers, and has a minimum at $r_* = {4 \pi L \over g^2}$, as already stated. 
 }
 
\begin{eqnarray}
\label{finiteTbion}
 V(x)_{BPS-\overline{KK}} &=& {8 \pi L \over g^2} \left( {1 \over x} + 2 \sum\limits_{n=1}^\infty \left[ {1 \over \sqrt{ x^2 + n^2 T^{-2} } }- {T \over n}\right] \right)  \\
 &&
 - 2 \log\left[ {L^2 \over x^2} - 2  \sum\limits_{n=1}^\infty {  (-1)^{n+1} x L^2 \over (x^2 + n^2 T^{-2})^{3\over 2}  }\right] - 2\log {x\over L}~. \nonumber
 \end{eqnarray}
Before proceeding, let us elucidate the origin of each term in the above potential: \begin{enumerate}\item The first term is due to the magnetic and scalar repulsion between the BPS and anti-KK monopole-instantons. This term is just the  thermal version of the bosonic propagator which can be obtained by summing an infinite number of image charges along the $x_0$ direction. Since the potential between two monopoles depends on the distance between them, we can place the monopoles in the $x_0=0$ plane, which, after regulating the sum by subtracting the terms $\sum_{n=1}T/n$, results in the expression given above.
  
\item The second term in (\ref{finiteTbion}) represents the attraction between the monopole constituents due the the fermionic zero mode hopping. In order to obtain this term, one has to calculate the thermal fermionic propagator which is given by 
\begin{eqnarray}
\nonumber
S(\vec x, x_0)&=&T\sum_{p=-\infty}^{\infty} \int \frac{d^2k}{(2\pi)^2}e^{i\omega_p x_0+i\vec k\cdot\vec x}\frac{\sigma^3\omega_p-\vec \sigma\cdot \vec k}{\omega_p^2+k^2}\\
\nonumber
&=&\frac{-i\sigma^3}{4\pi}\int_0^\infty dk k J_0(k|\vec x|)\left[(1-\tilde n(k))e^{-kx_0}+\tilde n(k)e^{kx_0} \right]\\
&&-\frac{\vec \sigma \cdot \hat x}{4\pi}\int_0^\infty dk k J_1(k|\vec x|)\left[(1-\tilde n(k))e^{-kx_0}-\tilde n(k)e^{kx_0} \right]\,,
\label{fermionic propagator}
\end{eqnarray}
where $\tilde n(k)=\frac{1}{e^{k/T}+1}$, $\omega_p=(2p+1) \pi T$, and $\{\vec \sigma, \sigma^3\}$ are the Pauli matrices. Again, we can place the monopoles in the $x_0=0$ plane to find that the term in the second line of (\ref{fermionic propagator}) (the contribution of the $\sigma^3$) is zero after regularization. Then, expanding $\frac{1}{e^{k/T}+1}=\sum_{n=0}(-1)^ne^{-nk/T}$, and integrating, we find:
\begin{eqnarray}
\nonumber
S(\vec x, 0)=-\frac{\vec \sigma \cdot \hat x}{4\pi}\int_0^\infty dk k J_1(k|\vec x|)(1-2\tilde n(k))=-\frac{\vec \sigma \cdot \hat x}{4\pi}\left[\frac{1}{|\vec x|^2}-2\sum_{n=1}^\infty\frac{(-1)^{n+1}|\vec x|}{\left[|\vec x|^2+n^2/T^2\right]^{3/2}}\right]\,.\\
\end{eqnarray} 
\item Finally, the  last term in (\ref{finiteTbion}) is due to the measure of integration over the quasi-zero mode $x$.
 \end{enumerate}
 The potential (\ref{finiteTbion}) can be studied numerically as a function of $T$ and $x$ (for small $g^2$). It is then easily seen that while the potential is linearly increasing at sufficiently large distances,\footnote{Rather than logarithmically, as at $T=0$, since due to the nonzero Matsubara mass of the gauginos  the long-distance propagator at $x T \gg 1$ is dominated by the $p=0$ mode in the first line of (\ref{fermionic propagator}). On the other hand, the re-summed expression (\ref{finiteTbion}) is useful to study the potential near $x T \sim 1$. We thank Tin Sulejmanpa\v si\' c for   pointing out a flaw of the discussion of this point in an early version.}  its form near the minimum  at  $x \sim r_*$  is unaffected,  for temperatures $T$ up to several times $g^2 \over 8 \pi L$. This minimum  corresponds to the correlated instanton--anti-instanton tunneling events,.

We now  continue our study and ignore the fermions in (\ref{full zero temperature Lagrangian}). We break the bosonic part of the Lagrangian into two parts: ${\cal L}_{\beta \rightarrow\infty}={\cal L}_{\phi\,,\beta \rightarrow\infty}+{\cal L}_{\sigma\,,\beta \rightarrow\infty}$ such that
\begin{eqnarray}
\nonumber
{\cal L}_{\phi\,,\beta \rightarrow\infty}&=&\frac{1}{2}\frac{g^2}{(4\pi)^2 L}\left(\partial_\mu \phi\right)^2+\frac{64\pi^2 L^3 e^{-\frac{8\pi^2}{g^2}} }{g^6}\cosh 2\phi\,,\\
{\cal L}_{\sigma\,,\beta \rightarrow\infty}&=&\frac{1}{2}\frac{g^2}{(4\pi)^2 L}\left(\partial_\mu \sigma\right)^2-\frac{64\pi^2 L^3 e^{-\frac{8\pi^2}{g^2}} }{g^6}\cos2\sigma\,.
\label{bosonicfinitetpotential}
\end{eqnarray}

We first argue that we can dimensionally reduce  ${\cal L}_{\phi\,,\beta \rightarrow\infty}$ and ${\cal L}_{\sigma\,,\beta \rightarrow\infty}$ near the transition point to $2$D. As in \cite{Dunne:2000vp,Simic:2010sv} and \cite{Anber:2011gn}, this is because the average distance between  two magnetic bions, represented by the $\cos 2 \sigma$ term in (\ref{bosonicfinitetpotential}), as well as between neutral bions, represented by the $\cosh 2 \phi$ term, is of order ${\Delta R}_{\rm bion} \sim L e^{\frac{4\pi^2}{3g^2}}$ and is thus much greater than the inverse temperature so the gas is essentially two-dimensional (see Fig. (\ref{fig:scales}) for a cartoon of the relevant scales). 
We could further regard $\sigma$ as a compact $2$D scalar. In this case, the field $\sigma$ will contain two parts: normal spin waves and vortices. The vortices can be thought of as $W$-bosons that are being liberated at any finite temperature $T$ and play a prominent role in understanding the phase transition.  The problem with such a description is that the fugacity of the $W$-bosons is implicit (as the energy of a vortex depends on the UV completion rather than being a free parameter).  As we will see below, the $W$-boson's fugacity is a $\phi$-dependent quantity, and hence directly reducing ${\cal L}_{\sigma\,,\beta \rightarrow\infty}$ to $2$D  can overlook important information encoded in it. For this reason, we choose below to follow a pedestrian but otherwise more transparent way to tackle this problem. 
 
Before continuing, let us also address the question about the perturbative contributions from the zero Kaluza-Klein modes, left over from the discussion of finite-$T$ perturbative effects, recall  section~\ref{Perturbative dynamics at non-zero temperature} and footnote~\ref{zeromodeatt}. At $T>0$, supersymmetry is broken by the boundary conditions on $\S^1_\beta$ and we expect that the potentials (\ref{bosonicfinitetpotential}) will receive $T$-dependent contributions.  Our goal is not a full calculation of the finite-$T$ loop corrections to (\ref{full zero temperature Lagrangian}) (see \cite{Paik:2009iz} for a calculation in Seiberg-Witten theory, where  the finite-$T$ contributions are the leading ones lifting the moduli space and are thus crucial), but rather an estimate of these corrections. The one-loop thermal correction to the potential\footnote{This estimate follows from the expression for the $T$-dependent part of the one-loop effective potential for a 3D scalar field $\phi$ of  mass $M(\bar\phi)$, which, for $T \gg M(\bar\phi)$, is easily seen to be $V_{\mbox{\scriptsize eff},T}(\bar\phi)=- {T M^2 (\bar\phi) \over 4 \pi} \log {M (\bar\phi) \over  T}$. We note  that these corrections are not included in our 2D simulations, as $V_{\mbox{\scriptsize eff},T}$ is the sum of the nonzero Matsubara modes on $\S^1_\beta$. Previous analytic and numerical studies \cite{Dunne:2000vp,Simic:2010sv,Anber:2011gn,Anber:2012ig} using the 2D reduction have also not included these effects (as in our case,   these contributions affect the pre-exponentials only). } of $\phi$, the second term in ${\cal L}_{\phi\,,\beta \rightarrow\infty}$,  has relative strength $\sim {16 \pi T L \over g^2}$ compared to the term already present in the  Lagrangian ${\cal L}_{\phi\,,\beta \rightarrow\infty}$. It does not, however, qualitatively change the behavior of the potential (even though, near $T_c$, it  gives an order unity correction to its pre-exponential coefficient). We shall ignore such corrections as they only change the pre-exponential factors, and taking these into account is much beyond the accuracy of our simulations either here or in  previous work \cite{Anber:2012ig}.
  
 The method we will use  to derive the electric-magnetic Coulomb gas representation of the thermal partition function  first appeared in Ref.~\cite{Anber:2013xfa}, applied to the case of the 3D Polyakov model at finite temperature. 
The idea is to go back to the original $U(1)$ field $F^3_{\mu\nu}$, instead of $\sigma$. One then adds to the perturbative photon fluctuations the magnetic field of arbitrary configurations of magnetic bion-instantons (and anti-instantons). The $W$-boson determinant at finite-$T$ is then evaluated in this multi instanton--anti-instanton background. In our case, we already evaluated the $W$-boson (and superpartner) determinant for constant backgrounds, with the result   (\ref{one loop approximate perturbative potential}). At distance scales $\gg L$, away from the cores of the solutions   (where our effective theory is valid), the background fields of the  magnetic bions are small  and it is a good approximation to use the varying backgrounds in the constant field potential $V_{\mbox{\scriptsize eff, pert}}\left(A_0^3,A_3^3\right)$; see also discussion after Eq.~(\ref{one loop approximate perturbative potential in terms of phi}). We will then show, following \cite{Anber:2013xfa}, that the partition function of our finite-$T$, long-distance theory, after a duality transformation, becomes  that of an electric-magnetic Coulomb gas coupled to the scalar $\phi$.

To put this into equations, we first take the field $F_{\mu\nu}^3$ (as well as the potential $A_\mu^3$)   not just as the fluctuations of the photon field, but as a superposition of two contributions: the long-range fields of the magnetic bions and the photon fields 
\begin{eqnarray}
\label{background}
 F^3_{\mu\nu}&=&{\cal F}_{\mu\nu}^3+F_{\mu\nu}^{3\,\rm ph} \nonumber \\
 A^3_\mu&=&{\cal A}_\mu^3+A_\mu^{3\,\rm ph}~.
 \end{eqnarray}  
Here, the magnetic bion background field ${\cal A}_\mu^3$ is  the field generated by the superposition of an arbitrary number of magnetic bions (and anti-bions) located at positions $x_a$ (in $\R^3$) and carrying charges $q_a$:
\begin{eqnarray}
\label{bionbckgd}
{\cal A}_\mu^3(x)=\sum_{a, q_a=\pm 1}q_a A_{\mu}^{3\,\rm bion}(x-x_a)\,,
\end{eqnarray}
where $A_{\mu}^{3\,\rm bion}$ is given by (\ref{magnetic bion field}) (note that, in the partition function, there will be a sum over arbitrary numbers of bions in (\ref{bionbckgd})  and an integral over their positions). At  finite temperature, one compactifies the theory over a circle of circumference $\beta$, and hence one  has to take into account the fact the bion field is the result of summing an infinite number of image charges along the compact dimension. Thus, we have
\begin{eqnarray}
\nonumber
{\cal A}_\mu^3(x)&=&\sum_{a, q_a=\pm 1}q_a A_{\mu}^{3\,(p)\,\rm bion}(x-x_a)\,,\\
A_{\mu}^{3\,(p)\,\rm bion}(x_a)&=&\sum_{n=-\infty}^{\infty}A_{\mu}^{3\,\rm bion}(\vec x-\vec x_a,x_0- x_{0a}+n\beta).
\label{total A field with periodicity}
\end{eqnarray}
Therefore, the total bosonic Lagrangian, written in terms of $F_{\mu\nu}^3$ and $A_3^3$ (i.e., including the field $\phi$), is given by
\begin{eqnarray}
\nonumber
{\cal S}_{\beta}=\int_{\mathbb R^2 \times \mathbb S^1_\beta} \frac{L}{4g^2}F^3_{\mu\nu}F^3_{\mu\nu}+\frac{g^2}{2(4\pi)^2L}\left(\partial_\mu\phi\right)^2+\frac{64\pi^2  e^{-\frac{8\pi^2}{g^2}} }{g^6 L^3}\cosh 2\phi+V_{\mbox{\scriptsize eff, pert}}(A_0^3,\phi)\,,\\
\label{the full 3D non perturbative action}
\end{eqnarray}
with the fields given in (\ref{background}) and 
 the one-loop perturbative potential given by (\ref{one loop approximate perturbative potential}), but with the replacement $\beta A_0^3 \rightarrow \int_0^\beta dx_0 A_0^3$:
\begin{eqnarray}
\nonumber
V_{\mbox{\scriptsize eff, pert}}\left(A_0^3,\phi\right)
&=&-\frac{4}{\pi \beta^3 }\sum_{n=-\infty}^{\infty}e^{-\beta \left|\frac{(2n+1)\pi}{L}+\frac{ g^2 \phi}{4\pi L}\right|}\left(1+\beta  \left|\frac{(2n+1)\pi }{L}+\frac{g^2 \phi}{4\pi L}  \right| \right)\\
&&\quad\quad\quad\quad\quad\quad\times\cos\left(\int_0^\beta dx_0 A_0^3(x_0,x_1,x_2)\right)\,.
\label{one loop approximate perturbative potential in terms of phi}
\end{eqnarray}
 Thus, instead of having a constant holonomy background, we now have a spatially varying holonomy due to the nonperturbative background monopole-instanton fields. Our one-loop effective potential $V_{\mbox{\scriptsize eff, pert}}$ is the leading (i.e., nonderivative) term in the derivative expansion of the $W$-boson determinant. Explicit expressions of the higher-derivative terms can be found in \cite{Megias:2003ui} and section 4.2 in \cite{Anber:2013xfa}. From these terms and the explicit form of the magnetic bion long-range fields (\ref{magnetic bion field}), we infer that higher derivatives of the holonomy along $\S^1_\beta$ ($A_0^3$) are suppressed at distances larger than $g^2/L$ and $1/M_W\sim L$, i.e., at distances larger than the sizes of the monopole-instanton and magnetic bion cores (the UV cutoff of our effective theory).

 Next, we come to the integral in the term $\cos\left(\int_0^\beta dx_0 A_0^3\right)$ that appears in (\ref{one loop approximate perturbative potential in terms of phi}). This integral can be split into two parts, corresponding to the photon and nonperturbative background:
\begin{eqnarray}
\int_0^\beta dx_0 A_0^{3\,\rm ph}+ \int_0^\beta dx_0 {\cal A}_0^{3}\,. \label{v1}
\end{eqnarray}
Using (\ref{total A field with periodicity}) and  (\ref{magnetic bion field}) we find
\begin{eqnarray}
\nonumber
\int_0^\beta dx_0 {\cal A}_0^{3}&=&\sum_{a, q_a=\pm 1} q_a\int_0^\beta \sum_{n=-\infty}^{\infty} A_0^{3,\rm bion}(\vec x-\vec x_a,x_0- x_{0a}+n\beta)\\
&=&\sum_{a, q_a=\pm 1} q_a\int_{-\infty}^\infty  A_0^{3,\rm bion}(\vec x-\vec x_a,x_0)= 4\sum_{a, q_a=\pm 1} q_a\Theta(\vec x-\vec x_a)\,, \label{v2}
\end{eqnarray}
where the $\Theta$ angle is defined as
\begin{eqnarray}
\Theta(\vec x)=-\mbox{sign}(x_1)\frac{\pi}{2}+\mbox{Arctan}\left(\frac{x_2}{x_1}\right)\,.\label{v3}
\end{eqnarray}

Finally, the grand partition function of the system is obtained as a path integral,  with an action given by ${\cal{S}}_\beta$ of Eq.~(\ref{the full 3D non perturbative action}), with $V_{\mbox{\scriptsize eff, pert}}$ given by (\ref{one loop approximate perturbative potential in terms of phi}, \ref{v1}, \ref{v2}, \ref{v3}). The path integral is  over the perturbative fluctuations  $A_\mu^{3\,\rm ph}$ and $\phi$, and also includes  a sum over  the possible nonperturbative backgrounds. These are represented as a sum over an arbitrary number of positive $N_{b+}$  and negative $N_{b-}$ magnetic bion-instantons and  integrals over their positions, while every bion comes with the appropriate fugacity given below in (\ref{bionfugacity}). Hence, the grand partition function reads 
\begin{eqnarray}
\nonumber
&&{\cal Z}{\mbox{\scriptsize grand}}=\nonumber \\
&&\sum_{N_{b\pm}, q_a=\pm 1 }\frac{\xi_{b}^{N_{b+} + N_{b-}}}{N_{b+}! N_{b-}!}\left(\prod_a^{N_{b+} + N_{b-}} \int d^3x_a\right) \int [{\cal D} A_\mu^{\mbox{\scriptsize ph}}]\int[{\cal D}\phi] \nonumber\\
\nonumber
& \times& \exp\left[ -\int_{\mathbb R^2 \times \mathbb S^1_\beta} \frac{L}{4g^2}\left(F^{3\,\rm ph}_{\mu\nu}+{\cal F}_{\mu\nu}^3\right)^2-2\xi_W(\phi)\beta^{-1}\cos\left(4\sum_{a, q_a=\pm 1} q_a\Theta(\vec x-\vec x_a)+\int_0^\beta dx_0A_0^{3\,\rm ph}\right)\right.\\
&&\left.\quad\quad\quad+\frac{g^2}{2(4\pi)^2L}\left(\partial_\mu\phi\right)^2+\frac{64\pi^2 e^{-\frac{8\pi^2}{g^2}} }{g^6  L^3}\cosh 2\phi\right]\,,  \label{total partition function}
\end{eqnarray}
where  $\xi_b$ is the magnetic bion fugacity 
\begin{equation}
\label{bionfugacity}
\xi_b=\frac{e^{-\frac{8\pi^2}{g^2}}}{g^6L^3}. 
\end{equation} 
When writing ${\cal Z}{\mbox{\scriptsize grand}}$, we have also defined the $\phi$-dependent quantity:
\begin{eqnarray}
\xi_W(\phi)&=&\frac{2}{\pi \beta^2 }\sum_{n=-\infty}^{\infty}e^{-\beta \left|\frac{(2n+1)\pi}{L}+\frac{ g^2 \phi}{4\pi L}\right|}\left(1+\beta  \left|\frac{(2n+1)\pi }{L}+\frac{g^2 \phi}{4\pi L}  \right| \right) \nonumber \\
&=&{2 \over \beta L \sinh  {\beta \pi \over L}}\left( \left[ 
{\coth {\beta \pi \over L} } +{L \over \pi \beta}\right]  \cosh{\beta g^2 \phi \over 4 \pi L} - {  g^2 \phi \over 4 \pi^2} \sinh {\beta g^2 \phi \over 4 \pi L}\right)
\label{the W fugacity as function of phi}
\end{eqnarray}
appearing in the $W$-boson (and superpartner) determinant and we used the fact that $\phi$ belongs to the Weyl chamber $-{ \pi  \over L} < {g^2 \phi\over 4 \pi L} \le {  \pi  \over L}$ in the second equality. 
 The quantity $\xi_W(\phi)$ will be interpreted as the $W$-boson fugacity, as we will shortly show. 
 We can already see that near $\phi=0$, the dominant contribution to $\xi_W(\phi)$ comes from the $n=0$ and $n=-1$ terms. Setting $\phi=0$, we obtain $\xi_W \simeq {4 \over \beta L}\; e^{- \beta \pi \over L} = \frac{4M_WT}{\pi}e^{-M_W/T}$, where we recall $\beta/L \sim M_W/T \gg 1$. In fact, this is four times what one expects to get from a single $W$-boson.\footnote{A single $W$-boson fugacity can also be obtained by integrating the Boltzmann distribution of a single non-relativistic $W$-boson $e^{-H/T}$, where $H=M_W+\frac{p^2}{2M_W}$, over the particle momenta:
\begin{eqnarray}
\xi_W=S_W\int \frac{d^2 p}{\left(2\pi\right)^2}e^{-\frac{M_W}{T}-\frac{p^2}{2M_WT}}=S_W\frac{TM_W}{2\pi}e^{-\frac{M_W}{T}}\,,
\end{eqnarray}
where $S_W=2$ is the spin degeneracy factor of the $W$-bosons.
}
The extra factor of $4$ comes  because both the zero and first excited Kaluza-Klein $W$-bosons have the same mass, thanks to the unbroken $\mathbb Z_2^{(L)}$ center symmetry. In addition, there are the superpartners of these $W$-bosons, which come with exactly the same mass and  contribute to the fugacity.  

At the boundary of the Weyl chamber, at $|{g^2 \phi\over 4 \pi L}|={  \pi  \over L}$, there is  a $W$-boson state that becomes massless. At this point on the Coulomb branch,    the full nonabelian $SU(2)$ gauge symmetry is restored and the weakly-coupled abelian  description is no longer valid. 
From either the top or bottom line of (\ref{the W fugacity as function of phi}), it can be seen that in the regime $LT \ll 1$ ($\beta/L \gg 1$), the dimensionless fugacity $\xi_W(\phi)/\xi_W(0)$  increases when $\phi$ approaches the boundary of the Weyl chamber (see also Fig.~\ref{fig:Fugacity} and the discussions at the end of section \ref{Simulations of the Coulomb gas} and section \ref{extrapolationsection}). This effect is, of course, countered by the fact that large values of $\phi$ are disfavored by the neutral bion induced potential  which dominates at low-$T$.

Now, we proceed with casting the partition function (\ref{total partition function}) as the partition function of a Coulomb gas of electrically charged particles with fugacity (\ref{the W fugacity as function of phi}) coupled to the nonperturbative magnetic bion sector as a gas of magnetically charged particles.
 We use  $\cos \alpha=\left(e^{i\alpha}+e^{-i\alpha}\right)/2$ to expand the term of the form $\exp\left(2\xi\int dx \cos\left( \alpha(x)\right)\right)$ as follows
\begin{eqnarray}
\exp\left(2\xi\int dx \cos\left( \alpha(x)\right)\right)=\sum_{n_+,n_-=0}^\infty\sum_{q_A=\pm 1}\frac{\xi^{n_++n_-}}{n_+!n_-!}\left(\prod_{A=1}^{n_++n_-} \int dx_A \right) e^{\sum_{A} iq_A \alpha(x_A)}\,.
\label{the main identity for exponential}
\end{eqnarray}
Using this trick in the partition function (\ref{total partition function}) we find
\begin{eqnarray}
\nonumber
{\cal Z}_{\mbox{\scriptsize grand}}&=&\sum_{N_{b\pm}, q_a=\pm 1 }\sum_{N_{W\pm}, q_A=\pm 1 }\left( \prod_a^{N_{b+} + N_{b-}} \int d^{2+1} x_a\right) \left(\prod_A^{N_{W+} + N_{W-}} \int d^{2+1}x_A \right)\\
\nonumber
&&\times \int [{\cal D}\phi] \; \frac{\xi_b^{N_{b+} + N_{b-}}}{N_{b+}! N_{b-}!} \; \frac{(T\xi_W(\phi))^{N_{W+} + N_{W-}}}{N_{W+}! N_{W-}!}  \exp\left[4i\sum_{aA}q_aq_A\Theta\left(\vec x_a-\vec x_A\right)\right]\\
\nonumber
&&\times \int \left[{\cal D} A_\mu^{3\,\mbox{\scriptsize ph}}\right] \exp\left\{- \int_{\mathbb R^2\times \mathbb S^1_\beta} \frac{L}{4g^2}\left( F^{3\,\mbox{\scriptsize ph}}_{\mu\nu}+{\cal F}^{3}_{\mu\nu}\right)^2-i\sum_A q_A A_0^{3\,\mbox{\scriptsize ph}}(\vec x, x_0)\delta(\vec x-\vec x_A)\right.\\
&&\left.\quad\quad\quad\quad\quad\quad\quad\quad\quad\quad+ \frac{g^2}{2(4\pi)^2L}\left(\partial_\mu\phi\right)^2+\frac{64\pi^2 e^{-\frac{8\pi^2}{g^2}} }{g^6  L^3}\cosh 2\phi   \right\}\,.
\label{semi final expression for Z}
\end{eqnarray}

The final step in our derivation---the path integral over $A_\mu^{3\,\rm ph}$---can be carried out using a duality transformation. The duality enables us to perform the integral without having to run into inconsistencies even in the presence of both electric and magnetic charges. This duality transformation was  considered before---see appendix C of Ref.~\cite{Anber:2013xfa} for a detailed description---and we do not repeat it here. 

Our final result for the partition function is:%
\begin{eqnarray}
\nonumber
{\cal Z}_{\mbox{\scriptsize grand}}&=&\sum_{N_{b\pm}, q_a=\pm 1 }\sum_{N_{W\pm}, q_A=\pm 1 }\int [{\cal D}\phi]\;\frac{(\beta\xi_b)^{N_{b+} + N_{b-}}}{N_{b+}! N_{b-}!}\frac{(\xi_W(\phi))^{N_{W+} + N_{W-}}}{N_{W+}! N_{W-}!}\left(\prod_a^{N_{b+} + N_{b-}} \int d^{2}x_a\right)\\
\nonumber
&&\times\left(\prod_A^{N_{W+} + N_{W-}} \int d^2x_A \right)\exp \left\{\frac{32 \pi L T}{g^2}\sum_{a>b}\log|\vec x_a-\vec x_b|+\frac{g^2}{2\pi LT}\sum_{A>B}\log|\vec x_A-\vec x_B|\right.\\
\nonumber
&&\left. +4i\sum_{aA}q_aq_A\Theta\left(\vec x_a-\vec x_A\right)+\int _{\mathbb R^2}\frac{g^2\beta}{2(4\pi)^2L}\left(\partial_\mu\phi\right)^2+\frac{64\pi^2\beta e^{-\frac{8\pi^2}{g^2}} }{g^6  L^3}\cosh 2\phi\right\}\,,\\
\label{final expression for Z}
\end{eqnarray}
where we have considered only the zero mode along the thermal circle, as this is the only important mode near the deconfinement transition (explained at the beginning of this section).  

The partition function ${\cal Z}_{\mbox{\scriptsize grand}}$ of Eq.~(\ref{final expression for Z}) is the main result of the first part of the paper. It represents a dual (i.e., electric and magnetic) Coulomb gas that consists of magnetic bions with constant fugacities, as well as $W$-bosons (and superpartners, as in the nonrelativistic limit the contributions of bosons and fermions to the thermal partition function is the same and only multiplicatively affects the fugacity) with fugacity $\xi_W(\phi)$ depending on the field $\phi$.
The partition function for $SU(2)$ nonsupersymmetric QCD(adj) has a form identical to our  ${\cal Z}_{\mbox{\scriptsize grand}}$ of Eq.~(\ref{final expression for Z}), except that the scalar field is absent. In particular, in the absence of the scalar field, the partition function has an electric-magnetic duality, which exchanges $W$-boson and bion fugacities as well as inverts the coupling (and temperature), i.e., $32 \pi LT/g^2 \leftrightarrow g^2/(2 \pi LT)$. Thus, it acts as a Kramers-Wannier duality and the self-dual value of $T$ is, naturally, the critical value. In the absence of scalars, the renormalization group equations for the Coulomb gas can be used to find the critical points and some of the critical exponents. Furthermore, also in the absence of scalars, the self-dual partition function ${\cal Z}_{\mbox{\scriptsize grand}}$ could be cast in the form of a self-dual sine-Gordon model; at the self-dual point, using bosonization, this model is exactly solvable and, as shown in \cite{Lecheminant:2002va}, is  equivalent to a free field theory.

The presence of the scalar $\phi$ and its coupling to the dual-Coulomb gas via the dependence of the $W$-boson fugacity on $\phi$ makes an analytic approach to studying the phase transition in (\ref{final expression for Z}) rather challenging. We have taken a numerical path towards the study of ${\cal Z}_{\mbox{\scriptsize grand}}$. In the rest of the paper, we will describe the Monte Carlo study of the partition function ${\cal Z}_{\mbox{\scriptsize grand}}$ using two different formulations:
\begin{enumerate}
\item Our first Monte Carlo study will be of  the dual-Coulomb gas, i.e., of the grand partition function ${\cal Z}_{\mbox{\scriptsize grand}}$ itself. This  most straightforward approach has the advantage that all parameters used in the simulation can (at least in principle) be made to take the values determined by the UV completion of the dual-Coulomb gas---the four-dimensional SYM theory. A drawback, however, is the presence of the Aharonov-Bohm phase interactions in (\ref{final expression for Z}). Thus, while ${\cal Z}_{\mbox{\scriptsize grand}}$ itself is real, near criticality, there is a substantial sign problem  precluding a detailed study of the  transition.
\item The second Monte Carlo study we will perform is free of a sign problem. We will recast the partition function ${\cal Z}_{\mbox{\scriptsize grand}}$ into the form of an  $XY$ model with a symmetry-breaking perturbation, whose coefficient depends on the scalar field $\phi$. This is similar to the system studied analytically in \cite{Anber:2011gn} and numerically in \cite{Anber:2012ig}, except for the coupling to the scalar field. The only disadvantage of the ``affine" XY-model approach is that the fugacity of magnetic bions is not a free parameter, as opposed to the dual-Coulomb gas. This disadvantage is common with \cite{Anber:2011gn,Anber:2012ig}; however, in the known cases, the qualitative properties of the phase transition have been seen to not  depend on this difference.
\end{enumerate}

In the following two sections, we will describe the results of our simulations. We will find that, at the qualitative level of our study, the results from the two approaches will agree. 
We end this section with a disclaimer regarding the rest of the paper. We would like to stress that the Monte Carlo simulations, whether in the first or second system described above, were not performed for values of the parameters as determined by the UV completion in a regime under theoretical control (small-$L$, small-$g$, $LT\ll 1$). This is, essentially, because the semiclassical exponentially small fugacities would make the generation of any nontrivial excitations by the Metropolis  algorithm prohibitively unlikely. One (weak) defense we have is based on previous experience with Coulomb gas systems in 2D, showing  that fugacities of order $1/e$ are  ``small" enough, so that lattice results have been seen to agree, even quantitatively, with results from analytic  small fugacity approximations. Further, at the end of the paper, in section \ref{extrapolationsection}, we offer some qualitative arguments in favour of the more general validity of our conclusions. Nonetheless, we stress that a simulation for the physically relevant small-$g$ regime is required in order to be decisive about the phase structure.
 
\section{Simulations of the dual-Coulomb gas}
\label{Simulations of the Coulomb gas}

In this section, we consider the simulations of the double Coulomb gas of $W$-bosons and magnetic bions in the background of the $\phi$ field. In order to perform the simulations, we will use a discrete version of the total action that appears in the partition function (\ref{final expression for Z}). Defining $\kappa=g^2/(2\pi)$, and taking the scale $L$ (the cutoff scale of our effective description (\ref{final expression for Z})) to be equal to the lattice spacing, hereafter  taken to unity, 
the discrete action of the dual-Coulomb gas reads:
\begin{eqnarray}
\nonumber
S&=&\underbrace{\sum_x\sum_\mu\left[ \frac{\kappa}{16 \pi T }\left(\nabla_\mu \phi_x\right)^2+\frac{8e^{-\frac{4\pi}{\kappa}}}{\pi T \kappa^3}\cosh(2 \phi_x)\right]}_{\mbox{$\phi$ field}}+\underbrace{\sum_{A}{\cal E}_W(A) -\frac{\kappa}{T}\sum_{A>B}q_Aq_B G(A,B)}_{\mbox{$W$-bosons}}\\
&&+\underbrace{\sum_{a}\frac{4\pi }{\kappa}-\frac{16 T}{\kappa}\sum_{a>b}q_aq_b G(a,b)}_{\mbox{magnetic bions}}\underbrace{-4i\sum_{aA}q_aq_A\Theta(a,A)\,.}_{\mbox{interaction between $W$-bosons and magnetic bions}}
\label{double C gas with phi field} 
\end{eqnarray}

Here, we have included the $W$-boson and magnetic bion fugacities as core energies: 
  ${\cal E}_W(A)$ is the core energy of the $W$-bosons (the negative of the logarithm of the fugacity), which depends  on the field $\phi_A$ at the position $A$ of the $W$-boson: 
\begin{eqnarray}
\label{core}
{\cal E}_W(A)&=&-\log \xi_W(\phi(x_A))\,,
\end{eqnarray}
with the fugacity given in Eq.~(\ref{the W fugacity as function of phi}), while the magnetic bion core energy is simply $4 \pi \over \kappa$. 
 The electric particles ($W$-bosons) and the field $\phi$ ``live" on the 2D lattice with points $(A,B...)$, while the magnetic bions reside on points of the dual lattice $(a, b,...)$ ($\nabla_\mu \phi_x = \phi_{x + \hat\mu} - \phi_x$ is the usual forward lattice derivative). 
The functions $G(a,b)$ and $\Theta(a,A)$ are the discrete versions of the corresponding continuum expressions.\footnote{$G(A,B)$ ($G(a,b)$) is  simply the massless propagator between points on a 2D (dual) lattice. The lattice version of the angle $\Theta(a,A)$, where $a$ and $A$ belong to dual lattices, has previously appeared in \cite{Kadanoff:1978ve} in the infinite volume limit. The finite volume lattice expressions of these functions  are given by (\ref{discrete log}) and (\ref{discrete theta}). In  appendix  \ref{The discrete dual Sine-Gordon model}, $\Theta(a,A)$ is derived  from a lattice action used to study a discrete  dual sine-Gordon model.} We use the Metropolis  algorithm to simulate the grand canonical dual-Coulomb gas (\ref{double C gas with phi field}). The details of the algorithm for the Coulomb gas are given in appendix (\ref{Monte-Carlo procedure for simulating the double Coulomb gas}). 

\begin{figure}[h]
\centering 
\includegraphics[width=.3\textwidth]{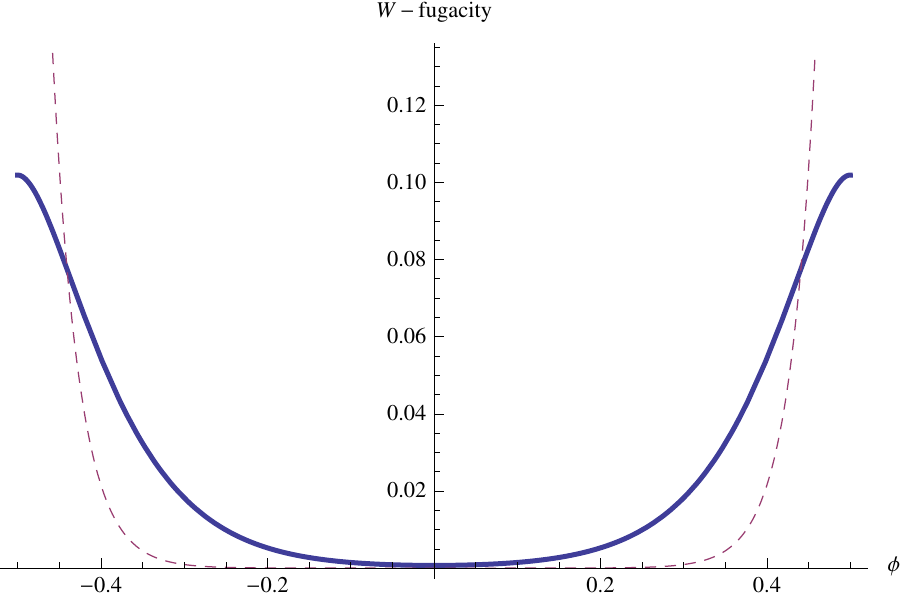}
\hfill
\includegraphics[width=.3\textwidth]{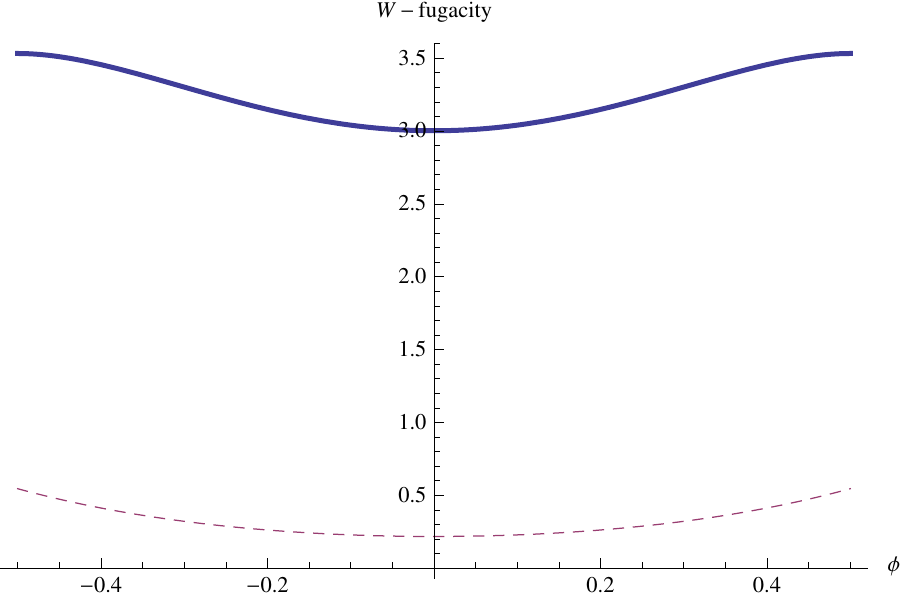}
\hfill
\includegraphics[width=.3\textwidth]{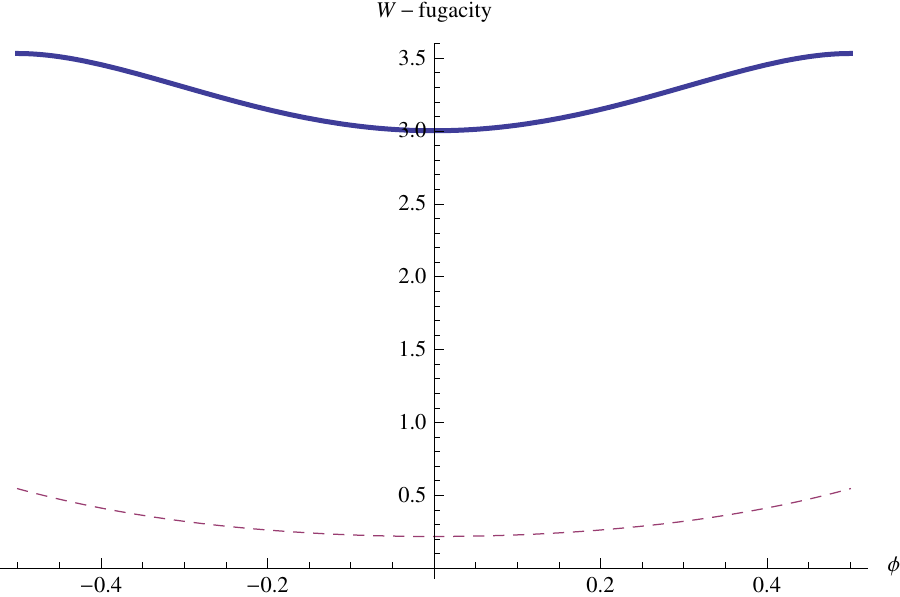}
\caption{\label{fig:Fugacity} $W$-boson fugacities $\xi_W(\phi)$ for $\kappa = {g^2 \over 2 \pi} = 4\pi$. LEFT:  $T=0.4$, MIDDLE: $T=2$, RIGHT: $T=5$. The one-loop expression of (\ref{the W fugacity as function of phi}) is shown by a thick line. The modified fugacity of Eq.~(\ref{positivecore}), used in the dual-Coulomb gas simulation to keep core energies positive, is shown by a dashed line. 
At low-$T$, the fugacities vary strongly with $\phi$---and would prefer values of $\phi$ near the edge of the Weyl chamber, where $W$-bosons become massless and our  abelian description is not appropriate. However $\phi$ is a dynamical variable, whose  value is determined by balancing the $W$-boson and neutral bion ($\cosh 2 \phi$) contribution. The latter is more important at low $T$ and favors $\phi \sim 0$, where the $W$'s are massive. At higher $T$, the fugacities flatten out as functions of $\phi$. See section \ref{extrapolationsection} for discussion of the weak coupling behavior of the fugacity.}
\end{figure}

Before we continue, we note that the fugacities of both magnetic bions and $W$-bosons are exponentially small numbers (in the semiclassical regime where the partition function is derived). A Monte Carlo simulation, with the resources available to us, will never generate any particles with exponentially small fugacities (of order $e^{- {4 \pi \over \kappa}} = e^{- {8 \pi^2 \over g^2}}$ as $g^2 \rightarrow 0$). Thus, we will, as in previous work \cite{Rastelli:2004bt,Anber:2012ig} on related theories, make the fugacities small, rather than exponentially small. More precisely, in our simulations of the Coulomb gas, we take $\kappa=4\pi$, so that the magnetic bion fugacities are $1/e$ (equivalently, core energies are equal to unity).  The electric $W$-boson fugacities, near $\phi=0$ and for $T \simeq \kappa/4$, are then of the same order, $\sim 1/e$.  In addition, another modification to (\ref{double C gas with phi field}) that we are forced to make is to drop the $1/\kappa^3$ term in the coefficient of the neutral bion potential.\footnote{At weak coupling,  $\kappa^{-3}$ represents  a relative enhancement, compared to the exponential suppression of $e^{- {4 \pi \over \kappa}}$, of the neutral bion amplitude, but at $\kappa \sim 4 \pi$ it generates a suppression making the $\phi$-potential completely irrelevant in the simulation; for $\kappa = 4 \pi$, the   suppression of nonzero $\phi_x$ in the path integral is thus roughly $e^{ - {\cal{O}}(1) {\cosh2 \phi_x \over T}}$ and we expect that at $T\sim{\cal O}(1)$  the field will take random values---this is, in fact, what we see in our simulations (which, of course, include the $W$ bosons and their back reaction on the $\phi$ field).} At $\kappa = 4\pi$, the fugacity (\ref{core}) also has to be modified. This is because, at these values of $\kappa$, the $W$-boson core energy can become negative. In turn, this dramatically shifts the transition point, since having negative core energies favors the liberation of the $W$-bosons at lower temperatures. Because we know that such a shift is an artifact of using large values of $\kappa$, and since negative core energies do not occur at  weak coupling, we modify (\ref{core}), see also (\ref{the W fugacity as function of phi}), to 
\begin{eqnarray}
\label{positivecore}
{\cal E}_{W_{\rm simulations}}=-\log\left[\frac{\cosh\left(\frac{g^2\phi}{4\pi T }\right)}{2\sinh( \beta \pi)}\right]\,,
\end{eqnarray} 
which amounts to removing the pre-exponential  factors in (\ref{the W fugacity as function of phi}) and guarantees the positivity of the core energies for temperatures beyond the transition point (see Fig.~\ref{fig:Fugacity} and further qualitative discussion towards the end of this section).

Our hope is that the modifications described above---while only justified by the practicality of the simulation---will lead to results  that are qualitatively similar to  the dynamics of the gauge theory at small $L$.

With the above discussion in mind, we now proceed to describe the parameters used for our dual-Coulomb gas simulations and a brief outline of the algorithm. We have only simulated two volumes, with lattice widths $N=16, 32$.  Our simulations involve gradually heating the system through a range of temperatures, performing 10000 sweeps at each temperature (where a sweep is defined as $N^2$ Metropolis iterations, and the first 500 sweeps were disregarded for equilibration). Data (such as the value of the action, the densities of the magnetic bions and $W$-bosons, and the mean value of both $\phi$ and $|\phi|$) was recorded at the end of every sweep. Simulations were initialized at low temperature in a configuration with no magnetic bions or $W$-bosons, with the $\phi$ field uniformly distributed in the range $[-{2\pi\over\kappa},{2\pi\over\kappa}]$. Each Metropolis iteration consisted of two processes. First, we attempt one of neutral pair creation, annihilation, or diffusion of either the $W$-boson or magnetic bion gas (with equal probability). Second, we attempt to change the value of the $\phi$ field at a random lattice site to a random value in the range $[-{2\pi\over\kappa},{2\pi\over\kappa}]$---such changes are accepted with the usual probability $p = \min(1,e^{-\Delta S})$ designed to produce configurations with probabilities that satisfy Boltzmann statistics (see appendix \ref{Monte-Carlo procedure for simulating the double Coulomb gas} for further detail).

\begin{figure}[tbp]
\centering 
\includegraphics[width=.45\textwidth]{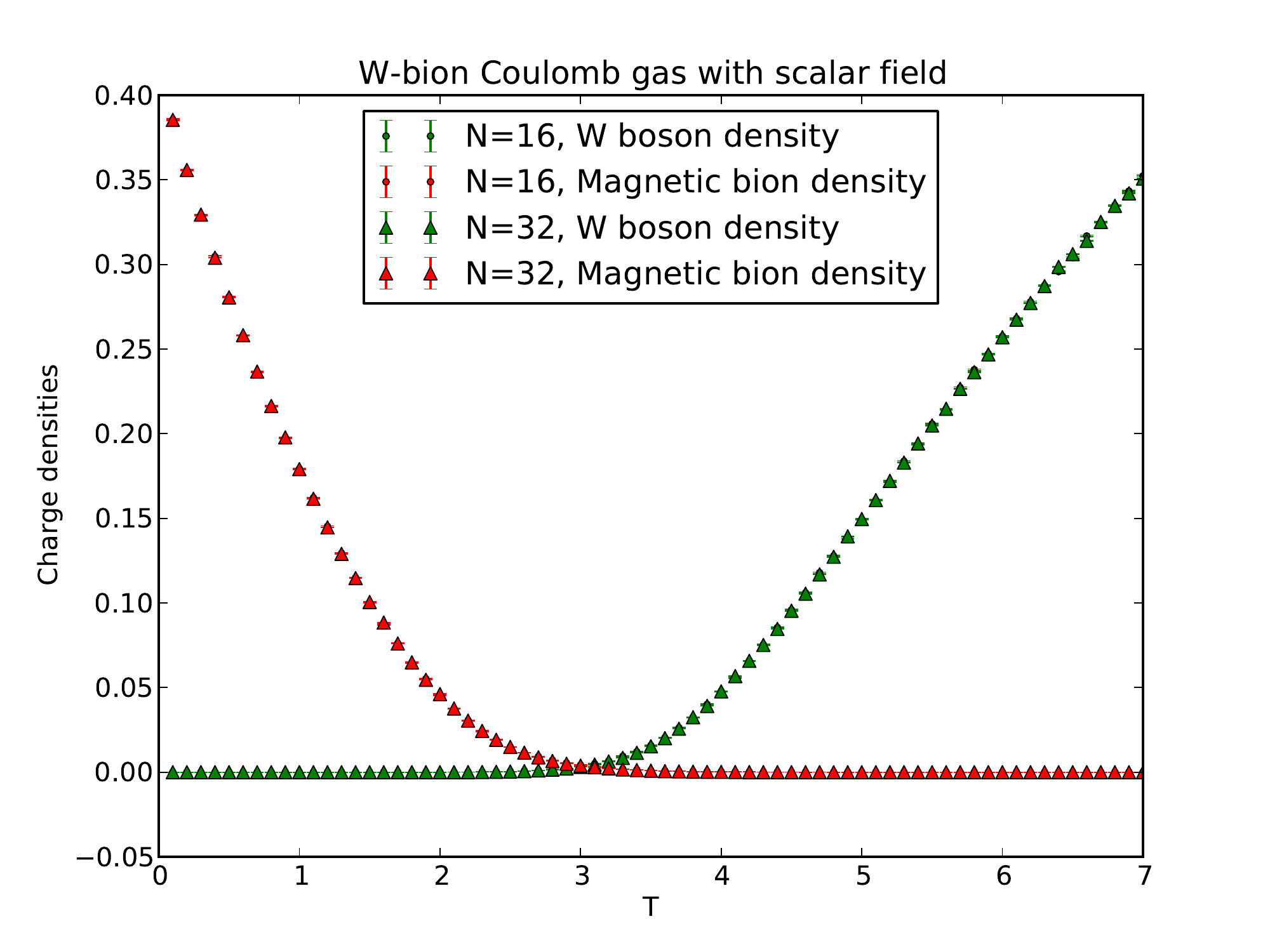} \hfill
\includegraphics[width=.45\textwidth]{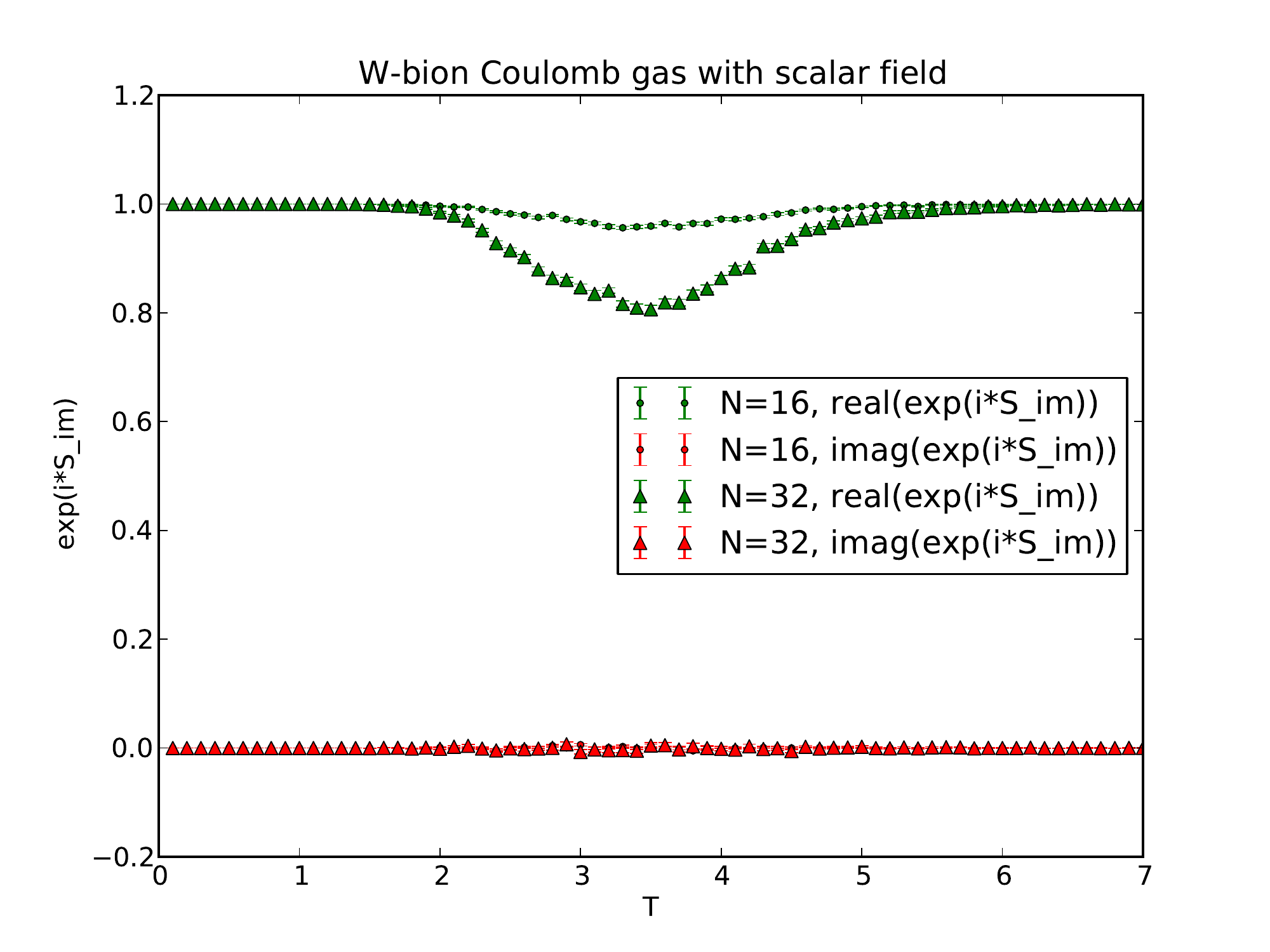}
\caption{\label{fig:CoulombCharges} LEFT: Magnetic and electric charge densities as a function of the temperature. RIGHT:  Aharonov-Bohm phase contribution to the partition function (see text for definition and interpretation). Two volumes, $N=16,32$ were only considered in the dual-Coulomb gas simulation. }
\end{figure}

On the left panel of Fig.~\ref{fig:CoulombCharges}, we show the magnetic and electric charge densities. The qualitative picture expected of a deconfinement transition---the dominance of magnetic charges at low temperature and of electric charges at high temperature---is evident. The nonzero density of magnetic charges at $T<3$  signifies the breaking of the $\Z_2^R$ discrete chiral symmetry, while the nonzero electric charge density at $T>3$ signifies the breaking of the center symmetry in the deconfined phase (we note that defining more precise order parameters in the Coulomb gas via correlation functions of half-magnetic bion operators and half-$W$-boson operators is possible, but their measurement is challenging and we have not attempted this). 

In most of the temperature range outside the `critical' temperature $T\simeq 3$, the two gases are completely decoupled from each other---as one of the densities is always too small (essentially zero) to have any appreciable effect. 
Only near the transition at $T \simeq 3$ do the gases interact significantly. This is reflected in the appearance of a sign problem, illustrated on the right panel of Fig.~\ref{fig:CoulombCharges}. There, we show the 
average (over the grand canonical ensemble of electric and magnetic particles and $\phi$) of the real and imaginary part of the Aharonov-Bohm factor in the partition function, $ e^{4i\sum_{aA}q_aq_A\Theta(a,A)}$. The bottom curve shows that  the re-weighting factor  is real, as expected. The upper curve shows that, as $N$ increases, the Aharonov-Bohm interaction becomes important  near the transition (when the two densities are comparable). For the temperature step used near the transition and for the  volumes we have studied, we do not see a serious sign problem yet---but such is expected to appear as the volume and resolution are increased. The value of  Re$\langle e^{4i\sum_{aA}q_aq_A\Theta(a,A)}\rangle$ in Fig.~\ref{fig:CoulombCharges} is seen to change  significantly upon doubling the volume and we expect that as $N$ further increases, the expectation value of Re$\langle e^{4i\sum_{aA}q_aq_A\Theta(a,A)}\rangle$ will approach zero; however,  
 studying this is beyond the task of this largely qualitative simulation. 

We note that the transition between the magnetically dominated and the electrically dominated phase occurs precisely when $T \sim \pi = {\kappa \over 4}$ (recall we use $\kappa = 4 \pi$), i.e., precisely when the electric (at $\phi=0$) and magnetic fugacities are the same. To ask about the order of the transition (in the dual-Coulomb gas picture, we have not studied the large-volume behavior of susceptibilities), 
 we studied the histograms of the distributions of the action as a function of temperature. Within our accuracy, we have not found any double-peaked distribution of the action that would indicate a first order transition (this is consistent with the study of the related $\Z_4$ model of \cite{Rastelli:2004bt,Anber:2011gn} and is in contrast with the $\Z_3 \times \Z_3$ models studied in \cite{Anber:2012ig} which exhibit a first order transition).

\begin{figure}[h]
\centering 
\includegraphics[width=.45\textwidth]{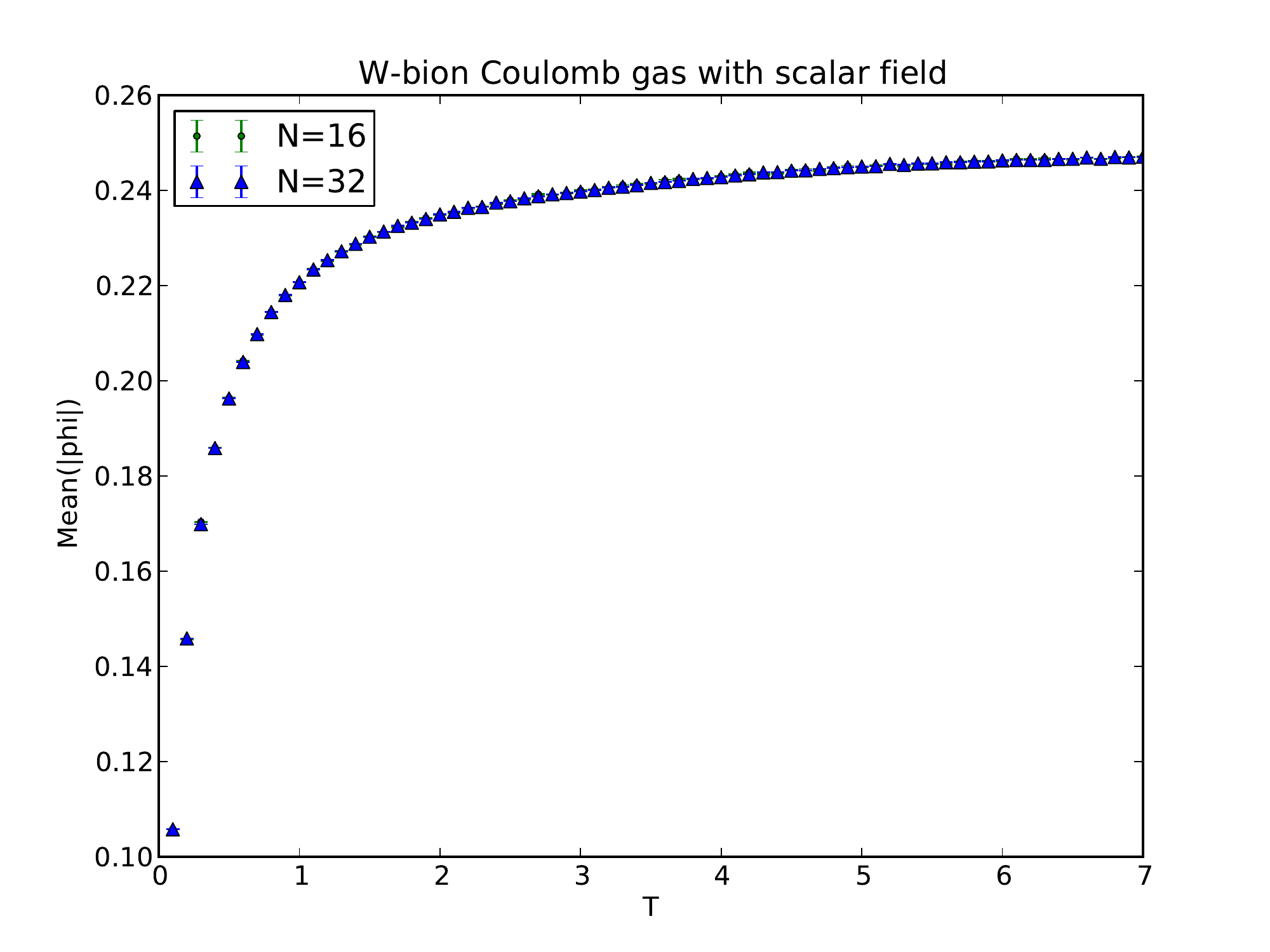}
\hfill
\includegraphics[width=.45\textwidth]{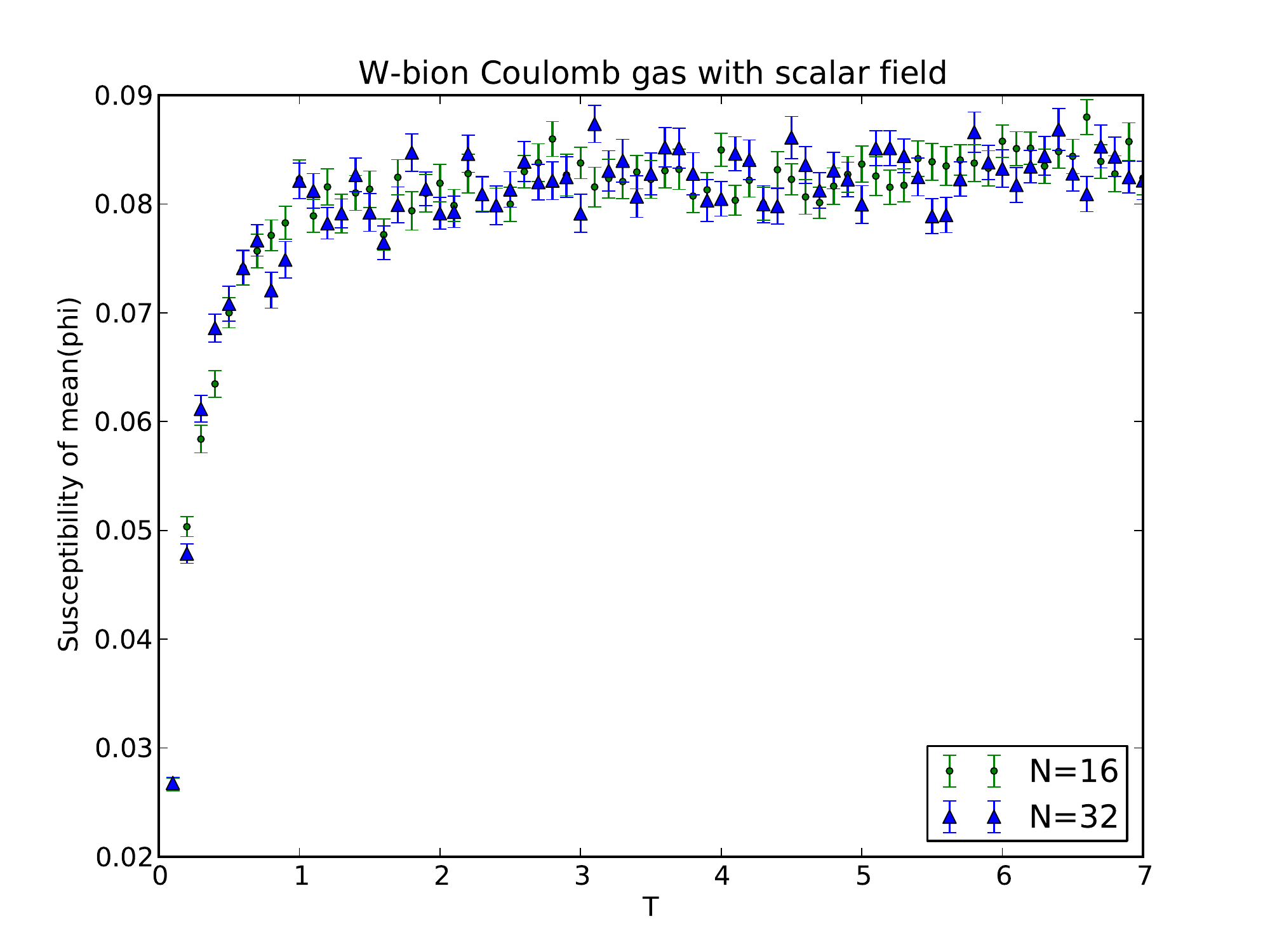}
\hfill
\caption{\label{fig:CoulombScalar} Dual Coulomb gas scalar field observables, Eq.~(\ref{scalarobservables}). LEFT: the average of $|\phi|$. RIGHT: susceptibility of $\phi$. We interpret these results as showing that, for $T>2$, the field strongly fluctuates around $\phi=0$; see text and Fig.~\ref{fig:CoulombScalar2}.}
\end{figure}

Another qualitative observation is that the fluctuations of $\phi$ do not seem to affect the transition in any significant manner. In order to study these, we have measured several quantities: the average of $\phi$, the average of $|\phi|$, and the susceptibility of $\phi$, defined as:
\begin{eqnarray}
\label{scalarobservables}
\overline\phi  &=& {1 \over N^2} \big\langle   \sum_x \phi(x)  \big\rangle \simeq 0\,, \nonumber \\
\overline{| \phi |} &=& {1 \over N^2} \big\langle   \sum_x |\phi(x)|  \big\rangle\,, \\
\chi(\phi) &=&{1 \over N^2} \big\langle \left(\sum_x \phi(x)\right)^2 \big\rangle - {1 \over N^2} \left(\big\langle \sum_x \phi(x)  \big\rangle\right)^2 \simeq \sum_x \big\langle\phi(x) \phi (0) \big\rangle\nonumber~,
\end{eqnarray}
where the $\langle ... \rangle$ denotes averaging with the grand canonical partition function with action (\ref{double C gas with phi field}). The $\simeq 0$ on the first line above indicates our finding  that the average value of $\phi$ is zero, at all temperatures (we do not display this result as all the corresponding plots show that the average value of $\phi$ is zero within errors of the simulation). Taking the vanishing of $\overline\phi$ into account, we have also indicated on the third line that the susceptibility $\chi(\phi)$ is essentially the zero-momentum Green's function of the scalar $\phi$ (in continuum field theory language---the inverse mass squared of the field in lattice units).

\begin{figure}[h]
\centering 

\includegraphics[width=.32\textwidth]{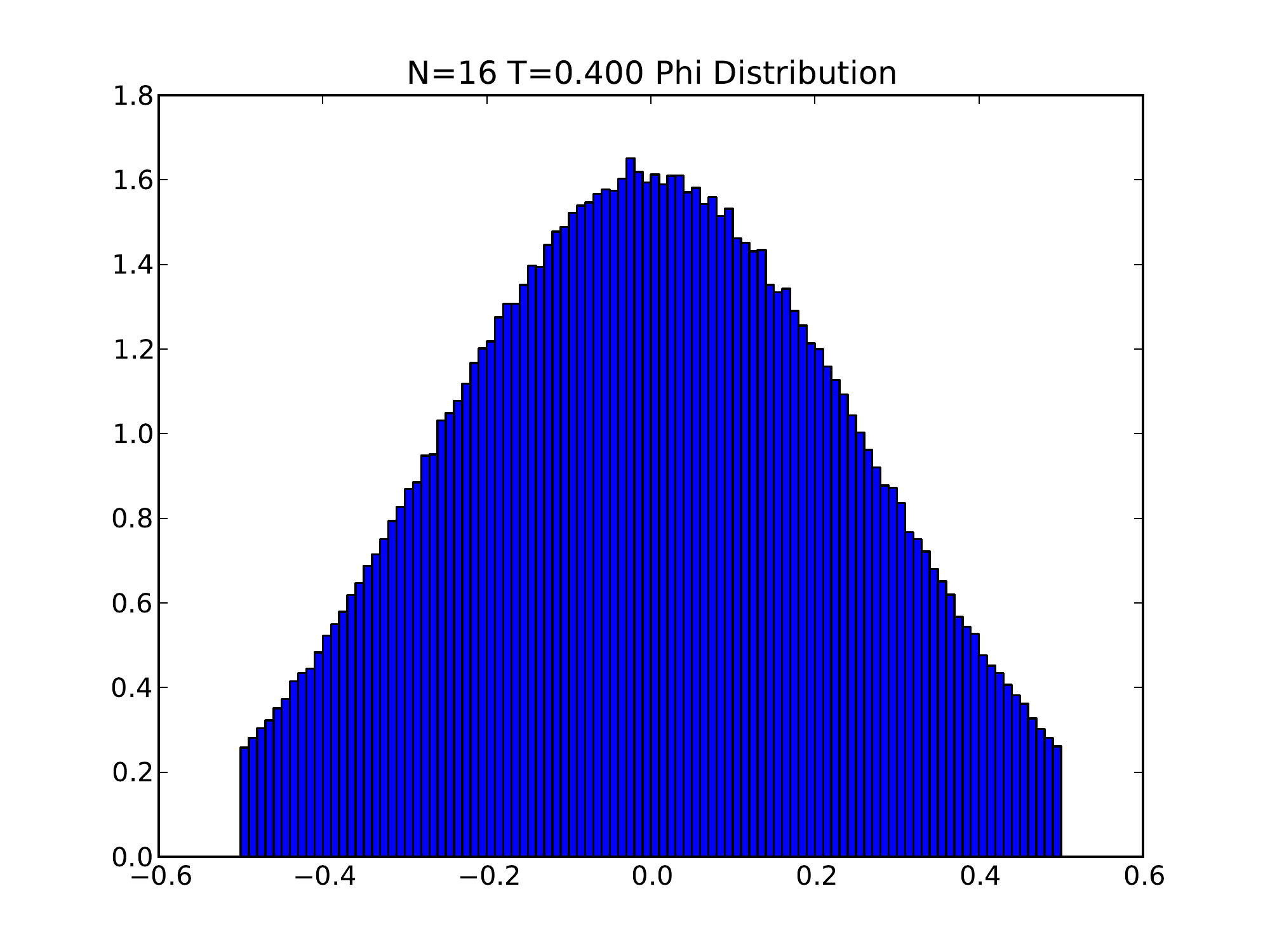}
\hfill
\includegraphics[width=.32\textwidth]{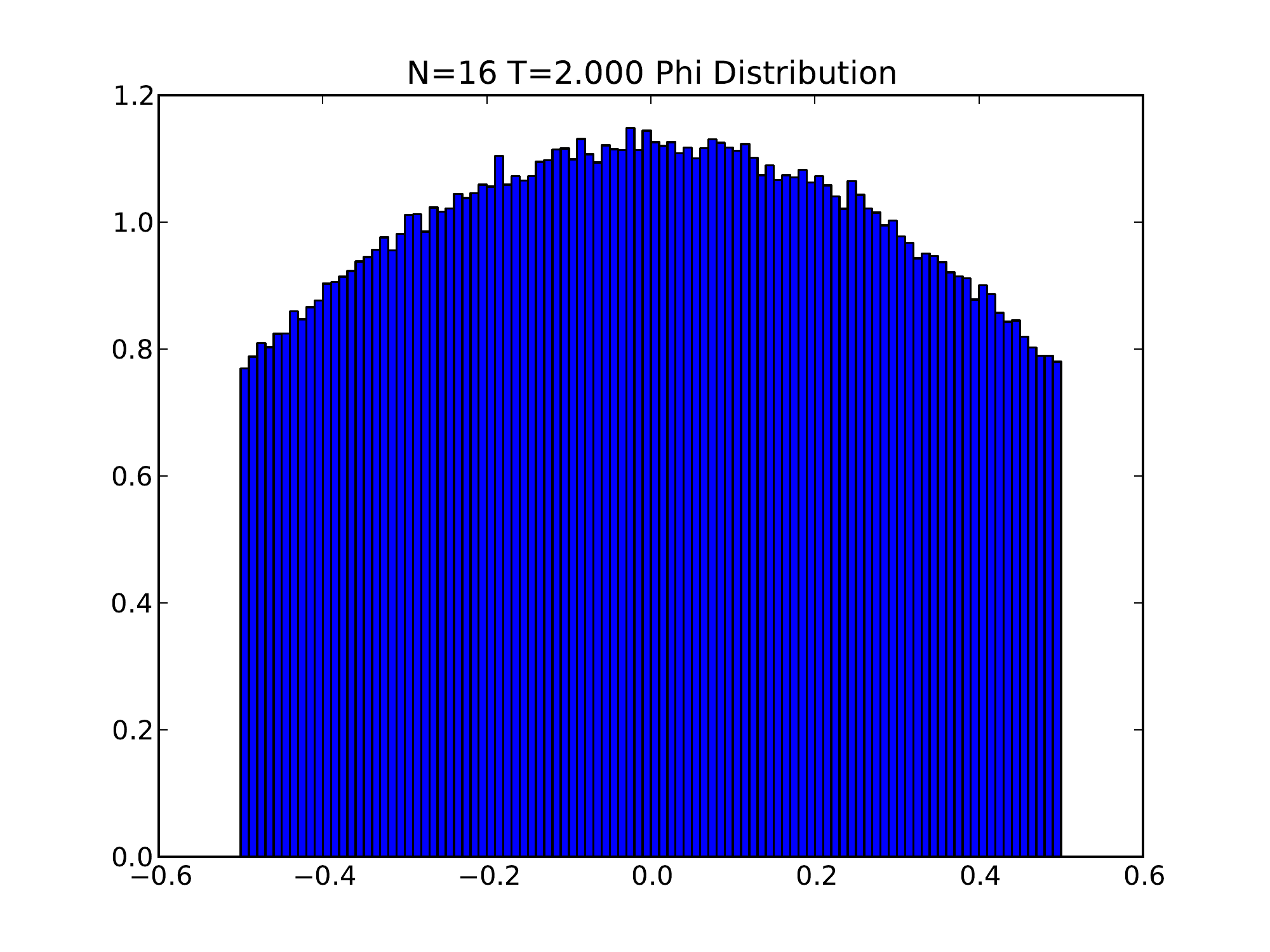}
\hfill
\includegraphics[width=.32\textwidth]{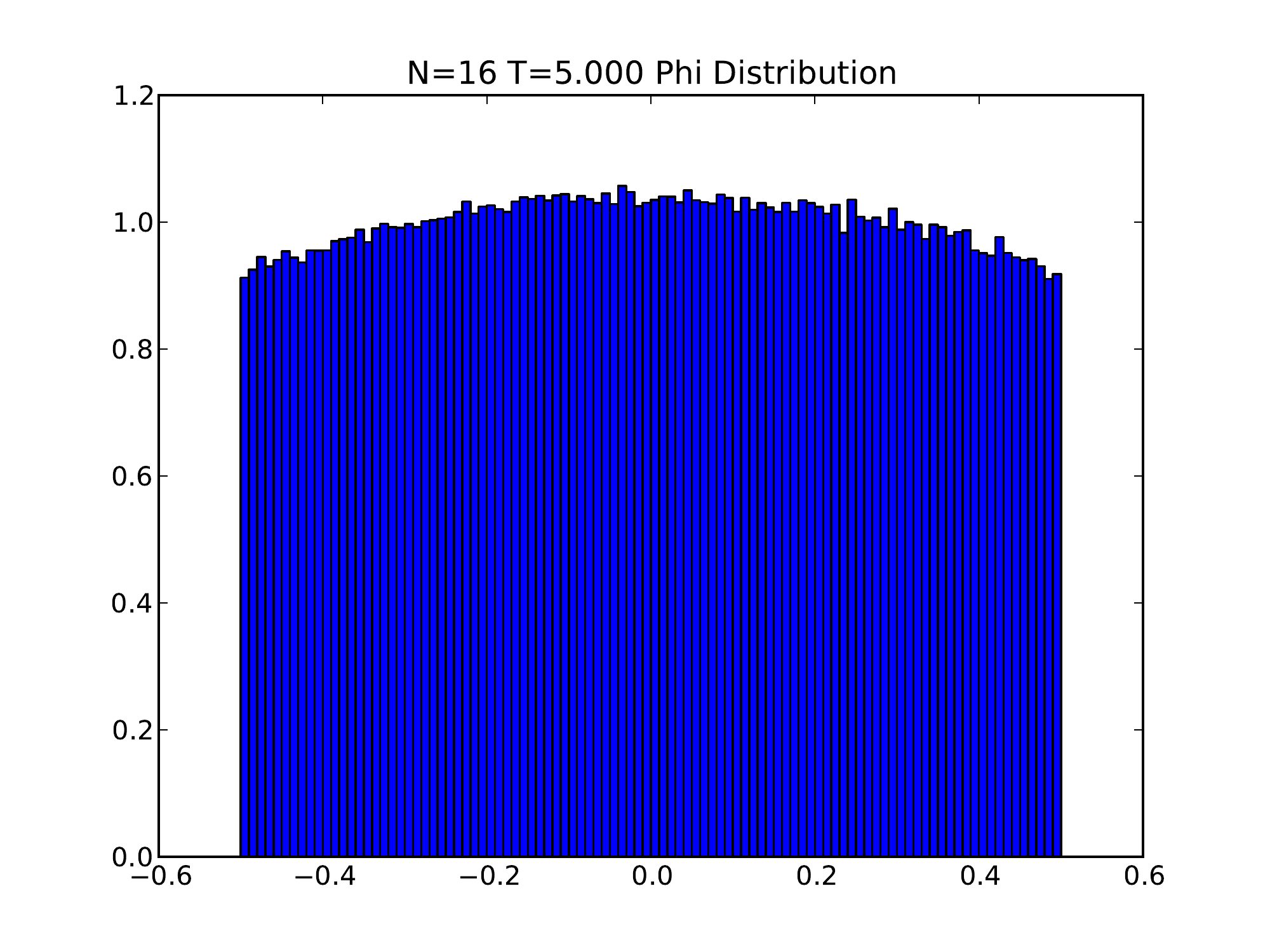}
\hfill
\caption{\label{fig:CoulombScalar2}  LEFT to RIGHT panel: Histograms of distributions of values of $\phi$, for $N=16$, for $T=0.4, 2$ and $5$. The histograms  for $N=32$ are identical. These confirm our interpretation of Fig.~\ref{fig:CoulombScalar}. (The histograms are normalized, i.e., the area under each curve equals one. In this and the following histograms, $10000$ Monte Carlo sweeps of the lattice were made at each temperature. Configurations were taken at every sweep, with the first 2000 neglected for equilibration.)  }
\end{figure}

On the left panel of Fig.~\ref{fig:CoulombScalar}, we show the results for  $\overline{|\phi|}$. At $T>2$, it can be seen to approach  $0.25$. For a scalar $\phi$ changing between $-1/2$ and $1/2$ (the appropriate values for $\kappa =4\pi$), this indicates that the field is uniformly distributed in this interval as the system is heated up (at smaller temperatures, the field is frozen near $\phi=0$). We have also produced histograms of the $\phi$-distributions showing approximately flat distributions with a slight peak near $\phi=0$, thus confirming this conclusion, see Fig.~\ref{fig:CoulombScalar2}. The data for the susceptibility of $\phi$ on the right panel of Fig.~\ref{fig:CoulombScalar} also shows that there is no indication of a  phase transition for $\phi$, as the susceptibility would be expected to grow with the volume (a study of the histograms also does not reveal phase coexistence).

We have also simulated the $W$-bion gas with $\phi$ outright set to zero and have found the results to be in qualitative agreement with the plots on Fig.~\ref{fig:CoulombCharges}. In a similar vein, we have also simulated the $\phi$-field alone, without $W$-bosons or magnetic bions and found a behavior qualitatively similar to that shown on Figs.~\ref{fig:CoulombScalar} and \ref{fig:CoulombScalar2}. 

The behavior described in the above  two paragraphs can be qualitatively explained as follows. It helps to look at the properties of the $W$-boson  fugacities (or core energies)---either the exact one-loop expression (\ref{the W fugacity as function of phi}) (shown by a thick line on   Fig.~\ref{fig:Fugacity}), or the modified one (\ref{positivecore}) (shown by a dashed line on   Fig.~\ref{fig:Fugacity}). As a function of $\phi$, both fugacities vary significantly at low $T$, as the leftmost panel on Fig.~\ref{fig:Fugacity} shows. However, at low temperatures, the $W$-boson fugacities are small (even for  $\kappa = 4\pi$) and the $\phi$-fluctuations are governed largely by the $\cosh 2 \phi$ neutral bion potential, which forces $\phi \sim 0$. At higher temperatures, $T>{\cal{O}}(1)$, however, the fugacities vary little with $\phi$ (for $\kappa =4\pi$ this variation is of order $15 \%$). Thus, at the temperatures where the electric charges are relevant, the fluctuations of the $\phi$-field essentially decouple  from the dynamics of the electric charges and thus have little influence on the qualitative properties of the transition. We will later, in section \ref{extrapolationsection}, argue that this result is likely to remain valid at weak coupling.
  
Let us now summarize the qualitative conclusions that can be drawn from our dual-Coulomb gas study for the values of parameters as indicated previously. There is a transition from a magnetic bion dominated low-temperature phase to a $W$-boson-dominated high-temperature phase. The transition appears qualitatively similar to the one studied for the $SU(2)$ QCD(adj) in \cite{Anber:2011gn} and is not qualitatively affected by the presence of the light scalar $\phi$. In particular, it appears that the $\Z_2^{(L)}$ center symmetry, for which $\phi$ is an order parameter, remains unbroken through the deconfinement transition. On the other hand, as $T$ increases past $T_c \sim \pi$,  $\Z_2^{(R)}$ is restored (by the vanishing of the magnetic charge density) and $\Z_2^{(\beta)}$ is broken (by the nonzero electric charge density).

In the next section, we shall study a related formulation of the system (\ref{double C gas with phi field}) in terms of an XY-spin model with a symmetry-breaking perturbation, coupled to the $\phi$ field. This system is free of a sign problem (however, the magnetic bion fugacity is not a free parameter). The results of this study will be in qualitative agreement with the findings made by using the dual-Coulomb gas.

\section{Simulations of the XY model dual to the Coulomb gas}
\label{XY model as a Coulomb gas}

The  XY model description of the partition function (\ref{final expression for Z}) has the action
\begin{eqnarray}
\label{affinexy}
S &&=\sum_{x,\mu} \;\;-\frac{8T}{\pi \kappa}\cos\nabla_\mu \theta_x+ \frac{\kappa}{16 \pi T }\left(\nabla_\mu \phi_x\right)^2  \nonumber\\
&& \; \;+\sum_x\;\;\frac{8e^{-\frac{4\pi}{\kappa}}}{\pi T \kappa^3  }\cosh(2 \phi_x)+2\xi_W(\phi_x)\cos(4\theta_x)\;. 
\end{eqnarray} 
where $\xi_W(\phi)$ is given by (\ref{the W fugacity as function of phi}).\footnote{Here, we are using the exact one loop value for $\xi_W$ from (\ref{the W fugacity as function of phi}), rather then the modified one (given by the argument of the logarithm in (\ref{positivecore})) used in the dual-Coulomb gas case (however, we still remove the large-$\kappa$ suppression term $\kappa^{-3}$ from the neutral bion potential). Thus, at $\kappa = 4\pi$ and for $T>{\cal{O}}(1)$, the coefficient of the $\cos$ term is large (an effect due to the strong coupling taken, discussed in the previous section around Eq.~(\ref{positivecore}), see also Fig.~\ref{fig:Fugacity}). However, the critical temperature in the XY model is largely governed by the relevance vs. irrelevance of the kinetic term for $\theta$, rather than the external field, while the critical exponents strongly depend on the strength of the external field perturbation. This is consistent with our previous findings of weak dependence of $T_c$ on the external field \cite{Anber:2012ig}, and with the study of \cite{Rastelli:2004bt}.} The theory (\ref{affinexy}) is formulated on the same lattice,  of spacing $L=1$. The field $\theta_x$ is a compact scalar field $\theta_x = \theta_x + 2 \pi$ and lives on the same lattice as $\phi_x$; $\nabla_\mu$ denotes the forward lattice derivative. 
The partition function involves a path integral of $e^{- S}$ over the fields $\theta$ and $\phi$. 

The qualitative map of (\ref{affinexy}) to the gauge theory can be described as follows. The field $\theta_x$ can be thought of as the original electric photon (the dual of the dual photon field $\sigma$, recall (\ref{sigmadef})). Its vortices are thus magnetically charged objects, the magnetic bions. Finally  the $\cos 4 \theta_x$ term represents the electric charges, $W$-bosons and partners, interacting via exchange of photons, $\theta_x$, as well as $\phi$.

The relation of (\ref{affinexy}) to the dual-Coulomb gas (\ref{double C gas with phi field}) can be established in more detail following the steps described  in \cite{Anber:2011gn}. We will not do these in detail and will only remind the reader of the correspondence.  
First, note that if $\phi=0$, Eq.~(\ref{affinexy}) is the $W$-boson/magnetic bion gas for $n_f > 1$ QCD(adj) in one of the several duality-frame versions formulated by two of us in \cite{Anber:2011gn}. The magnetic bions are the XY unit-charge vortices, and the $W$-bosons are represented by the $\sum_x\cos(4\theta_x)$ term (expanding this term and integrating over $\theta$ one finds $W$-$W$ interactions as well as $W$-magnetic bion interactions via the Aharonov-Bohm phase---see  \cite{Anber:2011gn} and references therein for a derivation). 
 When $\phi$-fluctuations are included, the treatment of the $\cos 4 \theta_x$ term is the same, except that the $W$-boson fugacity becomes $\phi$-dependent. Thus one obtains the dual-Coulomb gas of of (\ref{double C gas with phi field}). The only difference   is that the fugacity of  magnetic bions (the XY-model vortices) is not a free parameter and  is determined by the coefficient of the kinetic term for $\theta$ and the lattice spacing; one expects that the core energy is, roughly,  $\frac{8T}{\pi \kappa}$ times a number of order unity. However, the qualitative expectation is that vortices (magnetic bions) populate the system at low-$T$ (when fluctuations of $\theta$ are not suppressed) and are suppressed at high-$T$ (when $\theta$ fluctuations are strongly damped). Furthermore, in the continuum limit, the interactions between $W$-bosons and vortices (and between vortices themselves) obtained from Eq.~(\ref{affinexy}) exactly reproduce the ones from Eq.~(\ref{double C gas with phi field}) (see section 3.3.2 in \cite{Anber:2011gn}). 
 
  \begin{figure}[h]
\centering %
\includegraphics[width=.45\textwidth]{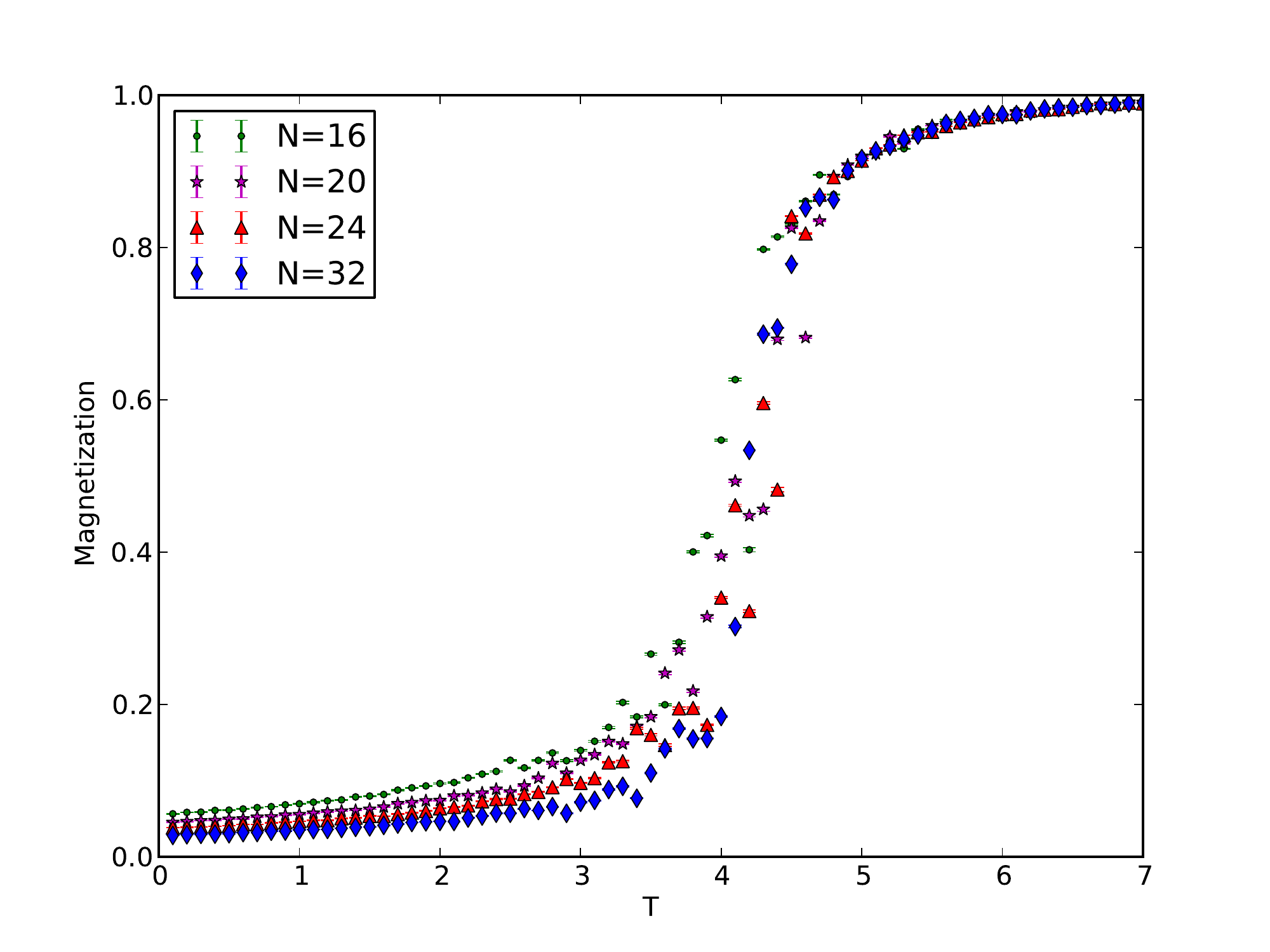}
\hfill
\includegraphics[width=.45\textwidth]{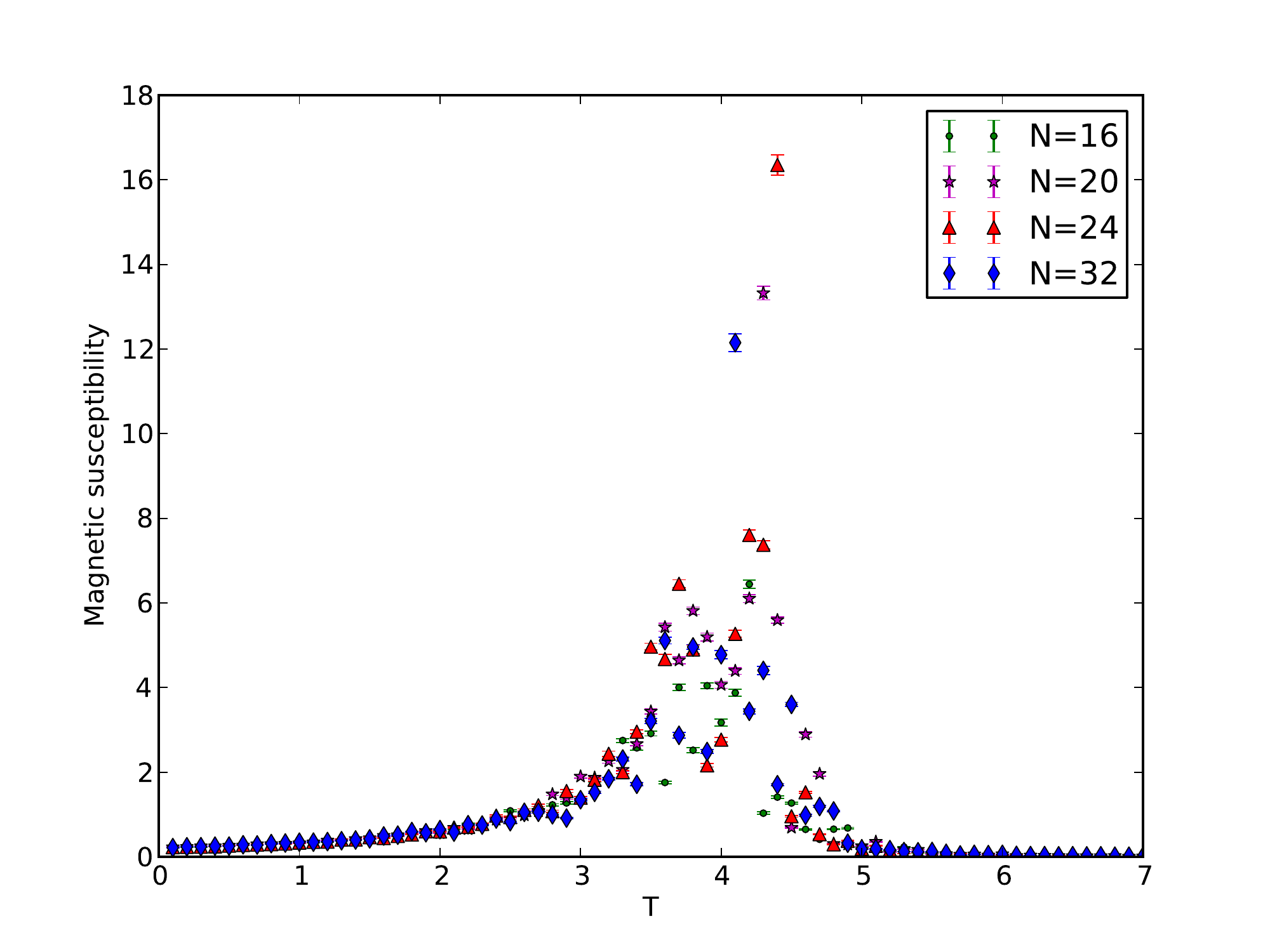}
\caption{\label{fig:XYMag1} LEFT: XY-model ``magnetization". RIGHT: susceptibility of ``magnetization"; see text and Eq.~(\ref{magnetization}). }
\end{figure}

The global $U(1)$ shift symmetry of the XY model is broken to $\Z_4$ by the  external-field  $\cos 4 \theta_x$ term; $\Z_4$ acts as $\theta_x \rightarrow \theta_x +  \pi/2$. A $\Z_2^{(\beta)}$ subgroup thereof can be identified with the center symmetry associated with $\S^1_\beta$ thermal circle. An insertion of $e^{i \theta_x}$ in the XY-model partition function is interpreted as the insertion of an external particle with electric charge one-quarter that of a $W$-boson (such probes do not exist in the $SU(2)$ SYM theory). However, an  insertion of $e^{2 i \theta_x}$ represents an electric probe with one-half the $W$-boson charge. Such probes are insertions of non-dynamical  fundamental quarks used to probe confinement. 
In our simulations, we will probe the $\Z_2^{(\beta)}$ symmetry 
realization by studying the corresponding order parameter (``magnetization") and its susceptibility, defined as
\begin{eqnarray}
\label{magnetization}
m &=& {1 \over N^2} \big\langle \vert \sum_x e^{i \theta_x} \vert \big\rangle = {\big\langle |M| \big\rangle \over N^2} \nonumber \\
\chi(m) &=&{ \big\langle |M|^2 \big\rangle - \big\langle|M|\big\rangle^2 \over N^2} = \sum_x \big\langle e^{i \theta_x} e^{- i \theta_0} \big\rangle_{(conn.)}~.\end{eqnarray}
In  the second line above, we have also shown $\chi(m)$ in terms of the usual Green's function (note that the connected correlator is computed).

\begin{figure}[h]
\centering %
\includegraphics[width=.32\textwidth]{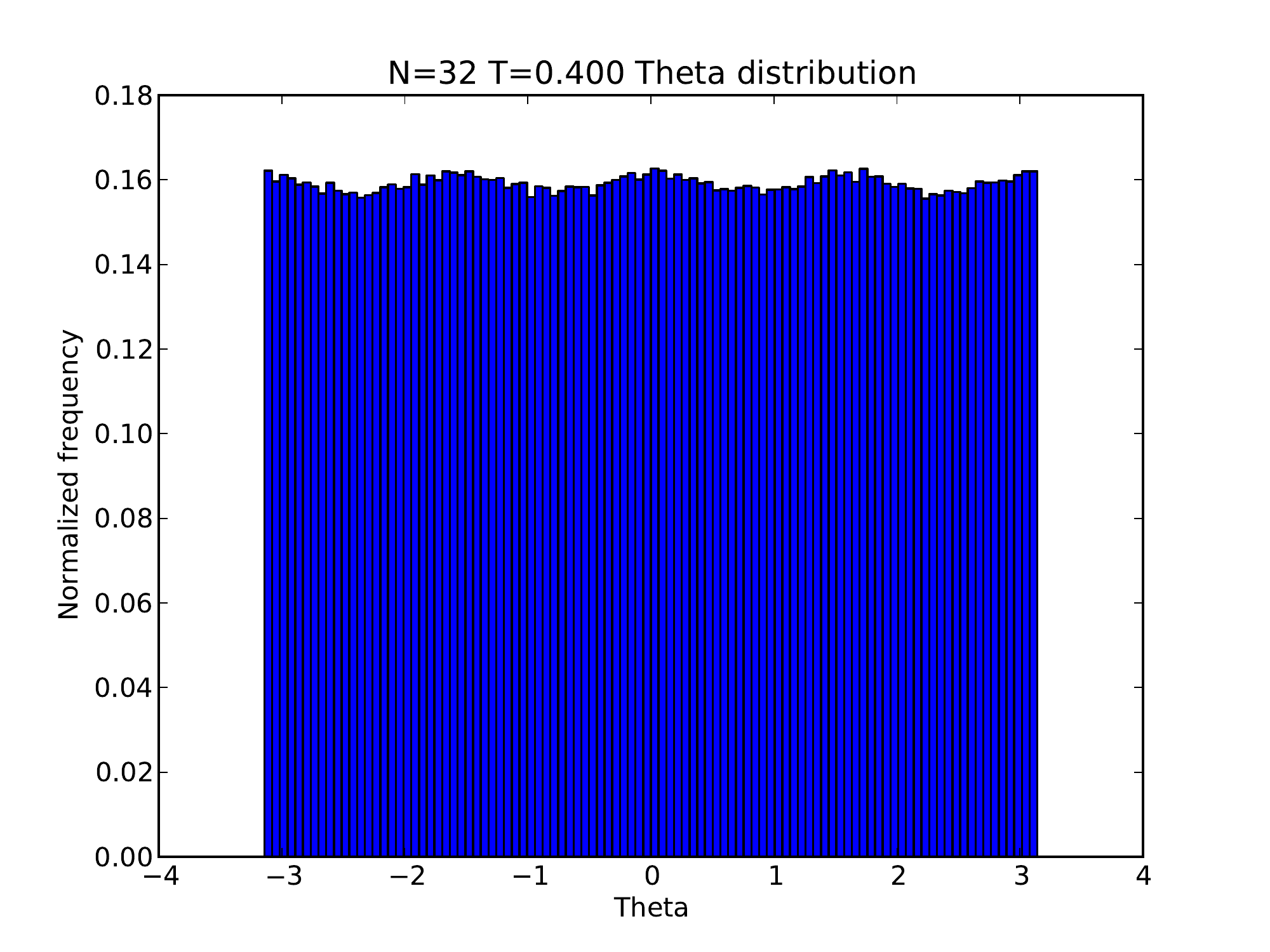}
\hfill
\includegraphics[width=.32\textwidth]{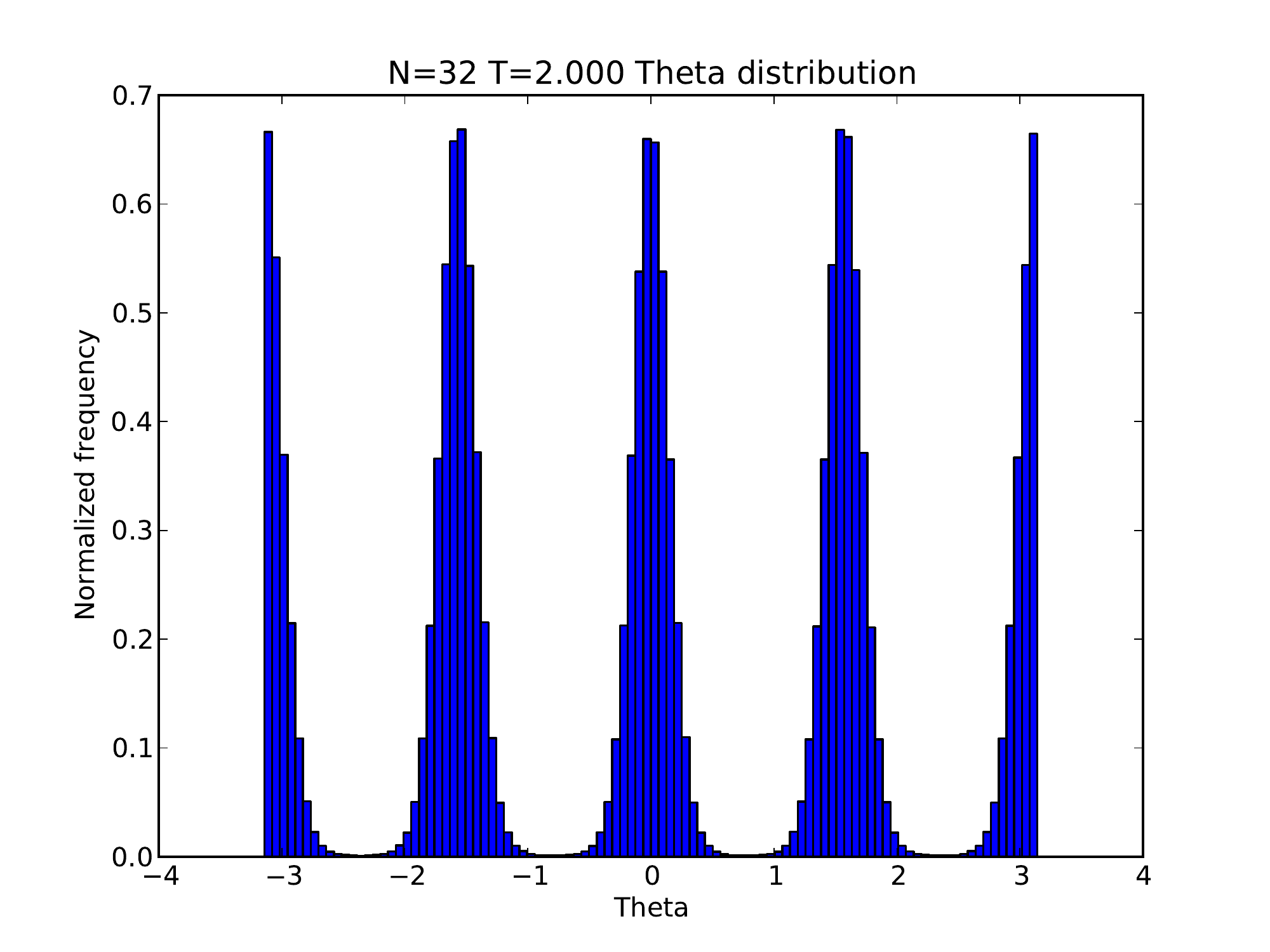}
\hfill
\includegraphics[width=.32\textwidth]{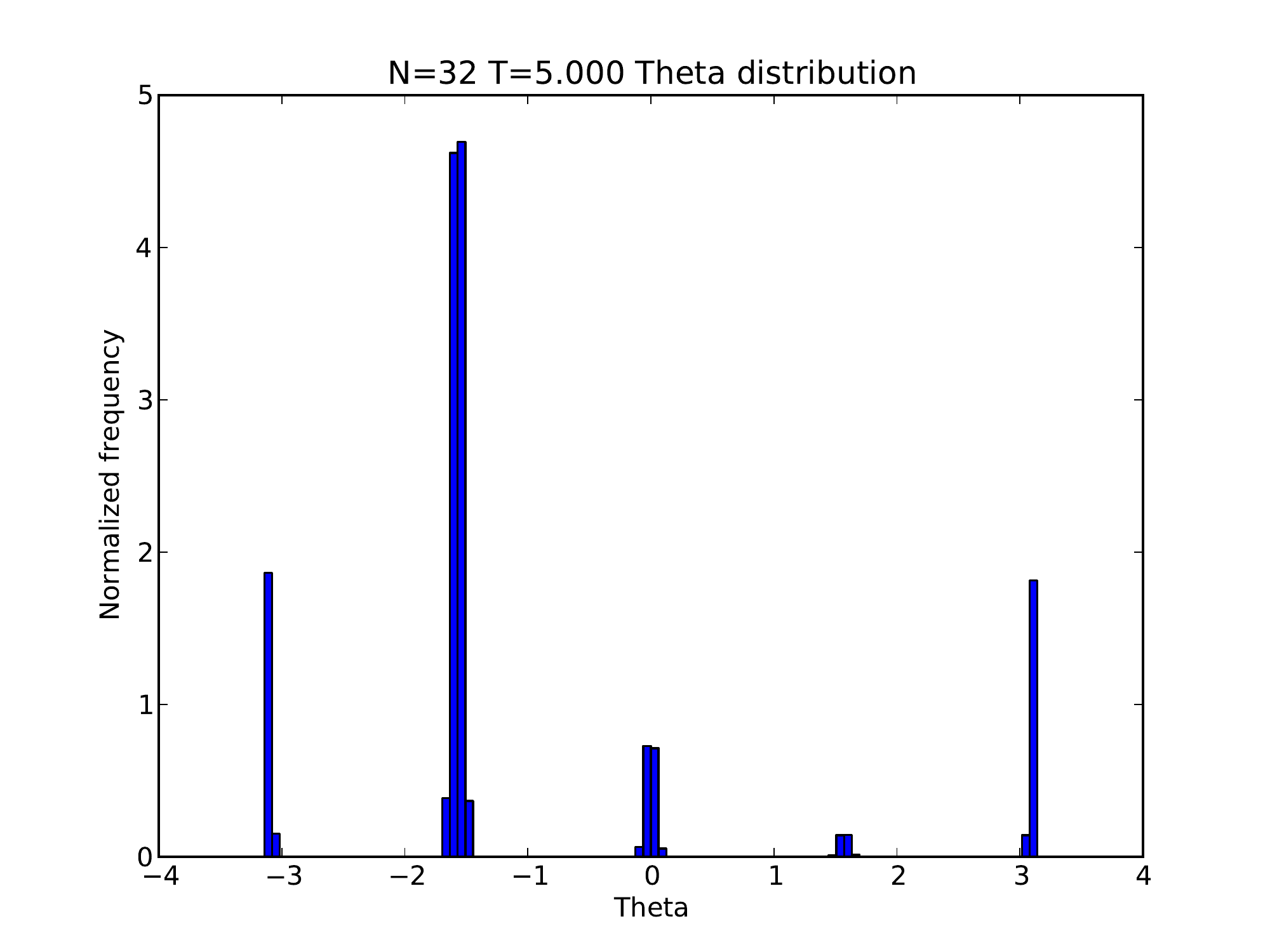}
\hfill
\caption{\label{fig:XYMag2}  LEFT to RIGHT panel: Histograms of distributions of $\theta_x$ for $T=0.4, 2$ and $5$. A clear indication of $\Z_4$ (center symmetry) breaking is seen.}
\end{figure}

 We have simulated the XY model for the same values of the parameters as for the dual-Coulomb gas. In order to get some qualitative idea about finite-size scaling, we have studied volumes with lattice widths $N=16,20,24, 32$.  Again, we simulate the system through a range of temperatures, performing 10000 sweeps at each temperature, with the first 500 sweeps  disregarded for equilibration. Data (such as the value of the action, the instantaneous magnetization, the density of vortices\footnote{Identified by the algorithm of \cite{Tobochnik:1979zz}.} of $\theta$, and the mean value of both $\phi$ and $|\phi|$) was recorded at the end of every sweep. Simulations were initialized at low temperature in a configuration with the $\theta$ field uniformly distributed in the range $[-\pi,\pi]$, and the $\phi$ field uniformly distributed in the range $[-{2\pi\over\kappa},{2\pi\over\kappa}]$. Each Metropolis iteration consisted of an attempt to change the value of the $\theta$ field at a random lattice site to a random value in the range $[-\pi,\pi]$, followed by an analogous attempt to change the value of the $\phi$ field at a random lattice site to a random value in the range $[-{2\pi\over\kappa},{2\pi\over\kappa}]$. In both cases, changes were accepted with the usual probability $p = \min(1,e^{-\Delta S})$ designed to produce configurations with probabilities that satisfy Boltzmann statistics.

\begin{figure}[h]
\centering %
\includegraphics[width=.50\textwidth]{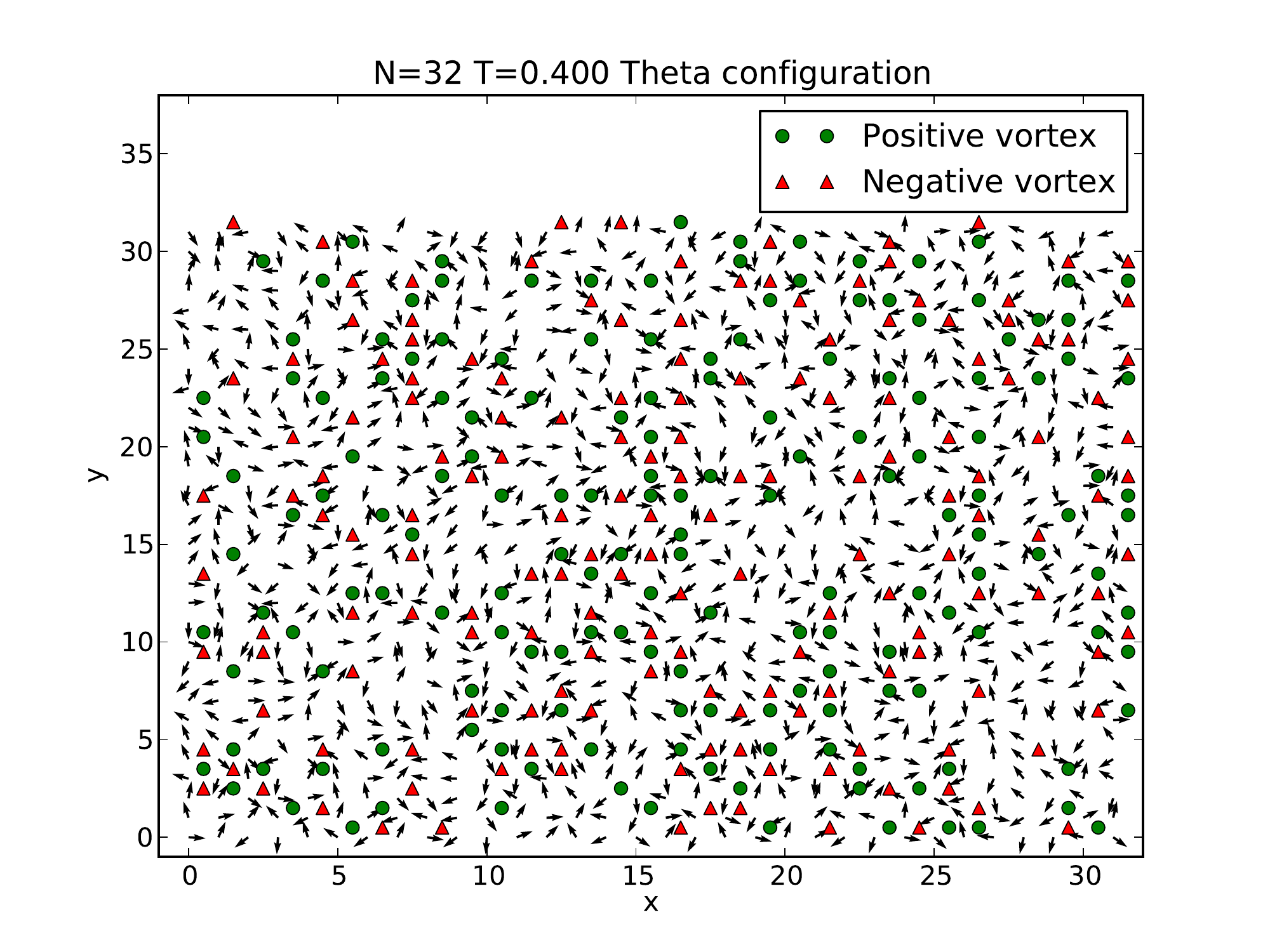}
\hfill
\includegraphics[width=.50\textwidth]{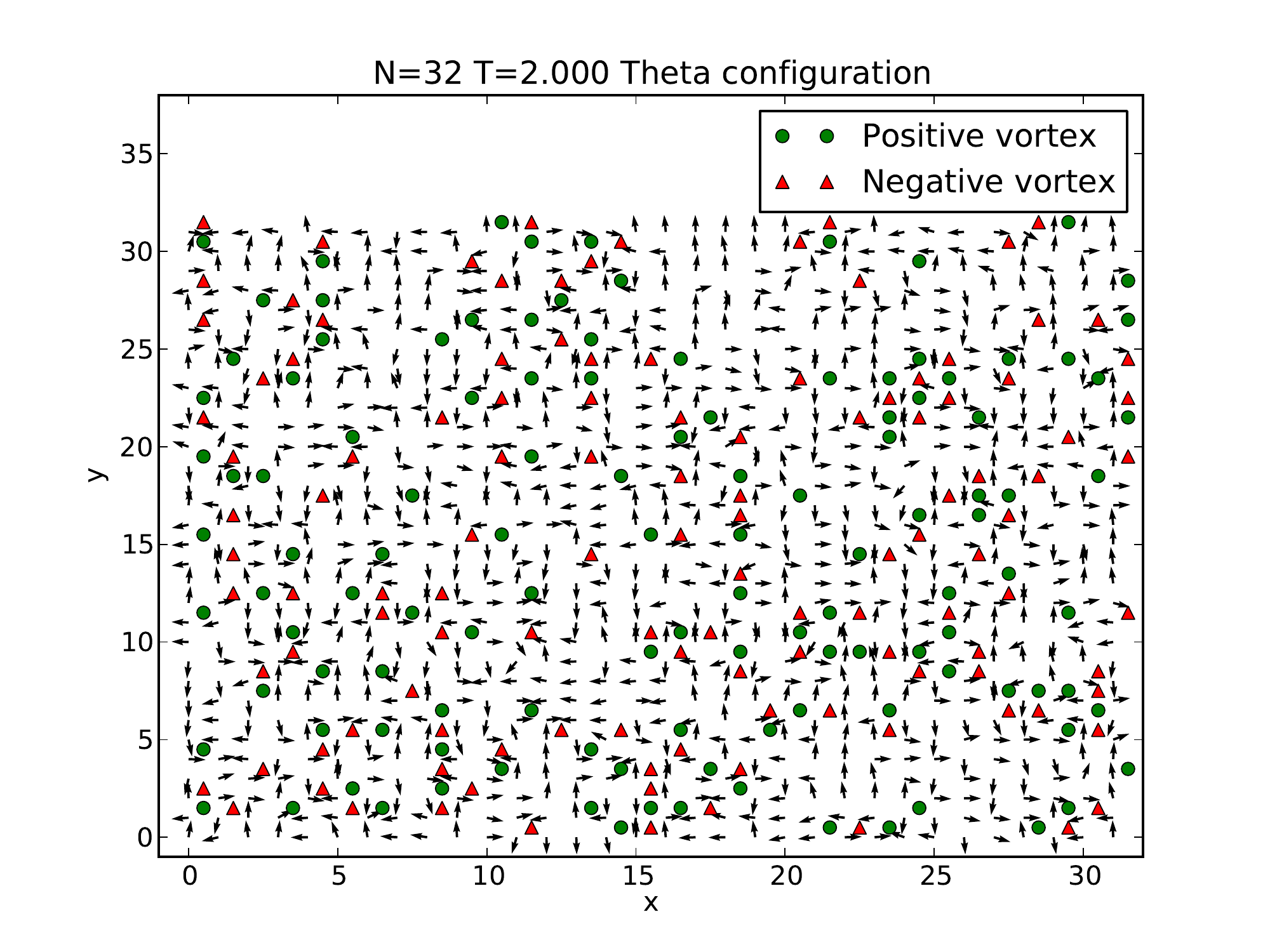}
\hfill
\includegraphics[width=.50\textwidth]{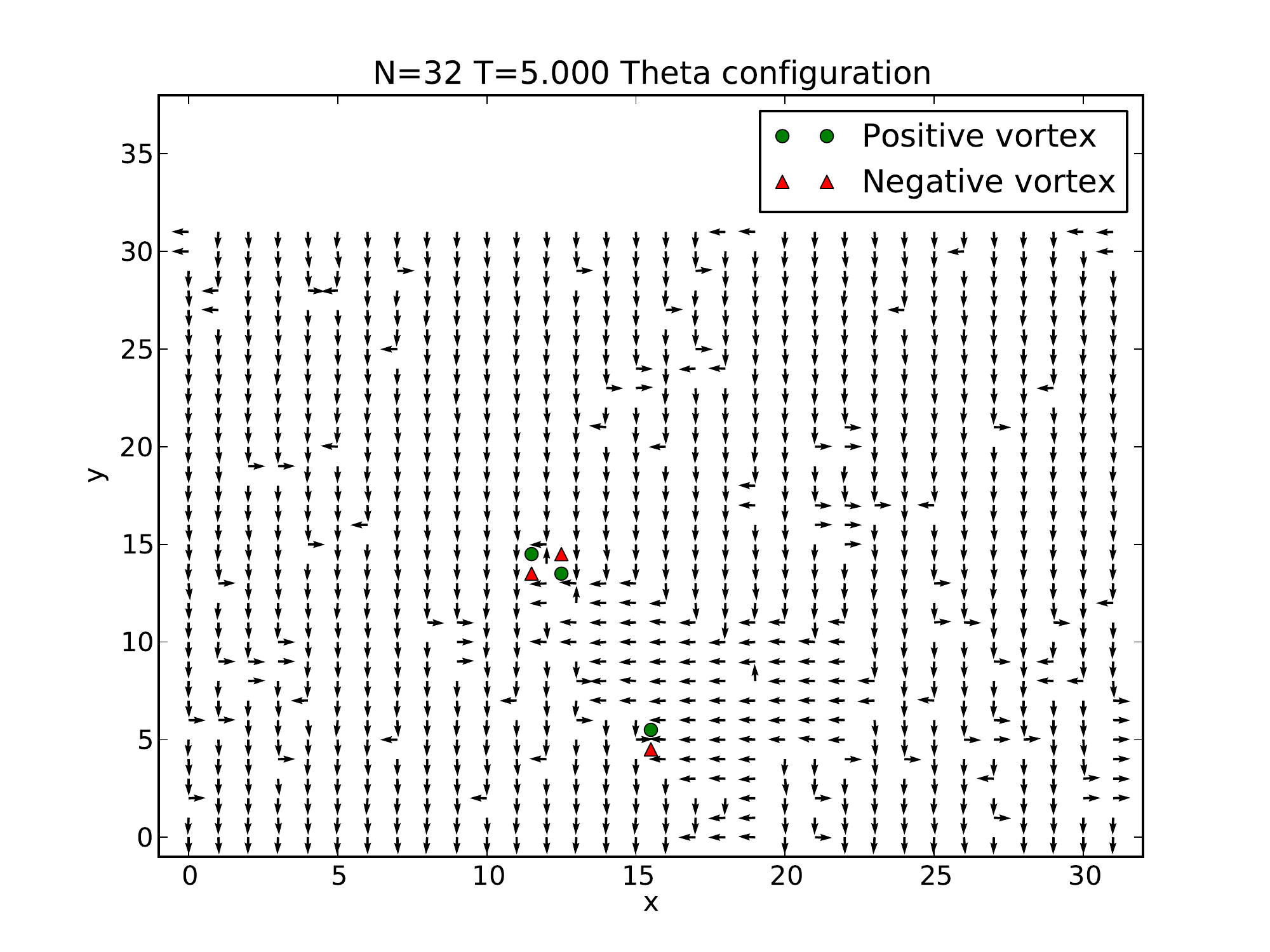}
\hfill
\caption{\label{fig:XYMag3}  TOP to BOTTOM panel: lattice configurations at $T=0.4, 2$ and $5$, showing the directions of $\theta_x$ as  arrows, as well as positive and negative vortices. See text for further discussion.}
\end{figure}

We begin by first discussing the results for the magnetization. On Fig.~\ref{fig:XYMag1}, we show the temperature dependence of the magnetization (\ref{magnetization})  and its susceptibility, for the four volumes. There is a clear transition (becoming sharper with increasing $N$), near $T\sim 4$, from a low-temperature phase with zero magnetization to a high-temperature phase with maximal magnetization. Furthermore, on the right panel, the susceptibility is seen to increase with the volume (we note that this is a qualitative observation; we have not performed  high-statistics and small temperature step simulations of the near-critical region). 
To corroborate the symmetry-breaking conclusion, on Fig.~\ref{fig:XYMag2}, we show histograms of the distributions of $\theta_x$ (which takes values in the range $[-\pi$, $\pi$]). A $\Z_4$ broken symmetry is clearly seen above the transition.

To end the discussion of the $\theta_x$ configurations, on Fig.~\ref{fig:XYMag3}, we show three typical lattice configurations at the same values (top to bottom) of $T=0.4, 2$ and $5$. The positions of the vortices (magnetic bions) are also indicated. It is clear that vortices disappear above $T_c$ and become bound in a small number of small dipole pairs, indicating that magnetic bions are confined above $T_c$. The ordering of $\theta$ at high-$T$ is also clearly seen.  At low $T$, on the other hand, the finite-density magnetic plasma  confines electric charges. It disorders the ``photon" (really, Polyakov loop) $\theta_x$,  which acquires a finite correlation length and thus restores the $\Z_2^{(\beta)}$ thermal center symmetry.

\begin{figure}[h]
\centering %
\includegraphics[width=.45\textwidth]{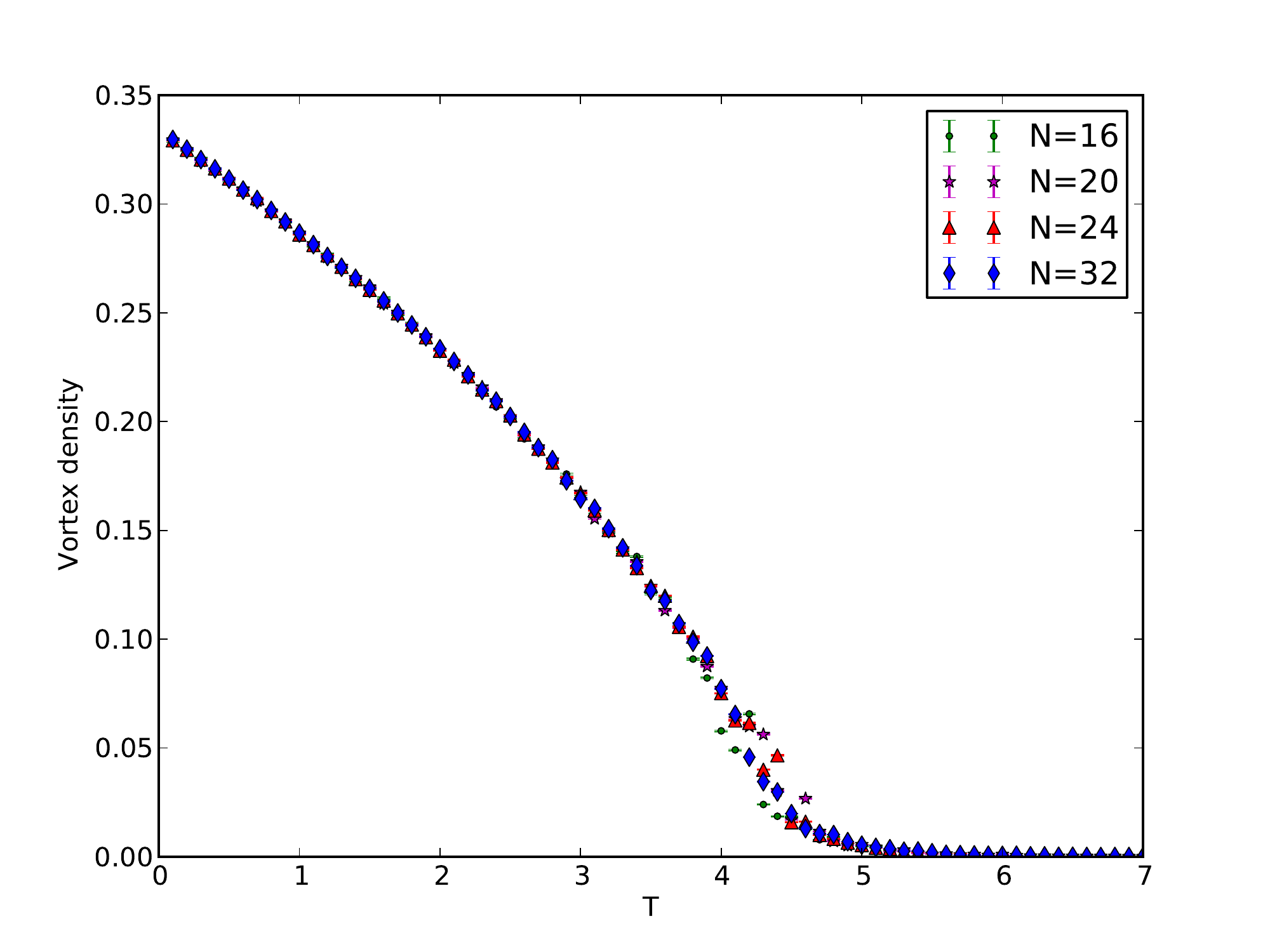}
\hfill
\includegraphics[width=.45\textwidth]{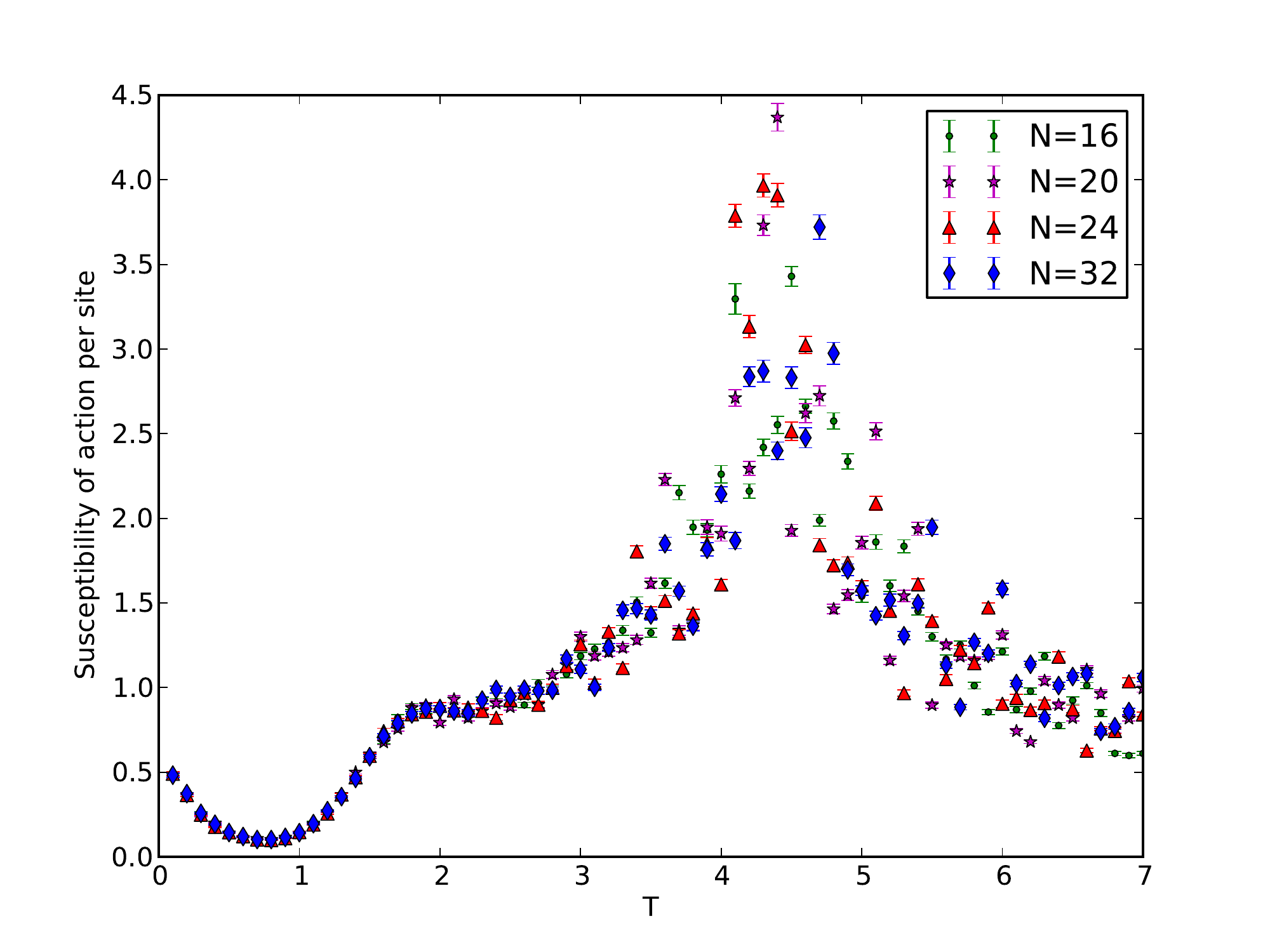}
\hfill
\caption{\label{fig:XY4}  LEFT: Vortex (magnetic bion) density. RIGHT: Susceptibility of action per site.  }
\end{figure}

The $\Z_2^{(R)}$ discrete chiral symmetry can not be identified in the formulation of the XY model in Eq.~(\ref{affinexy}). 
Recall that magnetic bions (the XY-model vortices) are neutral under $\Z_2^{(R)}$, but their constituent monopole-instantons are not (thus, one would need to define one-half vortex operators). We will instead study the vortex (i.e., magnetic bion) density and qualitatively associate the nonzero magnetic charge density with $\Z_2^{(R)}$ breaking, as expected at low-$T$. On the left panel of Fig.~\ref{fig:XY4}, we show the vortex  (magnetic bion) density, which clearly decreases sharply for $T>4$ (this is further corroborated by the lattice configurations on Fig.~\ref{fig:XYMag3}). On the right panel of the same figure, we show the susceptibility of the action per site; while it shows a peak around the same values $T \sim4$, the large-volume behavior is less pronounced as that for the magnetic susceptibility (a more detailed  study is beyond our qualitative goals here).

\begin{figure}[h]
\centering %
\includegraphics[width=.45\textwidth]{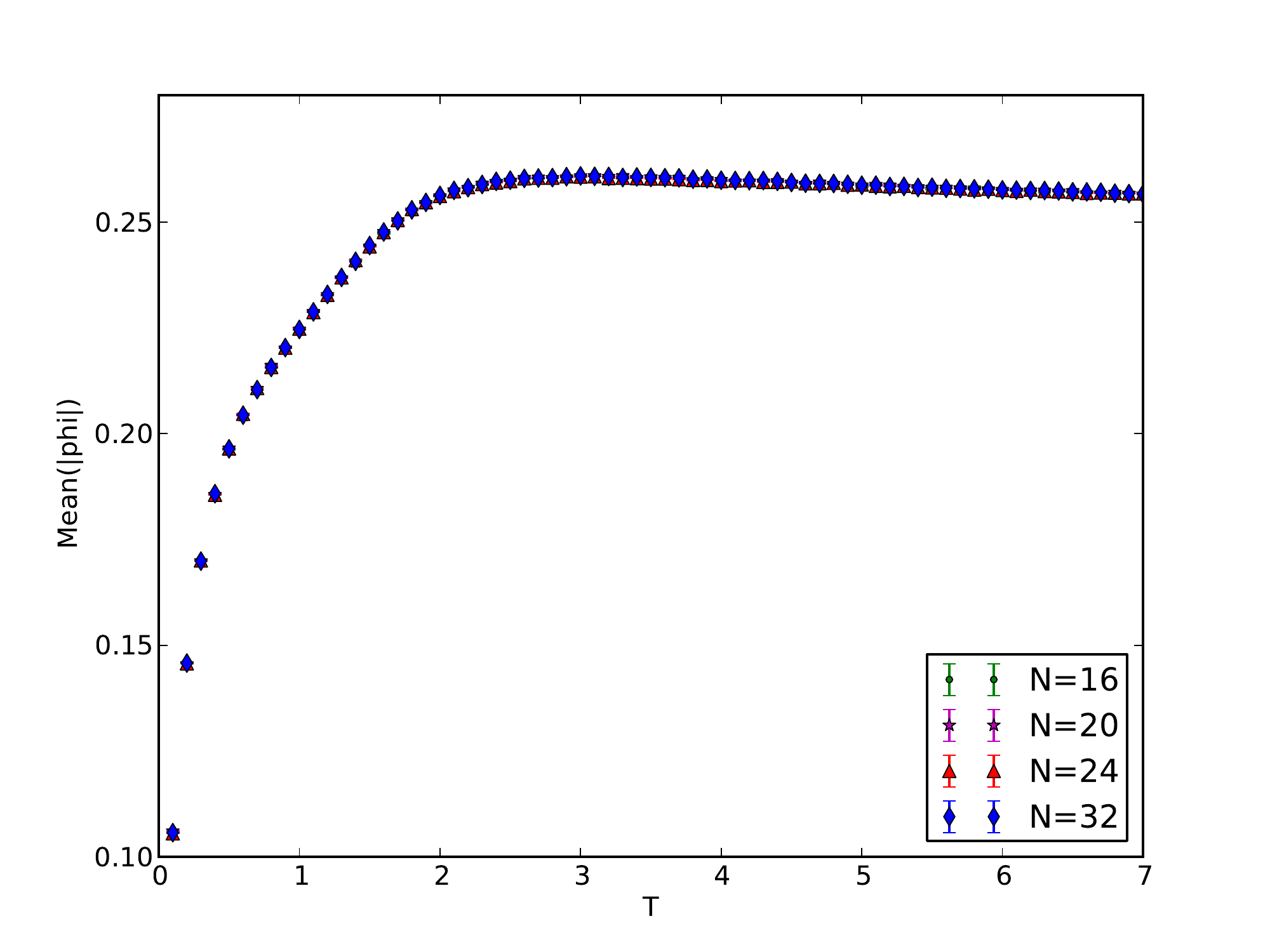}
\hfill
\includegraphics[width=.45\textwidth]{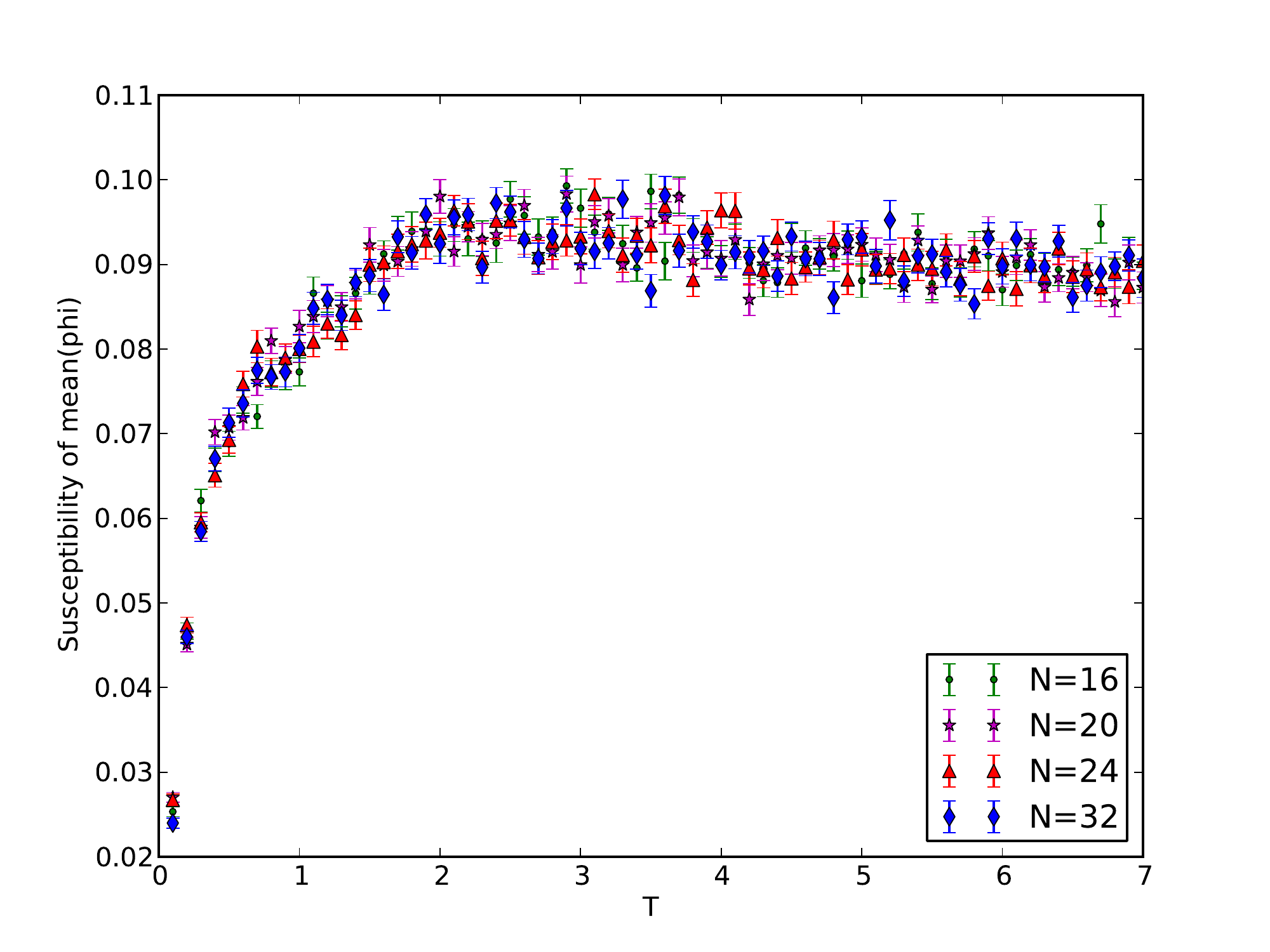}
\caption{\label{fig:XYScalar1} XY-model scalar field observables, Eq.~(\ref{scalarobservables}). LEFT: the average of $|\phi|$. RIGHT: susceptibility of $\phi$. We interpret these results as showing that, for $T>1.5$, the field strongly fluctuates around $\phi=0$, consistent with the dual-Coulomb gas formulation. No growth of $\chi(\phi)$ with $N$ is seen. }
\end{figure}

\begin{figure}[h]
\centering %
\includegraphics[width=.32\textwidth]{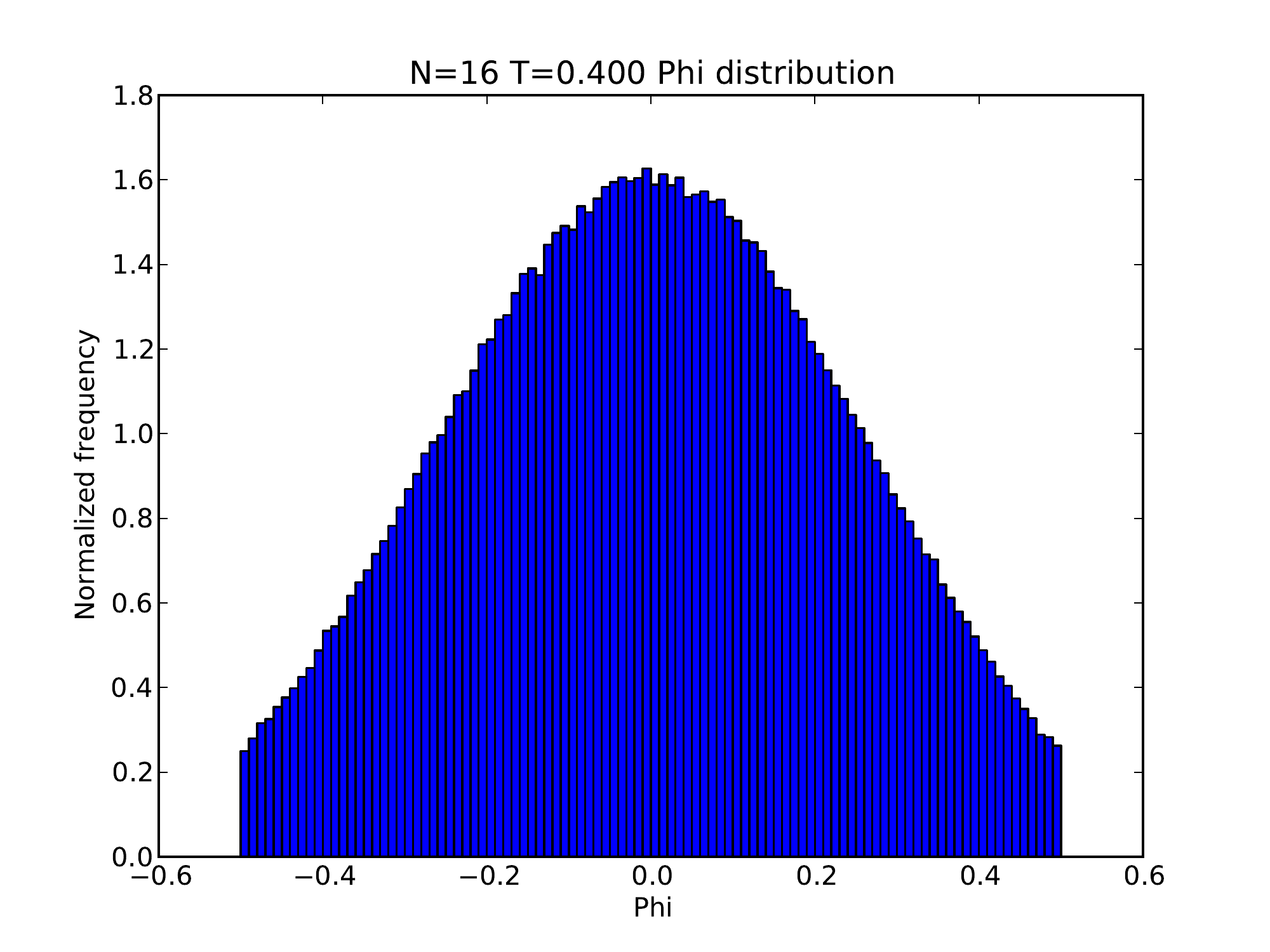}
\hfill
\includegraphics[width=.32\textwidth]{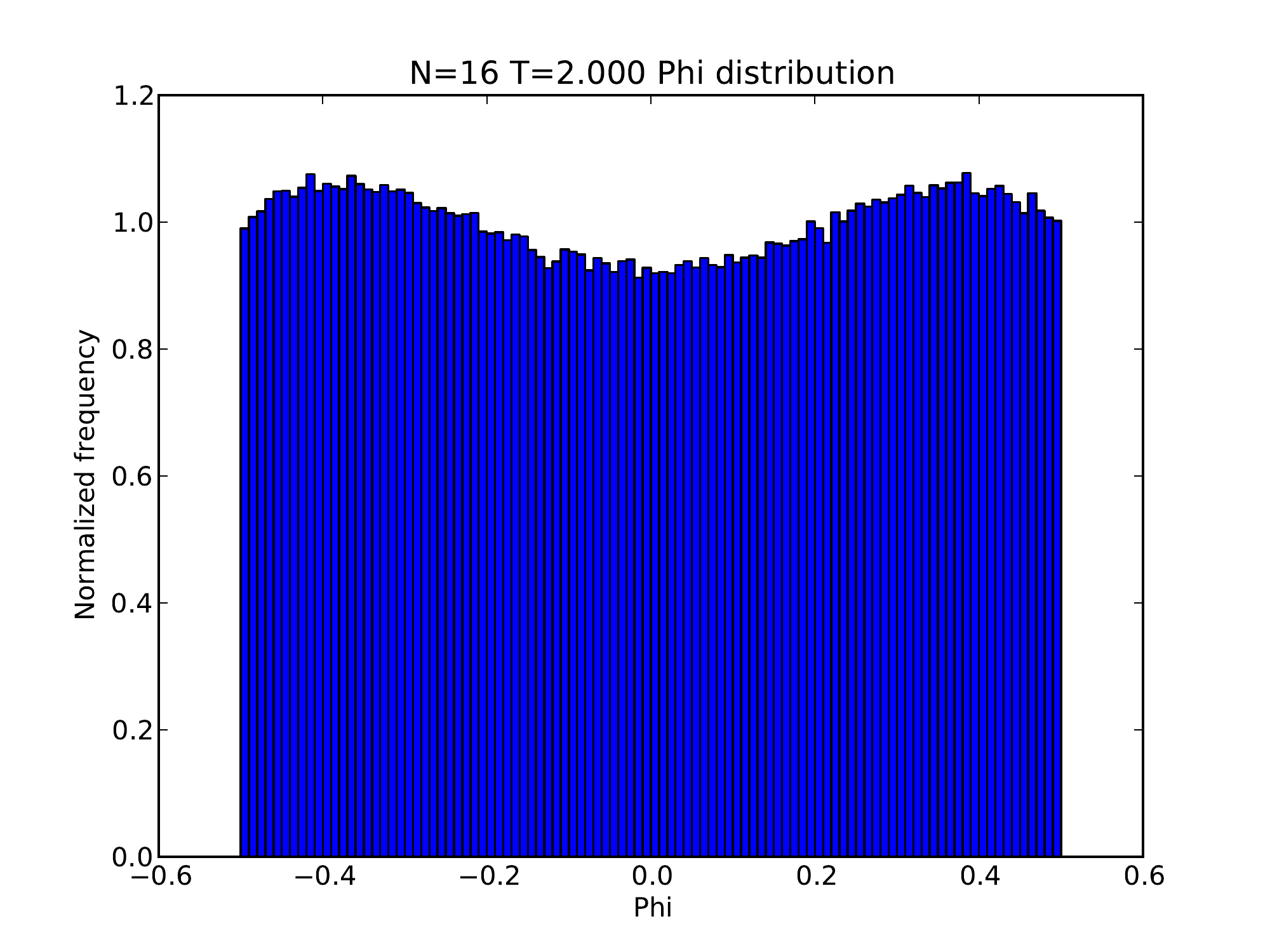}
\hfill
\includegraphics[width=.32\textwidth]{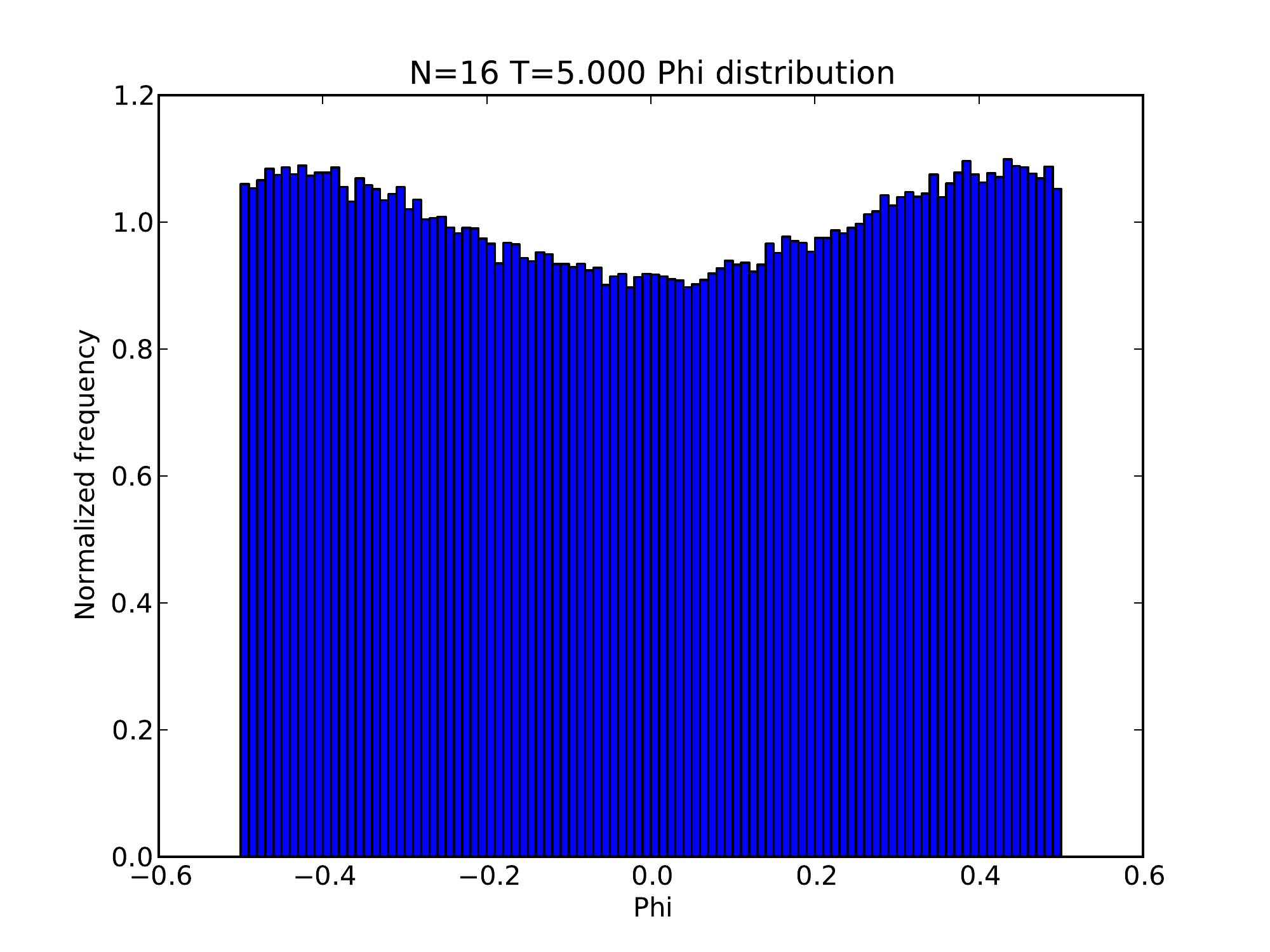}
\caption{\label{fig:XYScalar2}  LEFT to RIGHT panel: Histograms of distributions of values of $\phi$, for $N=16$, for $T=0.4, 2$ and $5$. Compared to the dual-Coulomb gas, a two-bump structure appears near $T=2$. Notice, however, that the susceptibility of $\phi$, Fig.~\ref{fig:XYScalar1}, does not grow with volume. The histograms  for all other values of $N$ are identical. No  pronounced peaks or growth of susceptibility $\chi(\phi)$, as the one associated with $\theta$ are seen. }
\end{figure}

We end with a discussion of the  scalar field observables.
As formulated, the XY-model (\ref{affinexy}) retains the $\Z_2^{(L)}$ center symmetry $\phi_x \rightarrow - \phi_x$ associated with the $\S^1_{L}$ circle in gauge theory.  We  probe the  $\Z_2^{(L)}$ realization by studying the same scalar field observables (\ref{scalarobservables}) as in the dual-Coulomb gas. In Fig.~\ref{fig:XYScalar1}, we show the same observables as in Fig.~\ref{fig:CoulombScalar}. The behavior of $\overline{|\phi|}$ is quantitatively similar to that in the Coulomb gas, i.e. indicates that the field tends to strongly fluctuate at high-$T$, taking all possible values (note that, this time, $0.25$ is approached from above). The behavior of the distributions of $\phi_x$, shown in the histograms on Fig.~\ref{fig:XYScalar2} confirms this expectation. We note that there is an indication of two small bumps at nonzero values of $\phi$. We believe that these are correlated with the two small bumps in the fugacity $\xi_W(\phi)$ at the same values of $T$---see the thick line plots on  Fig.~\ref{fig:Fugacity}.  However, the susceptibility of $\phi$ does not show an increase with volume and we see no indication of a phase transition (the right panel on Fig.~\ref{fig:XYScalar1}). Also, as in   the dual-Coulomb gas study, the study of the action histograms as a function of $T$  reveals no evidence of coexistence of broken- and unbroken-$\Z_2^{(L)}$  phases. We take the results for the scalar field observables as indicating that there is no phase transition breaking the $\Z_2^{(L)}$ center symmetry near the deconfinement transition, consistent with the finding in the dual-Coulomb gas picture. A study of the high-$T$ limit of the effective potential (see  Eq.~(\ref{full potential on T2}))  shows that at $T\gg 1/L$, the $\Z_2^{(L)}$ symmetry is broken. This temperature regime is beyond the goals of this study (and beyond the validity of our 2D effective dual-Coulomb gas  description).

\subsection{Comments on extrapolating simulation results to weak coupling}
\label{extrapolationsection}

As explained in section \ref{Simulations of the Coulomb gas}, all our simulations were performed for $\kappa = {g^2\over 2 \pi} = 4 \pi$. This value of $\kappa$ is far away from the $g^2({1\over L}) \rightarrow 0$ (and hence $\kappa \rightarrow 0$)  regime of small-$\Lambda L$, where the partition function (\ref{total partition function}) was reliably derived. In this section, we will contrast the qualitative behavior of the fugacity (\ref{the W fugacity as function of phi}) and the scalar potential in ${\cal Z}_{\mbox{\scriptsize grand}}$ at small and large $\kappa$.  We stress from the outset that we can not analytically prove that the small-$\kappa$ behavior will be exactly as found in our large-$\kappa$ simulations. However, we  shall observe that at the qualitative level, the behavior of the potential is similar. 

To state more precisely what we have in mind,   we consider the theory defined in (\ref{affinexy}).
 The main question we want to address is to what extent we expect the $\phi$ fluctuations to decouple from the dynamics of the transition in the weak-coupling (small-$\kappa$) limit. Naturally, minimizing the classical  potential  in (\ref{affinexy}) is not the way to  study the dynamics, as this is a 2D theory with strong fluctuations. 
 We can still learn something, however, from a closer look at the classical potential in (\ref{affinexy}).
 Recall that
 at small-$T$, fluctuations of $\phi$ away from $\phi \simeq 0$ are suppressed by the neutral bion potential. The $W$-boson induced potential, on the other hand, is relevant at temperatures near and above the transition. At these temperatures,  the $\theta_x$ variables essentially orient themselves in one of the minima of the $\cos 4 \theta_x$ potential. Thus, near and above the transition,  the  potential for $\phi_x$ can be found by simply taking $\cos 4 \theta_x = -1$: 
 \begin{eqnarray}
\label{affinexy1}
V(\phi)  = \frac{8e^{-\frac{4\pi}{\kappa}}}{\pi T \kappa^3 }\cosh(2 \phi) - 2\xi_W(\phi) \;,
\end{eqnarray} 
with $\xi_W$ given by (\ref{the W fugacity as function of phi}). Regarding the $\Z_2^{(L)}$ symmetry breaking, the scalar potential (\ref{affinexy1}) represents a ``worst case scenario", as it is the $W$-boson induced potential (the second term above) that favors symmetry breaking.

\begin{figure}[h]
\centering 
\includegraphics[width=.3\textwidth]{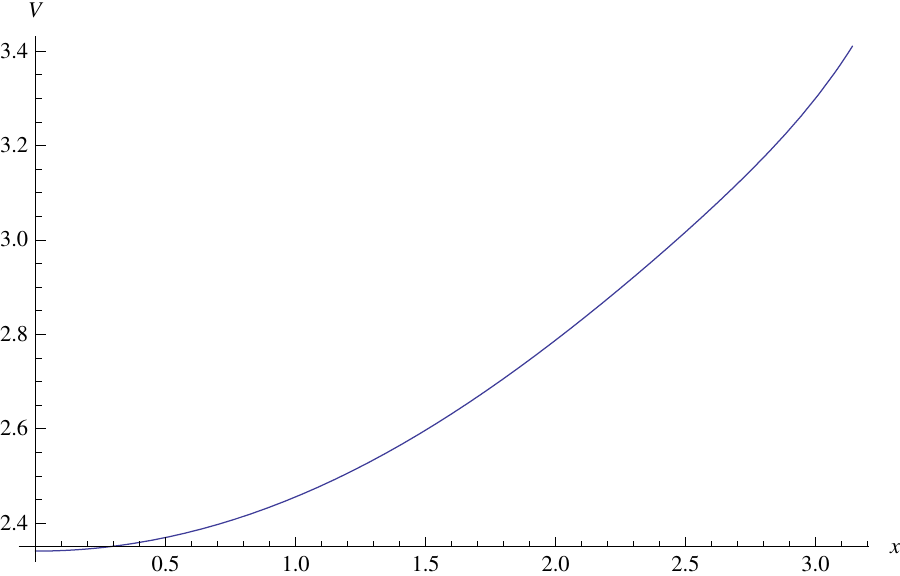}
\hfill
\includegraphics[width=.3\textwidth]{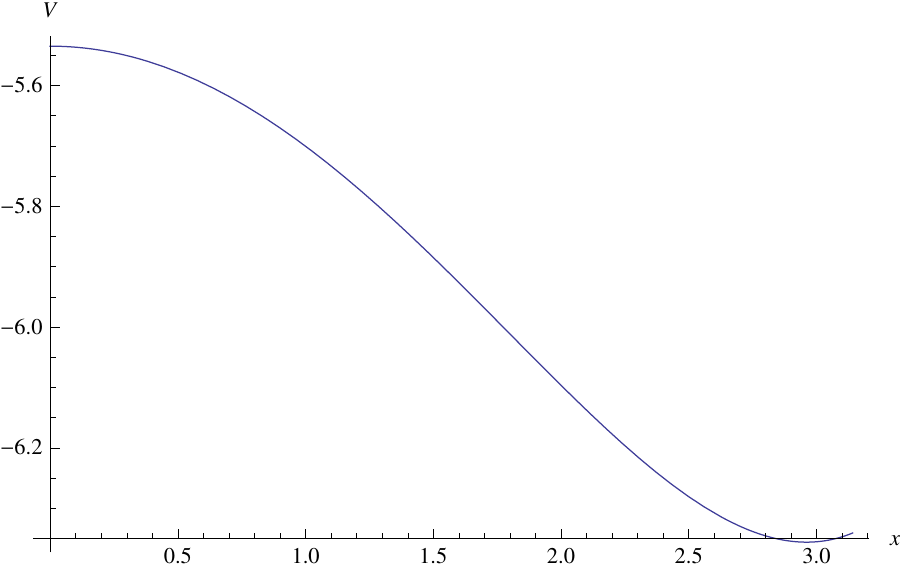}
\hfill
\includegraphics[width=.3\textwidth]{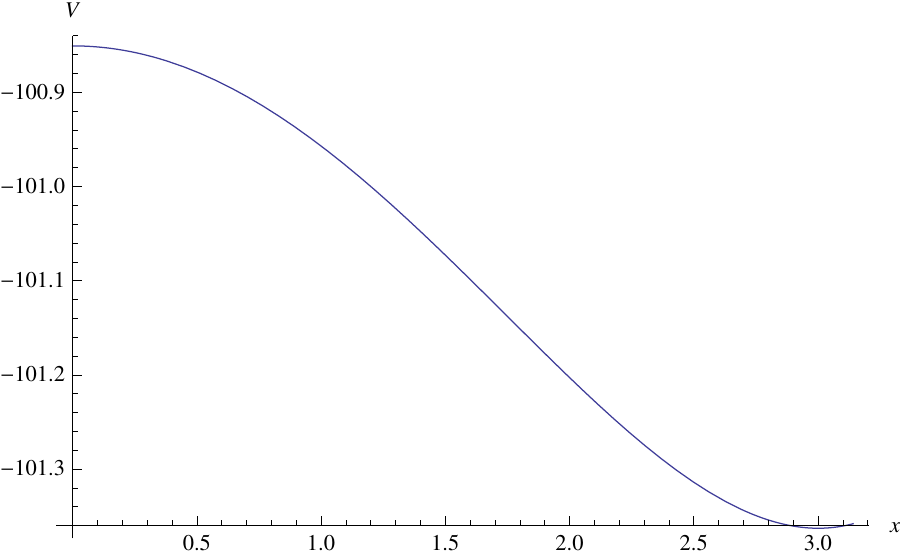}
\caption{\label{fig:Vstrong}The classical potential (\ref{affinexy1}) (modified as described in text) for $\kappa = 4\pi$ as a function of $x = {g^2 \phi \over 4 \pi}$ ($x=\pi$ is the boundary of the Weyl chamber). LEFT to RIGHT: $T=0.4$, $T=2$, $T=5$ (the corresponding ratios $\kappa \over T$ are $31, 6.3, 2.5$).}
\end{figure}

 The potential (\ref{affinexy1}) can be expressed as  function of $x = {g^2 \phi \over 4 \pi}$, which takes values $-\pi < x \le \pi$, as well as the temperature $T$, and the coupling $\kappa$.  At strong coupling $\kappa = 4\pi$, we multiplied  the first term in $V(\phi)$ by $\kappa^3$---as explained above (see section \ref{Simulations of the Coulomb gas}), this was done in order to remove the unphysical suppression of the neutral bion potential by this factor at strong coupling. Then,  on Fig.~\ref{fig:Vstrong}, we plot the potential thus obtained from Eq.~(\ref{affinexy})---as used in our simulation of the XY model---for several values of $T$, as a function of $x$, for $0\le x \le \pi$.  We chose $T=0.4, 2$, and $5$, the same values as in the histogram plots on Figs.~\ref{fig:CoulombScalar2} and \ref{fig:XYScalar2} for our simulations. Recall that the magnetic bion and $W$-boson fugacities (at $\phi=0$) are of the same order at  $T_c \sim {g^2 \over 8 \pi} = {\kappa\over 4} \sim \pi$, while, in our simulations, we found, roughly $T_c \sim 4$, a somewhat larger but comparable value. 
 
 We observe, from the middle and right panel of Fig.~\ref{fig:Vstrong}, that for $T=2$ and $5$ (near-critical and above-critical values),  the classical probability, $\sim e^{-V}$, for $x=0$ ($\phi =0$) is smaller than the probability for $x \sim 2.9$, which naively indicates a symmetry breaking taking place near the deconfinement temperature. However, recall that our simulations of the fluctuating theory, which take into account the $\phi$ and magnetic bion fluctuations not included in the classical potential (\ref{affinexy1}), found an essentially flat $\phi$ distribution and no evidence for symmetry breaking. As we stressed above, because we are dealing with a 2D theory, it is not possible to draw conclusions about symmetry breaking patterns based on the classical potential.

\begin{figure}[h]
\centering 
\includegraphics[width=.3\textwidth]{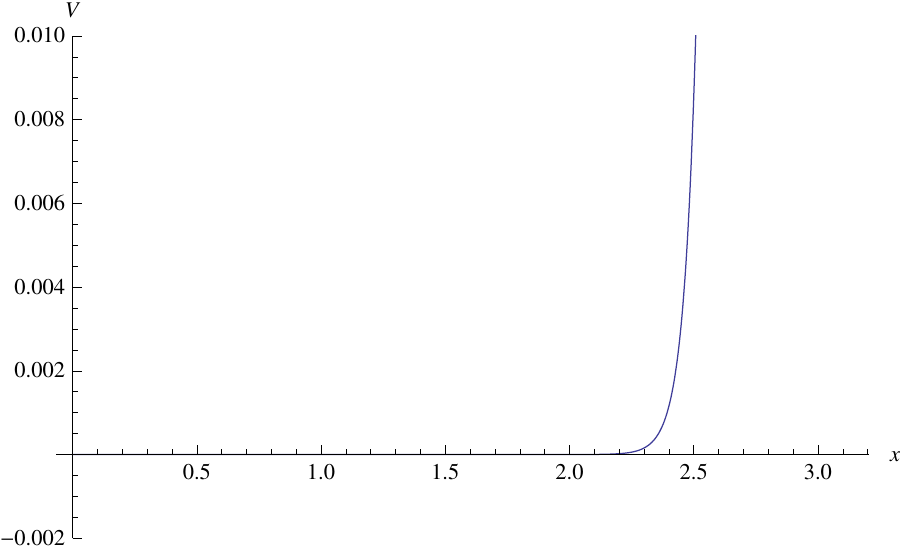}
\hfill
\includegraphics[width=.3\textwidth]{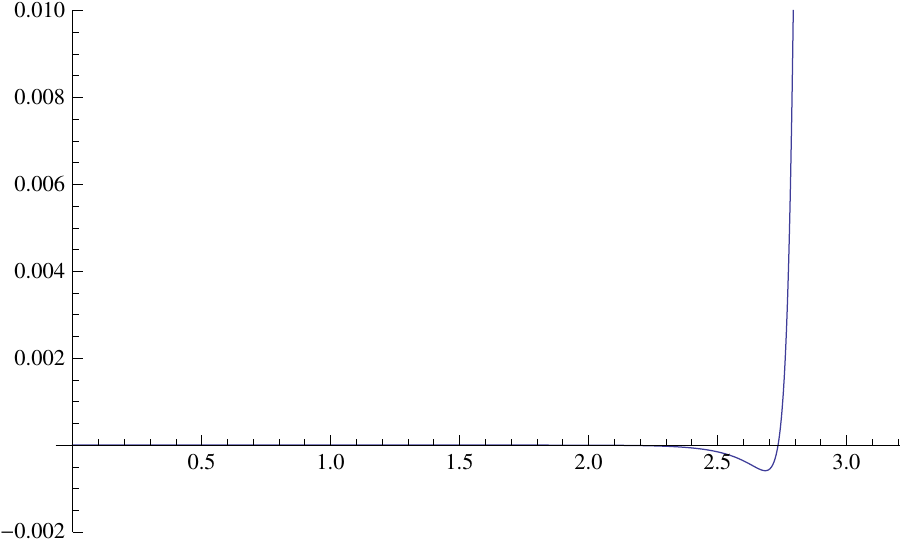}
\hfill
\includegraphics[width=.3\textwidth]{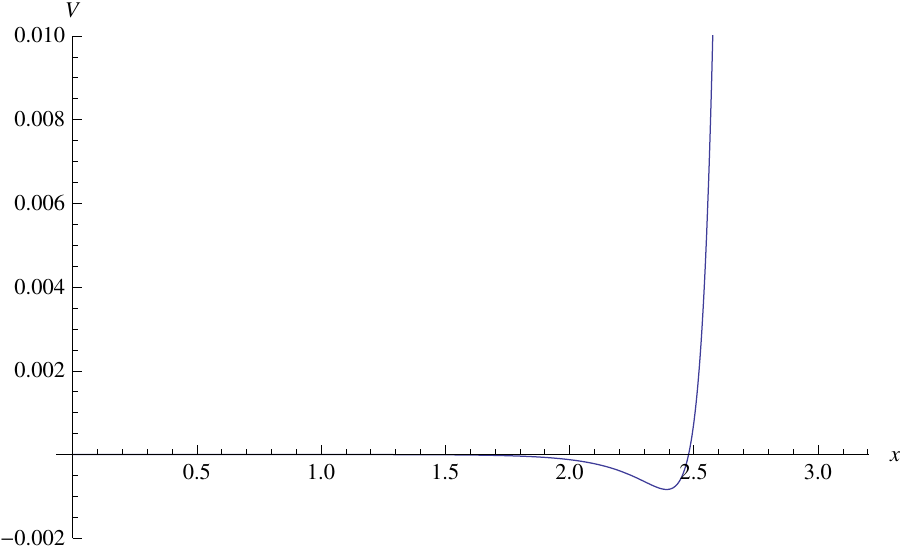}
\caption{\label{fig:Vweak} The classical potential (\ref{affinexy1}) at weak coupling, $\kappa = {g^2 \over 2 \pi} = 0.2$ as a function of $x$. LEFT to RIGHT: $T=0.05$, $T=0.1$, $T=0.15$, with $\kappa \over T$ $=4, 2,1.3$, respectively. Notice the exponentially small value of $V$ near $x = 0$ ($\phi=0$) and the suppression for $\phi$ near the edge of the Weyl chamber ($x=\pi)$ at weak coupling. }
\end{figure}

We now contrast the behavior of the potential (\ref{affinexy1}) for $\kappa = 4\pi$ on Fig.~\ref{fig:Vstrong} with $V(\phi)$ at weak  coupling, shown on Fig.~\ref{fig:Vweak}. To this end, we take  $\kappa = {g^2 \over 2\pi} = 0.2$ as a representative   ``weak coupling" value and plot the potential (\ref{affinexy1}) itself   for several values of $T$. This time, we took $T =0.05, 0.1$ and $0.15$; this choice of values of $T$ is, again, motivated by the naive expectation $T_c \sim {\kappa \over 4}$ and by the empirical observation that the   local minimum away from $x=0$ first appears  just below $T \sim 0.1$ for this value of $\kappa$.  We observe that for $T=0.1$ and $0.15$  the classical (i.e., without fluctuations)  probability for $x=0$ ($\phi =0$) is smaller than that for $x \sim 2.5$, the  value of $x$ near the local minimum,  similar to the strong coupling case of Fig.~\ref{fig:Vstrong}.
Thus, our study of  the classical potential  as a worst case scenario, not including $\phi$ and magnetic bion fluctuations (the latter could only decrease the effect of the $W$-boson term), indicates that  the $\Z_2^{(L)}$-symmetry realization may be similar in the weak-coupling regime. The true  behavior of $\phi$ and the $\Z_2^{(L)}$ symmetry, however, can  only be reliably determined after a simulation for the appropriate values  of parameters.

\section{Conclusions and outlook}
\label{conslusionsection}

In this paper, we studied the finite-temperature dynamics of SYM compactified on $\R^3 \times \S^1_L$, at small-$L$, for  temperatures  near the deconfinement transition. The present paper is a logical extension of our previous work on the thermal deconfinement transition in QCD(adj).  Studying the dynamics of this class of  theories is of interest, since they offer an arena where reliable theoretical tools can be used to identify both the perturbative and nonperturbative excitations relevant for the physics of confinement and the deconfinement transition.

The thermal SYM considered here is also an example of where the study of the supersymmetric theory is made significantly more complicated (compared to non-supersymmetric QCD(adj)) by the presence of the Coulomb-branch modulus scalar, still sufficiently protected by supersymmetry at the temperatures of interest.

 In the first part of the paper, we showed that the physics near $T_c \sim {g^2 \over 8 \pi L}$ can be mapped to a two-dimensional dual-Coulomb gas of electric and magnetic charges, where the electric $W$-bosons interact  with the scalar field.
We then studied the dynamics via Monte Carlo simulations, using two different representations of the dual-Coulomb gas partition function. These can be roughly described as an electric-magnetic Coulomb gas and an XY-model with appropriate ``external field" perturbations.  Admitting that performing the simulations for values of the parameters determined by the small-$L$ UV completion (i.e., the 4D gauge theory) of the dual-Coulomb gas is beyond our current abilities, we chose  parameters that enabled us to simulate the system with the resources available to us. At the same time, we attempted to keep important qualitative features intact. With this caveat in mind, 
we found results whose interpretation is as follows. The thermal $\Z_2^{(\beta)}$ center symmetry breaks, while the $R$-symmetry (discrete chiral $\Z_2^{(R)}$) is restored at the deconfinement transition. The transition appears continuous, as in $SU(2)$ QCD(adj), and, as in that theory, the $\Z_2^{(L)}$ center symmetry remains unbroken but is expected to break at even higher temperatures. Thus, we believe that the transition is qualitatively similar to the one in $SU(2)$ QCD(adj) and is also characterized by continuously varying critical exponents. 

Future studies of this theory, as well as the related QCD(adj) with $n_a >1$, could proceed in  the effective 2D description via dual-Coulomb gases (with more resources, one could come closer to the appropriate values of the parameters). Alternatively,   the full 4D gauge theory with sufficiently light adjoint fermions could be studied on the lattice. In fact,  lattice 
studies of QCD(adj) on $``\R^3" \times \S^1_L$, with a particular focus towards its abelianization at small-$L$ have  recently appeared, with promising results \cite{Cossu:2013ora}. Generally, we expect that the interplay of lattice and analytical methods will be crucial in the study of the transition from the small-$L$, weakly coupled abelian, to the large-$L$, strongly coupled nonabelian,  regime. In particular, such studies should be able to verify the results obtained in the  2D effective description both in this paper and  in \cite{Anber:2011gn, Anber:2012ig}.

We end by mentioning a topic that is presently under investigation. The attentive reader will have noticed that while we repeatedly mentioned  neutral bion ``molecules", in our simulations we always worked in terms of the Coulomb-branch modulus field $\phi$, along with its neutral bion-induced potential  $\cosh 2 \phi$. Now,  similar to representing a $\cos \sigma$ potential for a compact scalar field by a Coulomb gas of charged particles interacting by massless-$\sigma$ exchange, it is tempting to consider the $\cosh 2 \phi$ potential as giving rise to a gas of ``particles" interacting via massless $\phi$-exhange (see, e.g., \cite{Shuryak:2013tka}). These ``particles", however, differ in two essential ways from the ones in a usual  Coulomb gas: like-charge ``particles" attract while opposite-charge ``particles" repel; to boot, these ``particles" all have negative fugacity \cite{Poppitz:2011wy, Argyres:2012vv, Argyres:2012ka,Shuryak:2013tka}. Now, ``normal" (i.e., positive fugacity) gases of particles with like-charge attraction are known to have Dyson instabilities. However, in an overall neutral  system, there is no difference between positive and negative fugacity, as the total particle number is even, hence an instability is expected also for a neutral gas of ``neutral bion particles".\footnote{We have, in fact,  simulated charge-neutral  ``neutral bion" gases (even including  their coupling to $W$-bosons and the associated coupling to magnetic bions). An ``instability" is always seen to occur: at finite volume the system exhibits charge separation and positive and negative ``neutral bion particles" each fill half the volume (we allowed at most one particle per site). Upon scaling to the infinite-volume limit, it is seen that charge separation occurs  at any value of $T$.}
  On the other hand, the theory with a $\cosh 2 \phi$ potential is completely sensible---as follows, e.g., from the zero-temperature studies of SYM. In other words, we would like to gain a better  understanding of how (and if) a qualitative picture of the vacuum as a gas of neutral bions, with all the peculiar properties described above, is consistent with a theory with a stable  ground state. Since neutral bions are involved, the answer may involve an  appropriate analytic continuation. We think that it would be interesting to understand this in  detail and hope that this and further studies of SYM may reveal yet unappreciated features.
  
\section*{Acknowledgments}

We acknowledge discussions with Randy Lewis, Tin Sulejmanpa\v si\' c, and Mithat \" Unsal on various topics related to this paper. This work was partially supported by an NSERC Discovery Grant.  S.C. and S.S.-M. were also  supported by Undergraduate Student Research Awards (USRA) from NSERC.

\appendix

\section{Computation of the effective potential on $\mathbb R^2\times \mathbb S^1_L\times \mathbb S^1_\beta$}
\label{Computation of the effective potential}

We recall that the determinant of an operator ${\cal O}$ is given by
\begin{eqnarray}
\mbox{Det} {\cal O}=\prod_\lambda \lambda=\exp\left[\sum_\lambda \lambda\right]\,,
\end{eqnarray}
where $\{\lambda\}$ is the set of the eigenvalues ${\cal O}\psi_\lambda=\lambda\psi_\lambda$. Using the zeta function
\begin{eqnarray}
\zeta(s)=\sum_\lambda \lambda^{-s}\,,
\end{eqnarray}
we find
\begin{eqnarray}
\mbox{Det} {\cal O}=\exp[-\zeta'(s=0)]\,.
\end{eqnarray}
Therefore, the effective potential ${\cal V}$ is given in the case of $\R^2\times \S^1_L\times \S^1_\beta$ by ${\cal V}=\zeta'(0)/(L\beta)$ (for the fermion case) where $\beta=1/T$ is the inverse temperature. 

First, we are interested in evaluating the determinant $[\mbox{Det}{\cal O}]^{-1}$  of the bosonic operator ${\cal O}=D_M^2$ which is the operator of the gauge fluctuations. Notice that this operator is raised to power $-1$ instead of $-1/2$ because we have two degrees of polarization for the gauge field. Thus, we have $[\mbox{Det} {\cal O}]^{-1}=\exp[\zeta'(s=0)]$ and the effective potential is ${\cal V}_{\mbox{\scriptsize boson}}=-\zeta'(s=0)/(L\beta)$. The  eigenvalues $\lambda$ are given by the matrix-valued quantities
\begin{eqnarray}
\lambda_{n,m}=k^2+\left(\omega_n+\Phi^aT^a\right)^2+\left(\Omega_m+\Psi^aT^a\right)^2\,,
\end{eqnarray}
where $\vec k$ is a two-dimensional vector,  $\omega_n=2\pi n/L$ and $\Omega_m=2\pi m/\beta$ where $n,m \in Z$ are the Matsubara frequencies in the $\mathbb S^1_L$ and $\mathbb S^1_\beta$ directions respectively, and $T_a$ are the Lie generators in the adjoint representation. The variables $\Phi$ and $\Psi$ are respectively the third and zeroth components of the gauge potential $A_\mu$, i.e. $\Phi \equiv A_3$, and $\Psi\equiv A_0$. Since the torus $\mathbb T_2=\mathbb S^1_L \times \mathbb S^1_\beta$ is flat, the two fields $\Phi$ and $\Psi$ can always be chosen such that the holonomies $\Omega_L$ and $\Omega_\beta$ commute, i.e. $[\Omega_L, \Omega_\beta]=0$, such that the effective potential is minimized. In turn, we can take both $\Phi$ and $\Psi$ to lie along the third color direction without loss of generality. In addition, we can choose a gauge in which both $\Phi$ and $\Psi$ are independent of $x_0$ and $x_3$. Hence, we find that the Polyakov loops in the spatial and temporal directions read
\begin{eqnarray}
\nonumber
\Omega_L&=&e^{iL\Phi^aT^a}=\mbox{diag}\left(e^{iL\Phi^3},0, e^{-iL\Phi^3}  \right)\equiv \mbox{diag}\left(e^{iL\Phi},0, e^{-iL\Phi}  \right) \,,\\
\Omega_\beta&=&e^{i\beta\Psi^aT^a}=\mbox{diag}\left(e^{i\beta\Psi^3},0, e^{-i\beta\Psi^3}  \right)\equiv \mbox{diag}\left(e^{i\beta\Psi},0, e^{-i\beta\Psi}  \right)\,,
\end{eqnarray}
where we have suppressed the color superscript to reduce notational clutter. 
 
According to the above description, the determinant of ${D_M^2}$ is given by
\begin{eqnarray}
\zeta(s)=\int \frac{d^2 k}{\left(2\pi\right)^2}\sum_{n,m}\mbox{tr}_a\left[\frac{1}{\left[k^2+\left(\omega_n+\Phi^3T^3 \right)^2+\left(\Omega_m+\Psi^3T^3 \right)^2\right]^s}\right]\,,
\end{eqnarray}
where $\mbox{tr}_a$ denotes the trace in the color space. Performing the trace we find
\begin{eqnarray}
\zeta(s)=2\int \frac{d^2 k}{\left(2\pi\right)^2}\sum_{n,m}\frac{1}{\left[k^2+\left(\omega_n+\Phi \right)^2+\left(\Omega_m+\Psi \right)^2\right]^s}\,.
\end{eqnarray}
Next, we carry out the $k$ integration to obtain
\begin{eqnarray}
\nonumber
\zeta(s)&=&\frac{2}{4\pi (s-1)}\sum_{n,m}\frac{1}{\left[\left(\omega_n+\Phi \right)^2+\left(\Omega_m+\Psi \right)^2\right]^{s-1}}\\
&=&\frac{2}{4\pi(s-1)}\left(\frac{\beta}{2\pi}\right)^{2s-2}\sum_{n,m}\frac{1}{\left[\left(\frac{n\beta}{L}+\frac{\Phi\beta}{2\pi} \right)^2+\left(m+\frac{\beta\Psi}{2\pi} \right)^2\right]^{s-1}}\,.
\end{eqnarray}
Now, we perform the sum over $m$ by making use of the identity
\begin{eqnarray}
\nonumber
\sum_{m=-\infty}^{\infty}\frac{1}{\left[(m+a)^2+c^2\right]^s}=\frac{\sqrt{\pi}}{\Gamma(s)}|c|^{1-2s}\left[\Gamma\left(s-\frac{1}{2}\right) +4\sum_{p=1}\left(\pi p |c|\right)^{s-1/2}\cos\left(2\pi p a\right)K_{s-1/2}(2\pi p |c|) \right]\,,\\
\label{main identity}
\end{eqnarray}
to find 
\begin{eqnarray}
\nonumber
\zeta(s)&=&\frac{2}{4\pi(s-1)}\left(\frac{\beta}{2\pi}\right)^{2s-2}\frac{\sqrt{\pi}}{\Gamma(s-1)}\left\{\sum_{n=-\infty}\frac{\Gamma\left(s-\frac{3}{2}\right)}{\left[\epsilon^2+\left(\frac{n\beta}{L}+\frac{\Phi\beta}{2\pi}\right)^2 \right]^{s-3/2} }\right.\\
\nonumber
&&\left. +4\sum_{n=-\infty, p=1}\left(\pi p\right)^{s-3/2}\left|\frac{n\beta}{L}+\frac{\Phi\beta}{2\pi}\right|^{-s+3/2}\cos\left(p\beta \Psi\right)K_{s-3/2}\left(2\pi p\left|\frac{n\beta}{L}+\frac{\Phi\beta}{2\pi}\right|\right)\right\}\,.\\
\end{eqnarray}
Notice that we introduced a small parameter $\epsilon$ in the first term above, where the limit $\epsilon \rightarrow 0$ is to be understood. The sum in this term takes the same form of the identity (\ref{main identity}). Hence, we find
\begin{eqnarray}
\nonumber
&&\sum_{n=-\infty}\frac{1}{\left[\epsilon^2+\left(\frac{n\beta}{L}+\frac{\Phi\beta}{2\pi}\right)^2 \right]^{s-3/2}}=\left(\frac{L}{\beta}\right)^{2s-3}\sum_n \frac{1}{\left[\left(\frac{\epsilon L}{\beta}\right)^2+\left(n+\frac{\Phi L}{2\pi}\right)^2 \right]^{s-3/2}}\\
\nonumber
&&= \left(\frac{L}{\beta}\right)^{2s-3}\frac{\sqrt{\pi}}{\Gamma\left(s-\frac{3}{2}\right)} \left(\frac{\epsilon L}{\beta} \right)^{-2s+4}\left[\Gamma(s-2)+4\sum_{p=1}\left(\pi p \frac{\epsilon L}{\beta}\right)^{s-2}\cos\left(p \Phi L\right)K_{s-2}\left(2\pi p \frac{\epsilon L}{\beta}\right)  \right]\,.\\
\end{eqnarray}
Putting things together, we obtain, apart from a trivial constant,
\begin{eqnarray}
\nonumber
\zeta(s)&=&\frac{\sqrt{\pi}}{2\pi (s-1)\Gamma(s-1)}\left(\frac{\beta}{2\pi}\right)^{2s-2}\\
\nonumber
&&\times\left\{4\sqrt{\pi} \left(\frac{L}{\beta}\right)^{2s-3}\left(\frac{\epsilon L}{\beta} \right)^{-2s+4}\sum_{p=1}\left(\pi p \frac{\epsilon L}{\beta}\right)^{s-2}\cos\left(p \Phi L\right)K_{s-2}\left(2\pi p \frac{\epsilon L}{\beta}\right)\right. \\
\nonumber
&&+\left. 4\sum_{n=-\infty, p=1}\left(\pi p\right)^{s-3/2}\left|\frac{n\beta}{L}+\frac{\Phi\beta}{2\pi}\right|^{-s+3/2}\cos\left(p\beta \Psi\right)K_{s-3/2}\left(2\pi p\left|\frac{n\beta}{L}+\frac{\Phi\beta}{2\pi}\right|\right)\right\}\,.\\
\end{eqnarray}
Finally, we take the derivative of $\zeta$ with respect to $s$ and set $s=\epsilon=0$ to find
\begin{eqnarray}
\nonumber
{\cal V}_{\mbox{\scriptsize boson}}&=&\frac{-1}{L\beta }\zeta'(s=0)\\
\nonumber
&=&-\frac{4}{\pi^2 L^4}\sum_{p=1}\frac{\cos(pL\Phi)}{p^4}-2\sum_{n=-\infty, p=1}\frac{e^{-2\pi p\left|\frac{n\beta}{L}+\frac{\Phi \beta}{2\pi}\right|}}{\pi \beta^3 L p^3}\left(1+2\pi p \left|\frac{n\beta }{L}+\frac{\Phi\beta}{2\pi}  \right| \right)\cos\left(p\beta \Psi\right)\,.\\
\label{perturbative bosonic part}
\end{eqnarray}
The first term in (\ref{perturbative bosonic part}) is the Gross-Pisarski-Yaffe (GPY)  standard result for the effective potential on $\mathbb \R^3\times \mathbb \S^1_L$. The second term in (\ref{perturbative bosonic part}) is the finite temperature correction part to the $\S^1_L$ GPY potential (i.e., the correction due to the presence of an extra $\S^1_\beta$).

Now, we come to the fermionic determinant. One needs to calculate $\mbox{Det} {\cal O}$, where in this case the fermionic operator ${\cal O}=\slashed{D}$ is a non-Hermitian operator. To avoid calculating the spectrum of a non-Hermitian operator, we instead make use of the property $\log \slashed{D}=\frac{1}{2}\log \slashed{D}^2$ to calculate $\mbox{Det} {\cal O}$ with ${\cal O}=\slashed{D}^2=D_M^2-\sigma_{MN} F_{MN}/2$. We notice that $F_{MN}=0$ along the directions of holonomies, and that the fermions have two degrees of freedom. Hence, the fermionic potential is given by ${\cal V}_f=\zeta'(s=0)/(L\beta)$, and we basically need to perform the same calculations we carried out above taking into account the fact that fermions satisfy anti-periodic boundary conditions along the thermal direction, i.e. $\Omega_m=\pi(2m+1)/\beta$. Hence, we have
\begin{eqnarray}
\zeta(s)=\frac{1}{4\pi(s-1)}\left(\frac{\beta}{2\pi}\right)^{2s-2}\sum_{n,m}\mbox{tr}_a\left[\frac{1}{\left[\left(\frac{n\beta}{L}+\frac{\Phi\beta}{2\pi} \right)^2+\left(m+\frac{\beta\Psi}{2\pi} +\frac{1}{2}\right)^2\right]^{s-1}}\right]\,.
\end{eqnarray}
So, the calculations proceed as above with the substitution: $\Psi \rightarrow \Psi+\pi/\beta$. Then, we find that the fermionic potential is given by
\begin{eqnarray}
\nonumber
{\cal V}_{\mbox{\scriptsize fermion}}=\frac{1}{L\beta }\zeta'(0)&=&\frac{4}{\pi^2 L^4}\sum_{p=1}\frac{\cos(pL\Phi)}{p^4}\\
\nonumber
&&+2\sum_{n=-\infty, p=1}(-1)^p\frac{e^{-2\pi p\left|\frac{n\beta}{L}+\frac{ \beta\Phi}{2\pi}\right|}}{\pi \beta^3 L p^3}\left(1+2\pi p \left|\frac{n\beta }{L}+\frac{\beta\Phi}{2\pi}  \right| \right)\cos\left(p\beta\Psi\right)\,.\\
\label{perturbative fermionic part}
\end{eqnarray}
We notice that the first term in (\ref{perturbative fermionic part}) is exactly the negative of the first term in (\ref{perturbative bosonic part}) and corresponds to the cancellation of the $\S^1_L$ GPY potentials due to supersymmetry at zero temperature (this term is insensitive to the $\S^1_\beta$ boundary conditions).

Upon adding the bosonic (\ref{perturbative bosonic part}) and fermionic (\ref{perturbative fermionic part})  parts we obtain the final result
\begin{eqnarray}
\nonumber
V_{\mbox{\scriptsize eff, pert}}&=&{\cal V}_{\mbox{\scriptsize boson}}+{\cal V}_{\mbox{\scriptsize fermion}}\\
\nonumber
&=&-2\sum_{n=-\infty, p=1}\left[1-(-1)^p\right]\frac{e^{-2\pi p\left|\frac{n\beta}{L}+\frac{ \beta\Phi}{2\pi}\right|}}{\pi \beta^3 L p^3}\left(1+2\pi p \left|\frac{n\beta }{L}+\frac{\beta\Psi}{2\pi}  \right| \right)\cos\left(p\beta \Psi\right)\,.\\
\end{eqnarray}
%

\section{The discrete dual Sine-Gordon model and the lattice Aharonov-Bohm interaction}
\label{The discrete dual Sine-Gordon model}

In this appendix, we derive the discrete form of the  $\Theta$-angle used in the simulations. We start with the action used to study  the dual Sine-Gordon model in the continuum, see, e.g., the appendix of Ref.~\cite{Kovchegov:2002vi}, $S=\int d^2x {\cal L}$, where
\begin{eqnarray}
{\cal L}=\frac{1}{2}\left(\partial_x \phi\right)^2+\frac{1}{2}\left(\partial_x \chi\right)^2-i\partial_x\phi\partial_\tau\chi+J_\phi\phi+J_\chi\chi\,,
\end{eqnarray} 
and $J_\phi$ and $J_\chi$ are external currents. The discrete version of the above Lagrangian can be obtained by putting it on a lattice. Since the fields $\phi$ and $\chi$ are the dual of each other, i.e. $\partial_i \phi=\epsilon_{ij}\partial_j \chi$ where $\epsilon_{x\tau}=1$, it is natural to put one of the fields, say $\phi$, on the lattice, and the other on the dual lattice.  We define the forward derivatives as $\partial_x \phi=\phi_{x+\hat 1}-\phi_{x}$, and $\partial_\tau\chi=\chi_{x^*}-\chi_{x^*-\hat 2}$, where $x$ and $x^*$ are points on the lattice and its dual, and $\hat 1$, $\hat 2$ are unit vectors in the direction of the two axis. Now, let us consider the discretization of the different terms. We start with $\int d^2 x\left(\partial_x \phi\right)^2/2$ which takes the discrete form  
\begin{eqnarray}
\int d^2 x\frac{1}{2}\left(\partial_x \phi\right)^2 \rightarrow \frac{1}{2}\sum_{x} \left(\phi_{x+\hat 1}-\phi_{x} \right)^2=\frac{1}{2}\sum_x \left(\phi^2_{x+\hat 1}+\phi^2_x-2\phi_x\phi_{x+\hat 1}\right)\,.
\end{eqnarray}
However, $\sum_x \phi^2_{x+1}=\sum_x \phi^2_{x}$ and hence we find
\begin{eqnarray}
\int d^2 x\frac{1}{2}\left(\partial_x \phi\right)^2 \rightarrow \frac{1}{2} \sum_x \left(2\phi^2_x-2\phi_x\phi_{x+\hat 1}\right)\,.
\end{eqnarray}
Taking the functional derivative w.r.t. $\phi_y$ we find
\begin{eqnarray}
\nonumber
\frac{\delta}{\delta \phi_y}\left[\frac{1}{2} \sum_x \left(2\phi^2_x-2\phi_x\phi_{x+\hat 1}\right)\right]&=&2\delta_{x,y}\phi_x-\delta_{x,y}\phi_{x+\hat 1}-\delta_{x+\hat 1,y}\phi_x \\
&=&2\phi_y-\phi_{y+\hat 1}-\phi_{y-\hat 1}\,.
\label{first term}
\end{eqnarray}

Now, we turn to the term $-i \int d^2 x \partial_x \phi \partial_\tau \chi$ which takes the discrete form
\begin{eqnarray}
\nonumber
-i\int d^2 x  \partial_x \phi \partial_\tau \chi &\rightarrow& -i\sum_x   \left(\phi_{x+\hat 1}-\phi_x\right)\left(\chi_{x^*}-\chi_{x^*-\hat 2}\right)\\
&=&-i \sum_x \phi_x \left(\chi_{x^*-\hat 2}-\chi_{x^*}\right)-i\sum_x \phi_{x+\hat 1} \left(\chi_{x^*}-\chi_{x^*-\hat 2}\right)\,.
\end{eqnarray}
Using $-i\sum_x \phi_{x+\hat 1} \left(\chi_{x^*}-\chi_{x^*-\hat 2}\right)=-i\sum_{x'}\phi_{x'}\left(\chi_{x'^*-\hat 1}-\chi_{x'^*-\hat 1-\hat2}\right)$, where we have used $x'=x+\hat1$ (which in turn implies $x'^*=x^*+\hat1$) we find after changing the dummy variable $x'$ to $x$:
\begin{eqnarray}
-i\int d^2 x  \partial_x \phi \partial_\tau \chi \rightarrow i\phi_x \left(\chi_{x^*}+\chi_{x^*-\hat 1-\hat 2}-\chi_{x^*-\hat 2}-\chi_{x^*-\hat1}\right)\,.
\label{mixed term}
\end{eqnarray}
Taking the derivative of $i\phi_x \left(\chi_{x^*}+\chi_{x^*-\hat 1-\hat 2}-\chi_{x^*-\hat 2}-\chi_{x^*-\hat1}\right)$  w.r.t. $\phi_y$ we obtain
\begin{eqnarray}
\frac{\delta}{\delta \phi_y} \left[i\phi_x \left(\chi_{x^*}+\chi_{x^*-\hat 1-\hat 2}-\chi_{x^*-\hat 2}-\chi_{x^*-\hat1}\right)\right]= i\left(\chi_{y^*}+\chi_{y^*-\hat 1-\hat 2}-\chi_{y^*-\hat 2}-\chi_{y^*-\hat1}\right)\,,
\label{the second term}
\end{eqnarray}
where we have used $\delta_{x,y}=\delta_{x^*,y}$. Then, adding (\ref{first term}) and (\ref{the second term}) we find
\begin{eqnarray}
-(\phi_{x+\hat 1}+\phi_{x-\hat 1}-2\phi_x)+i\left(\chi_{x^*}+\chi_{x^*-\hat 1-\hat 2}-\chi_{x^*-\hat1}-\chi_{x^*-\hat 2}\right)+J_{\phi x}=0\,.
\label{first EOM}
\end{eqnarray}
Repeating the same steps we have
\begin{eqnarray}
\int d^2 x\frac{1}{2}\left(\partial_x \chi\right)^2 \rightarrow \frac{1}{2}\sum_{x} \left(\chi_{x^*+\hat 1}-\chi_{x^*} \right)^2=\frac{1}{2}\sum_x \left(\chi^2_{x^*+\hat 1}+\chi^2_{x^*}-2\chi_{x^*}\chi_{x^*+\hat 1}\right)\,.
\label{chi term}
\end{eqnarray}
Also (\ref{mixed term}) can be written in the form
\begin{eqnarray}
-i\int d^2 x  \partial_x \phi \partial_\tau \chi \rightarrow i\chi_{x^*} \left(\phi_{x}+\phi_{x+\hat1+\hat 2}-\phi_{x+\hat1}-\phi_{x+\hat 2}\right)\,.
\label{mixed term other form}
\end{eqnarray}
Taking the derivative of (\ref{chi term}) and (\ref{mixed term other form}) w.r.t. $\chi_y$ we obtain
\begin{eqnarray}
-\left(\chi_{x^*+\hat1}+\chi_{x^*-\hat1}-2\chi_{x^*}  \right)+i\left(\phi_x+\phi_{x+\hat1+\hat2}-\phi_{x+\hat 1}-\phi_{x+\hat 2} \right)+J_{\chi x^*}=0\,.
\label{second EOM}
\end{eqnarray} 

Next, we use the discrete Fourier transform on a compact lattice of dimensions $N\times N$
\begin{eqnarray}
\left[\begin{array}{c} \phi_x \\\chi_{x^*} \end{array}  \right]=\frac{1}{N^2}\sum_{\vec p=1}^{N} \left[\begin{array}{c} \phi_{\vec p} \\\chi_{\vec p} \end{array}  \right]e^{-\frac{2\pi i}{N}(p_1 x_1+p_2 x_2)}\,.
\label{DFT} 
\end{eqnarray}
Substituting (\ref{DFT}) in (\ref{first EOM}) and (\ref{second EOM}) we obtain
\begin{eqnarray}
\nonumber
&&A\phi_{\vec p}+iB\chi_{\vec p}=-J_{\phi \vec p}\,,\\
&&A\chi_{\vec p}+iB^*\phi_{\vec p}=-J_{\chi \vec p}\,,
\label{EOM in fourier space}
\end{eqnarray}
where 
\begin{eqnarray}
\nonumber
A&=&2-e^{-\frac{2\pi i}{N}p_1}-e^{\frac{2\pi i}{N}p_1}=2-2\cos\left(\frac{2\pi p_1}{N}\right)\\
B&=&1+e^{\frac{2\pi i}{N}(p_1+p_2)}-e^{\frac{2\pi i}{N}p_1}-e^{\frac{2\pi i}{N}p_2}\,.
\end{eqnarray}
Defining $\omega=e^{\frac{2\pi i}{N}p_1}$ we find
\begin{eqnarray}
\nonumber
B&=&1+\omega^{p_1+p_2}-\omega^{p_1}-\omega^{p_2}=\left(\omega^{p_1}-1\right)\left(\omega^{p_2}-1\right)\\
\nonumber
&=&\omega^{p_1/2}\omega^{p_2/2}\left(\omega^{p_1/2}-\omega^{-p_1/2}\right)\left(\omega^{p_2/2}-\omega^{-p_2/2}\right)\\
&=&-4\omega^{(p_1+p_2)/2}\sin\left(\frac{\pi p_1}{N}\right)\sin\left(\frac{\pi p_2}{N}\right).
\end{eqnarray}
The solution of (\ref{EOM in fourier space}) is
\begin{eqnarray}
\left[\begin{array}{c} \phi_{\vec p} \\\chi_{\vec p} \end{array}  \right]=-\frac{1}{A^2+BB^*}\left[\begin{array}{cc} A & -iB\\ -iB^* & A\end{array}\right] \left[\begin{array}{c} J_{\phi\vec p} \\J_{\chi\vec p} \end{array}  \right]\,,
\label{the solution}
\end{eqnarray}
where the denominator reads
\begin{eqnarray}
A^2+BB^*=16\sin^2\left(\frac{\pi p_1}{N}\right)\left[\sin^2\left(\frac{\pi p_1}{N}\right)+\sin^2\left(\frac{\pi p_2}{N}\right)   \right]\,.
\end{eqnarray}
At this stage, we can substitute (\ref{the solution}) into the Lagrangian and  proceed to get the propagator, as is usually done in any textbook on quantum field theory. However, it is much easier to read the propagators directly  from (\ref{the solution}) by inspecting the diagonal component, which gives $\langle \phi(\vec p)\phi(-\vec p) \rangle $, and the off-diagonal component, which gives $\langle \phi(\vec p)\chi(-\vec p) \rangle $. This leaves some normalization factors which can be fixed by comparing with the continuous propagators.  Using the definition (\ref{DFT}) we find
\begin{eqnarray}
\nonumber
\langle \phi(\vec x)\phi(0) \rangle&=&-\frac{1}{N^2}\sum_{\vec p=0}\frac{Ae^{-\frac{2\pi i}{N}(p_1 x_1+p_2 x_2)}}{A^2+BB^*}=\frac{-1}{4N^2}\sum_{p_1=1, p_2=1}^{N-1, N}\frac{e^{-\frac{2\pi i}{N}(p_1 x_1+p_2 x_2)}}{\sin^2\left(\frac{\pi p_1}{N}\right)+\sin^2\left(\frac{\pi p_2}{N}\right)  }\\
&=&\frac{-1}{N^2}\sum_{p_1=1, p_2=1}^{N-1, N}\frac{e^{-\frac{2\pi i}{N}(p_1 x_1+p_2 x_2)}}{4-2\cos\left(\frac{2\pi p_1}{N}\right)-2\cos\left(\frac{2\pi p_2}{N}\right)}\,,
\end{eqnarray}
where we have excluded the strip $p_1=N$ to avoid singularities. Then, the Green's function $G(x_1,x_2)$ is given by multiplying by the normalization factor $2\pi$:
\begin{eqnarray}
G(x_1,x_2)=\frac{2\pi}{N^2}\sum_{p_1=1, p_2=1}^{N-1, N}\frac{e^{-\frac{2\pi i}{N}(p_1 x_1+p_2 x_2)}}{-4+2\cos\left(\frac{2\pi p_1}{N}\right)+2\cos\left(\frac{2\pi p_2}{N}\right)}\,.
\label{discrete log}
\end{eqnarray}

Similarly, we have for $\langle \chi(\vec x)\phi(0) \rangle$
\begin{eqnarray}
\nonumber
\langle \chi(\vec x)\phi(0) \rangle&=&-\frac{1}{N^2}\sum_{\vec p=0}\frac{iBe^{-\frac{2\pi i}{N}(p_1 x_1+p_2 x_2)}}{A^2+BB^*}\\
&=&\frac{i}{4N^2}\sum_{p_1=1, p_2=1}^{N-1, N}\frac{\sin\left(\frac{\pi p_2}{N}\right)e^{\left[-\frac{2\pi ip_1}{N}\left(-\frac{1}{2}+x_1\right)-\frac{2\pi ip_2}{N}\left(-\frac{1}{2}+x_2\right)\right]}}{\sin\left(\frac{\pi p_1}{N}\right)\left[\sin^2\left(\frac{\pi p_1}{N}\right)+\sin^2\left(\frac{\pi p_2}{N}\right) \right]}\,.
\label{chi phi prop}
\end{eqnarray}
Now, one can insert the normalization factors. We simply multiply (\ref{chi phi prop}) by $-2\pi$ and ignore the factor of $i$ to get
\begin{eqnarray}
\Theta(x_1,x_2)=-\frac{2\pi}{4N^2}\sum_{p_1=1, p_2=1}^{N-1, N}\frac{\sin\left(\frac{\pi p_2}{N}\right)e^{\left[-\frac{2\pi ip_1}{N}\left(-\frac{1}{2}+x_1\right)-\frac{2\pi ip_2}{N}\left(-\frac{1}{2}+x_2\right)\right]}}{\sin\left(\frac{\pi p_1}{N}\right)\left[\sin^2\left(\frac{\pi p_1}{N}\right)+\sin^2\left(\frac{\pi p_2}{N}\right) \right]}\,.
\label{discrete theta}
\end{eqnarray}
As mentioned in the main text, the infinite-volume lattice version of (\ref{discrete theta}) appears in \cite{Kadanoff:1978ve}. 
%

\section{Monte-Carlo procedure for simulating the dual-Coulomb gas}
\label{Monte-Carlo procedure for simulating the double Coulomb gas}

Here we first describe in some detail the Monte Carlo algorithm for simulating a basic Coulomb gas model (not discussed in this paper), and go on to discuss how this algorithm can be generalized to  the simulation of the dual-Coulomb gas discussed in section \ref{Simulations of the Coulomb gas}.

\subsection{Coulomb gas lattice models}

We are interested in Coulomb gas models on the lattice with Hamiltonians of the following form:
\begin{equation}
    \label{Hamiltonian}
	\mathcal{H} = E_c\sum_a q_a^2 -\zeta\sum_{\{a<b\}}q_a q_b G_{ab},
\end{equation}
where $G_{ab}$ is the Green's function for the Coulomb interaction on a two-dimensional lattice, as given in (\ref{discrete log}), $E_c$ is the `core energy', while $\zeta$ represents the strength of the Coulomb interaction. In (\ref{Hamiltonian}), the sum is over all $N^2$ lattice sites labeled by $a, b...$ and $q_a \in \{-1,0,1\}$ is the charge at the $a$-th site.

We implement the Coulomb gas models  by defining a `Lattice' class. Lattice objects store the state of the system in lists of the locations of positive charges, negative charges, and empty sites on the lattice. These lists are updated after every iteration of the Monte-Carlo procedure.

Data is obtained by performing a number of sweeps of a Lattice object through a range of temperatures (taking measurements at the end of every sweep), where a sweep consists of $N^2$ Monte-Carlo iterations. The algorithm described below is inspired by \cite{Cohen}.

At every Monte-Carlo iteration, one of three things can happen: creation of a neutral pair (attempted with probability $P$), annihilation of a neutral pair (attempted with probability $P$), and diffusion of a single charge (attempted with probability $1-2P$). For the Monte-Carlo procedure to satisfy detailed balance, we require that 
\begin{equation}\label{DB}
p_{ij}\pi_i = p_{ji}\pi_j,
\end{equation}
where $p_{ij}$ is the probability for state $i$ to transition to state $j$ in a single step, and 
\begin{equation}\label{GCP}
\pi_i = {1\over{{\cal Z}_{\text{grand}}}}e^{-\beta E_i}
\end{equation}
is the probability of state $i$ in the grand canonical ensemble (note that the chemical potential $\mu = -E_c$ is absorbed into the definition of the configuration energy (\ref{Hamiltonian}).) Given that the transition probability is
 \begin{equation}\label{Prob}
 p_{ij} = q_{ij}A_{ij}\,,
 \end{equation}
where $q_{ij}$ is the probability to consider a trial move from state $i\rightarrow j$ and $A_{ij}$ is the probability that such a move will be accepted, we must carefully constrain the acceptance probabilities in our Monte-Carlo procedure to ensure that detailed balance is satisfied.

\subsection*{Addition and annihilation}
\label{additionAnnihilation}
Addition and annihilation of neutral pairs is attempted with equal probability $P$. Consider a state $j$ which is obtained from state $i$ (with $N_i = N_i^+ + N_i^-$ charges) via creation of a neutral pair of charges, so that $N_j = N_i + 2$. 

When creation is attempted, two sites are randomly chosen from the Lattice object's list of empty sites. Clearly, the number of distinct ways in which to choose two sites from $N_{\text{empty}} = N^2 - N_i$ openings is equal to $\binom{N_\text{empty}}{2} = {1\over{2}}(N^2-N_i)(N^2-N_i-1)$. However, given that the two sites will be assigned different charges upon creation of a neutral pair, we must instead consider the number of distinct \emph{permutations} of 2 out of $N_\text{empty}$ sites, which is equal to $(N^2-N_i)(N^2-N_i-1)$. Thus we obtain the probability that we will attempt a move from state $i\rightarrow j$:
\begin{equation}
q_{ij} = {P\over(N^2-N_i)(N^2-N_i-1)} = {P\over(N^2-N_j+2)(N^2-N_j+1)}.
\end{equation}

Similarly, state $j$ is obtained from $i$ through annihilation of a neutral pair. When annihilation is attempted, one site is randomly chosen from each of the list of positive charges and the list of negative charges. It automatically follows that the number of distinct ways to choose two such charges for annihilation is given by $N_j^+ N_j^- = N_j^2/4$. Thus the probability to attempt a move to state $i$ from $j$ is given by
\begin{equation}
q_{ji} = {4P\over N_j^2} = {4P\over (N_i+2)^2}.
\end{equation}
We then combine (\ref{DB}) and (\ref{Prob}) to arrive at an expression for the acceptance ratio:
\begin{equation}\label{AR}
{A_{ij}\over A_{ji}} = {p_{ij}q_{ji}\over p_{ji}q_{ij}} = {\pi_j q_{ji}\over\pi_i q_{ij}} = {4(N^2-N_i)(N^2-N_i-1)\over(N_i+2)^2}e^{-\beta(E_j-E_i)}.
\end{equation}

To ensure that detailed balance is satisfied, we can define the acceptance probabilities $A_{ij}$ and $A_{ji}$ in the following way. Defining
\begin{align}\label{Adef}
A &\equiv {4(N^2-N_i)(N^2-N_i-1)\over(N_i+2)^2}e^{-\beta(E_j-E_i)}\,,\nonumber\\ 
A^{-1} &= {N_j^2\over 4(N^2-N_j+1)(N^2-N_j+2)}e^{-\beta(E_i-E_j)}\,,
\end{align}
we make the choice
\begin{align}\label{ARs}
A_{ij} &= \text{min}(1,A)\text{ for creation}\,,\nonumber\\
A_{ji} &= \text{min}(1,A^{-1})\text{ for annihilation.}
\end{align}

\subsection*{Diffusion}
\label{diff}
Particle diffusion is attempted with probability $1-2P$. Diffusion consists of choosing a random charge (positive or negative), and selecting a nearest neighbour of that charge. If the neighbouring site is unoccupied, then diffusion is attempted. It is straightforward to see that diffusion is a manifestly symmetric process: if state $j$ is reached via diffusion of one particle from state $i$, then it follows that 
\begin{equation}
q_{ij} = {1-2P\over{4 N_i}},
\end{equation}
while the reverse process has the trial probability
\begin{equation}
q_{ji} = {1-2P\over{4 N_j}} = q_{ij}.
\end{equation}
As a result, the choice
\begin{equation}
A_{ij} = \text{min}(1,e^{-\beta(E_j-E_i)})\text{ for diffusion}
\end{equation}
suffices to guarantee detailed balance.

\subsection{Dual-Coulomb gas simulation}

We now go on to describe the simulation of the dual-Coulomb gas (\ref{double C gas with phi field}) on the lattice. We perform the simulation using only the real part of the action, then use the technique of re-weighting (outlined below) to measure observables for the full system.
The gas contains three components: the $W$-bosons, the magnetic bions, and the $\phi$ field. The $W$s and bions live on the lattice and dual lattice respectively, so particles of different types can have the same indices. For example, a $W$-boson and  bion both at (1,1) is allowed. The $\phi$ field takes on values in the range $(-2\pi/\kappa,2\pi/\kappa)$ at each point on the lattice (recall that $\kappa = g^2/(2 \pi)$ is taken $4\pi$ in our simulations).

In a single Monte Carlo step, one of the three components is first chosen randomly. For the $W$'s and bions, one of addition, annihilation, or diffusion is attempted with probability $P$. The algorithm and acceptance ratios for these steps are outlined in section \ref{additionAnnihilation}. The only modification is that now $N_i$ corresponds to the number of $W$'s on the lattice if a $W$ move is being attempted, or the number of bions on the dual lattice if a bion move is being attempted. Similarly, $N_{empty}$ is now the number of empty sites on the lattice or dual lattice depending on the move being attempted.

The algorithm for a single $\phi$ step moving from state $i \rightarrow j$ is as follows: first select a random site on the lattice, then choose a random number $\phi_{new} \in (-2\pi/\kappa,2\pi/\kappa)$. Next, compute the change in action $\Delta S = S_j - S_i$ and accept the move with probability $p = e^{-\Delta S}$. If the move is accepted, assign the value $\phi_{new}$ to the selected site. This is the standard metropolis algorithm.

In both the $W$ and $\phi$ moves, care must be taken when computing the change in action to account for the $\phi$-dependent $W$ fugacity. In particular, when a change in $\phi$ is attempted at a site where a $W$ boson is located, then the change in action must include the change in fugacity of this $W$.
\subsection{Re-weighting}

As mentioned above, the action used during the simulation is the real part of the full action (\ref{double C gas with phi field}). This means that states are generated with probability proportional to 

\begin{equation}
\label{prob}
p = e^{-S_{\text{Re}}}, 
\end{equation}
where $S = S_{\text{Re}} + iS_{\text{Im}}$ is the full action of the system. To compute the average of an observable $Q$ of the full system, we use the `Estimator' equation:

\begin{equation}
\langle Q \rangle = {\sum_{i} {p_i}^{-1} e^{-S_i} Q_i \over \sum_{j} {p_j}^{-1} e^{-S_j}}\,,
\end{equation}
where the sums are over configurations of the system during the simulation, and $p_i$ is the probability that configuration $i$ is generated during the simulation. Using (\ref{prob}) for $p_i$, this simplifies to

\begin{equation}
\langle Q \rangle = {\sum_{i} e^{-iS_{\text{Im},i}} Q_i \over \sum_{j} e^{-iS_{\text{Im},j}}}\, .
\end{equation}
For the sake of completeness, we recall the imaginary part of the action is
\begin{equation}
S_{\text{Im}} = -4i \sum_{a=0, A=0}^{N, N} q_a q_A \Theta(x_{1,a} - x_{1,A}, x_{2,a} - x_{2,A})\,,
\end{equation}
where $a$ and $A$ denote the locations of the bions and $W$-bosons, respectively, and $\Theta$ is given by (\ref{discrete theta}).

\bibliography{SusyDeconfarxiv}
\bibliographystyle{JHEP}

\end{document}